\documentclass{aa}

\usepackage{txfonts}
\usepackage{tabularx}
\usepackage{subfigure}  % Subfigures
\usepackage{graphicx,amsmath}
\usepackage{amssymb}
\usepackage{color}
\usepackage{natbib}		%for A&A : reimplements the LaTeX \cite command 
\usepackage{url}
\usepackage{multirow}
\usepackage[para,online,flushleft]{threeparttable}   %follow the A&A style
\usepackage{soul,mathabx}
\bibpunct{(}{)}{;}{a}{}{,} % to f
\usepackage{lscape}
\usepackage{adjustbox,textcomp}
\usepackage{dblfloatfix}
\usepackage{caption}

\usepackage{hyperref}   % Including hyper-references
\usepackage{braket}     % package for brakets
\usepackage{upgreek}    % Upright Greek letters
\usepackage{array,multirow,booktabs}

\hypersetup{colorlinks=true,citecolor=blue}

\def\approxinf{%
  \def\p{%
    \setbox0=\vbox{\hbox{$<$}}%
    \ht0=0.6ex \box0 }%
  \def\s{%
    \vbox{\hbox{$\sim$}}%
  }%
  \mathrel{\raisebox{0.7ex}{%
      \mbox{$\underset{\s}{\p}$}%
    }}%
}

\begin{document}

   \title{DREAM I. Orbital architecture orrery}

\author{   
V.~Bourrier\inst{\ref{inst:1}},
O.~Attia\inst{\ref{inst:1}},
M.~Mallonn\inst{\ref{inst:8}}, 		
A.~Marret\inst{\ref{inst:1}}, 		
M.~Lendl\inst{\ref{inst:1}}, 	
P.-C.~Konig\inst{\ref{inst:13},\ref{inst:9}}, 
A.~Krenn\inst{\ref{inst:1},\ref{inst:6}},  
M.~Cretignier\inst{\ref{inst:1}}, 		
R.~Allart\inst{\ref{inst:12}}, 			
G.~Henry\inst{\ref{inst:7}}, 		
E.~Bryant\inst{\ref{inst:15}},		
A.~Leleu\inst{\ref{inst:1}}, 		
L.~Nielsen\inst{\ref{inst:1},\ref{inst:9}},    
G.~Hebrard\inst{\ref{inst:13},\ref{inst:14}}, 
N.~Hara\inst{\ref{inst:1}}, 	
D.~Ehrenreich\inst{\ref{inst:1}},  
J.~Seidel\inst{\ref{inst:1},\ref{inst:10}}, 
L.~dos Santos\inst{\ref{inst:11}}, 		
C.~Lovis\inst{\ref{inst:1}},  		 
D.~Bayliss\inst{\ref{inst:4}}, 					
H.~M.~Cegla\inst{\ref{inst:5},\ref{inst:4}}\thanks{UKRI Future Leaders Fellow}, 	
X.~Dumusque\inst{\ref{inst:1}}, 		
I.~Boisse\inst{\ref{inst:3}},		
A.~Boucher\inst{\ref{inst:2}},		  
F.~Bouchy\inst{\ref{inst:1}},	    
F.~Pepe\inst{\ref{inst:1}},			
B.~Lavie\inst{\ref{inst:1}}, 		
J.~Rey~Cerda\inst{\ref{inst:1}},		  
D.~S\'egransan\inst{\ref{inst:1}},	
S.~Udry\inst{\ref{inst:1}},			
T.~Vrignaud\inst{\ref{inst:1}}	    	
}

\authorrunning{V.~Bourrier et al.}
\titlerunning{DREAM I}
\offprints{V.B. (\email{vincent.bourrier@unige.ch})}

\institute{Observatoire Astronomique de l'Universit\'e de Gen\`eve, Chemin Pegasi 51b, CH-1290 Versoix, Switzerland\label{inst:1}\and
Leibniz Institute for Astrophysics Potsdam, An der Sternwarte 16, D-14482 Potsdam, Germany\label{inst:8}\and
Institut d'astrophysique de Paris, UMR7095 CNRS, Universit\'e Pierre \& Marie Curie, 98bis boulevard Arago, 75014 Paris, France\label{inst:13}\and
European Southern Observatory, Karl-Schwarzschildstr. 2, D-85748 Garching bei M{\"u}nchen, Germany\label{inst:9}\and
Space Research Institute, Austrian Academy of Sciences, Schmiedlstraße 6, A-8042 Graz\label{inst:6}\and
Department of Physics, and Institute for Research on Exoplanets, Universit\'e de Montr\'eal, Montr\'eal, H3T 1J4, Canada\label{inst:12}\and
Center of Excellence in Information Systems, Tennessee State University, Nashville, TN 37209 USA\label{inst:7}\and
Mullard Space Science Laboratory, University College London, Holmbury St Mary, Dorking, Surrey RH5 6NT, UK\label{inst:15}\and
Observatoire de Haute-Provence, CNRS, Universit\'e d'Aix-Marseille, 04870 Saint-Michel l'Observatoire, France\label{inst:14}\and
European Southern Observatory, Alonso de Co\'rdova 3107, Vitacura, Regi\'on Metropolitana, Chile\label{inst:10}\and
Space Telescope Science Institute, 3700 San Martin Drive, Baltimore, MD 21218, USA\label{inst:11}\and
Department of Physics, University of Warwick, Gibbet Hill Road, Coventry CV4 7AL, UK\label{inst:4}\and
Centre for Exoplanets and Habitability, University of Warwick, Gibbet Hill Road, Coventry, CV4 7AL, UK\label{inst:5}\and
Aix Marseille Univ, CNRS, CNES, LAM, 38 rue Fr\'ed\'eric Joliot-Curie, 13388 Marseille, France\label{inst:3}\and
Institut de Recherche sur les Exoplan\`etes, Universit\'e de Montr\'eal, D\'epartement de Physique, C.P. 6128 Succ. Centre-ville, Montr\'eal, QC, H3C 3J7, Canada\label{inst:2}}

   \date{Received XXX; accepted XXX}

% \abstract{}{}{}{}{} 
% 5 {} token are mandatory
 
  \abstract
   {The distribution of close-in exoplanets is shaped by a complex interplay between atmospheric and dynamical processes. The Desert-Rim Exoplanets Atmosphere and Migration (DREAM) program aims at disentangling those processes through the study of the hot Neptune desert, whose rim hosts planets that are undergoing, or survived, atmospheric evaporation and orbital migration. In this first paper, we use the Rossiter-McLaughlin Revolutions (RMR) technique to investigate the orbital architecture of 14 close-in planets ranging from mini-Neptune to Jupiter-size and covering a broad range of orbital distances. While no signal is detected for the two smallest planets, we were able to constrain the sky-projected spin--orbit angle of six planets for the first time, to revise its value for six others, and, thanks to constraints on the stellar inclination, to derive the 3D orbital architecture in seven systems. These results reveal a striking three-quarters of polar orbits in our sample, all being systems with a single close-in planet but of various stellar and planetary types. High-eccentricity migration is favored to explain such orbits for several evaporating warm Neptunes, supporting the role of late migration in shaping the desert and populating its rim. Putting our measurements in the wider context of the close-in planet population will be useful to investigate the various processes shaping their architectures. }


   \maketitle
%
%-------------------------------------------------------------------

\section{Introduction}
\label{sec:intro}

\begin{figure*}
\begin{minipage}[tbh!]{\textwidth}
\includegraphics[trim=0cm 0cm 0cm 0cm,clip=true,width=\columnwidth]{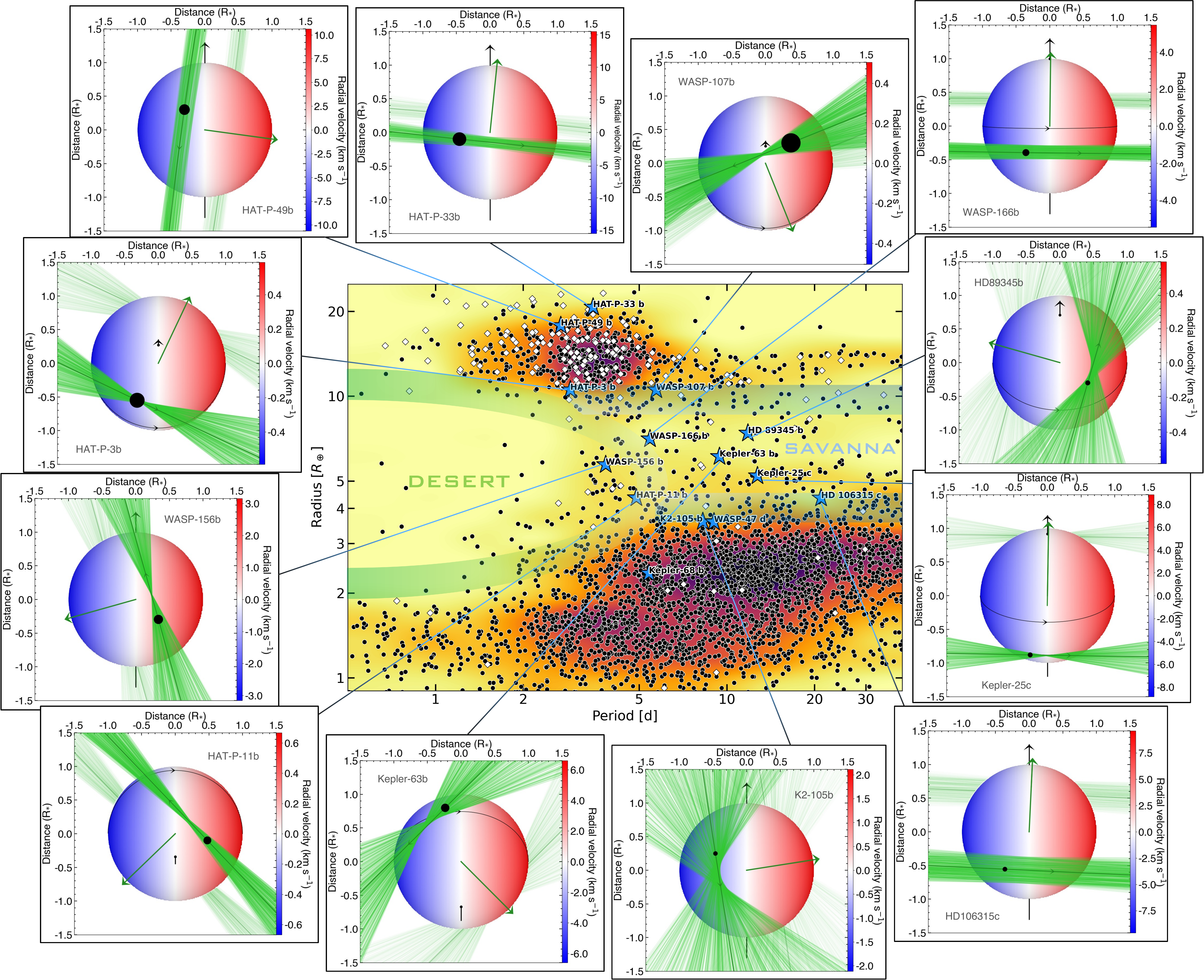}  %plus leger 
\centering
\end{minipage}
\caption[]{Distribution of close-in exoplanets as a function of their radius and orbital period. Green and blue contours show the approximate boundaries of the Neptunian desert and savanna. White squares indicate exoplanets with measured spin--orbit angles. Blue stars highlight planets in our sample, whose projections on the sky plane are displayed for the best-fit orbital architectures. By default, we show the configuration where the stellar spin-axis (shown as a black arrow extending from the north pole) is pointing toward the Earth, except for HAT-P-11 and Kepler-63 for which the degeneracy on $i_\star$ is broken and favors the configuration where their south pole is visible. The stellar equator, plotted as a solid black line, is shown only in systems where the stellar inclination (and thus the 3D spin-orbit angle) is constrained. The stellar disk is colored as a function of its surface RV field. The normal to the planetary orbital plane is shown as a green arrow extending from the star center. The green solid curve represents the best-fit orbital trajectory. The thinner lines surrounding it show orbits obtained for orbital inclination, semi-major axis, and sky-projected spin--orbit angle values drawn randomly within 1 $\sigma$ from their probability distributions. The star, planet (black disk), and orbit are to scale for a given system.}
\label{fig:Systems_view}
\end{figure*}

Exoplanets ranging from half the size of the Earth to twice the size of Jupiter can be found in orbits shorter than 30\,days around their star (Fig.~\ref{fig:Systems_view}). Far from being homogeneous, the distribution of these close-in planets traces the variety of formation and evolution processes that shapes the nature and orbits of exoplanets. One of the main features in this distribution is the hot Neptune desert, which is a lack of planets in between $\sim 2 - 10$ $R_\oplus$ and $P \approxinf 3$ days that has been progressively mapped out over the last decade \citep[e.g.,][]{Lecav2007,Davis2009,Szabo2011,Beauge2013,Lundkvist2016}.

Atmospheric escape is thought to play a major role in sculpting the desert \citep{Lecav2004,Owen2012,Owen2019}, eroding Neptune-size planets into mini-Neptunes or bare rocky cores \citep[e.g.,][]{Ehrenreich_desert2011,Lopez2013,Pezzotti2021}. The extreme X-ray and ultra-violet (UV) stellar irradiation received by these planets can indeed lead to the hydrodynamical expansion of their atmosphere and its dramatic escape into space \citep{Lammer2003,VM2003}. However, it is not clear what stage of their life evaporation affects the different classes of planets. Super-Earths and possibly mini-Neptunes can form in situ \citep{Chiang2013}, while close-in Neptune- and Jupiter-size planets are thought to migrate from their birthplace beyond the ice line \citep[e.g.,][]{Rafikov2006,Dawson2018}. Most studies accounting for long-term atmospheric escape thus assume early atmospheric erosion, kindled during formation or after disk-driven migration \citep[e.g.,][]{Jin2014}. Yet, gaseous planets may avoid the strongest irradiation from the young host star if they migrate long after their formation. Late dynamical migration was indeed proposed as one of the processes shaping the desert \citep{Matsakos2016, Mazeh2016}, but its coupling with atmospheric evolution needs to be explored further \citep{Owen2018,Vissapragada2022}.

Interestingly, the desert opens up into a milder deficit of Neptune-size planets at longer periods and lower irradiation levels (Fig.~\ref{fig:Systems_view}), which we propose to name the Neptunian ``savanna". Runaway core accretion is thought to be responsible for the mass gap between mini-Neptunes and Jupiter-mass planets \citep[e.g.,][]{Mordasini2015,Batygin2016}. However, this formation process occurs beyond the ice line, and it is not clear how much the savanna reproduces the primordial distribution of Neptune-size planets that formed at larger orbital distances. Among the questions that need to be addressed are whether Neptunes migrate into the desert and savanna through different processes, and whether the transition from quiescent to hydrodynamical escape (e.g., \citealt{Koskinen2007}) occurs at the edge of the desert or further out into the savanna, which again requires investigating the coupling between atmospheric and dynamical evolution.

Our ability to disentangle this interplay, and to determine how it depends on stellar and planetary properties, has been limited by a lack of observational constraints. Until recently, only a small number of evaporating planets could be probed through UV spectroscopy, preventing the validation of atmospheric escape models and the derivation of a sample of mass-loss rates \citep[e.g.,][]{Owen2019}. These limitations have recently been alleviated by the rediscovery of helium as a tracer of escape \citep{Oklopcic2018,Spake2018,Allart2018}. Meanwhile, formation and dynamical processes can be traced by the present-day orbital architecture of planetary systems, in particular the angle between the stellar spin-axis and the normal to its planets' orbital plane \citep[see review by][and references therein]{Triaud2018}. 

Disk-driven migration \citep{Goldreich1979,Lin1996,Baruteau2016} is expected to conserve the alignment between the angular momenta of the protoplanetary disk and the star \citep[e.g.,][]{Palle2020b,Zhou2020,Mann2020}, although primordial misalignments can originate from the star (chaotic formation, \citealt{Bate2010,Thies2011,Fielding2015}; internal gravity waves, \citealt{Rogers2012}; magnetic torques, \citealt{Lai2011}; gravitational torques from companions, \citealt{Tremaine1991,Batygin2011,Storch2014}) or the disk \citep{Batygin2012,Lai2014,Zanazzi2018}. The primordial angle between the stellar spin-axis and planetary orbits can then evolve at later stages, in particular through high-eccentricity migration processes (planet--planet scattering, \citealt{Ford2008,Chatterjee2008,Nagasawa2008,Nagasawa2011,Gratia2017}; Kozai--Lidov migration \citealt{Wu2003,Fabrycky2007,Naoz2011,Teyssandier2013}; secular chaos, \citealt{Wu2011}). Measurements of spin--orbit angles for hot Jupiters revealed that many of them live on misaligned orbits, which could naturally result from high-eccentricity migration \citep{Naoz2012,Albrecht2012,Nelson2017}. The dynamical history of smaller and cooler planets is difficult to study due to the lack of alignment constraints, but a fraction of warm Jupiters \citep{Petrovich2016} and warm Neptunes \citep{Correia2020} have moderately eccentric orbits that could trace the circularization phase following high-eccentricity migration. This is particularly interesting for warm Neptunes, whose evaporation could be delayed by a late high-eccentricity migration, allowing them to survive the erosion of their hot siblings that migrated early on. GJ 436 b and GJ 3470 b may be the prototypes of these late Neptunian migrators, as their present location at the edge of the desert, their eccentric and misaligned orbits \citep{Bourrier_2018_Nat,Stefansson2021}, and their ongoing evaporation \citep{Kulow2014,Ehrenreich2015,Bourrier2016,Lavie2017,Bourrier2018_GJ3470b,DosSantos2019,Palle2020,Ninan2020} would be natural outcomes of a late-stage Kozai--Lidov migration \citep{Bourrier_2018_Nat,Attia2021}.

This highlights the interest of extending spin--orbit angle measurements to a wider range of systems. Until recently, most measurements, obtained through transit spectroscopy \citep[e.g.,][]{Queloz2000,cameron2010a}, were limited to hot Jupiters around early-type stars \citep[although see e.g.,][]{SanchisOjeda2012,huber2013,VanEylen2014}. Improvements in spectrographs and analysis techniques \citep[e.g.,][]{Cegla2016} opened the way to build architecture samples for smaller planets \citep{Kunovac2021,Bourrier2021} around cooler stars \citep{Bourrier_2018_Nat}.

In this context, we initiate the DREAM (Desert-Rim Exoplanets Atmosphere and Migration) series, as part of the SPICE DUNE (SpectroPhotometric Inquiry of Close-in Exoplanets around the Desert to Understand their Nature and Evolution) project. Its objectives are to better understand the origins and evolution of close-in planets, in particular the fraction of planets whose history was influenced by high-eccentricity migration. On the observational side, we aim at gathering tracers of atmospheric escape and orbital architecture for exoplanets representative of the different formation and evolution mechanisms. These tracers will inform models developed to describe the upper atmosphere of evaporating planets and to simulate the secular, coupled atmospheric--dynamical evolution of close-in planets. Our studies focus on planets located around and within the Neptunian desert, because it bears the imprint of the evolutionary processes that shaped close-in exoplanets \citep{Mazeh2016,Zahnle2017}. Planets at the rim of the desert are either transitioning into it because they undergo migration and erosion, or arrived at this location at the end of these processes, and are thus ideal targets to study their workings. 

The goal of this first DREAM paper is to determine the orbital architectures of planets sampling the Neptunian desert and savanna and whose past dynamical evolution is of particular interest to the understanding of these features. Our sample consists of 14 planets, most of which (HAT-P-3 b, HAT-P-33 b, HAT-P-49 b, HD 89345 b, K2-105 b, Kepler-25 c, Kepler-63 b, Kepler-68 b, WASP-47 d) were observed in three programs obtained with GIARPS as part of SPICE DUNE. We complete this sample with HARPS, HARPS-N, and CARMENES data of HAT-P-11 b, HD 106315 c, WASP-107 b, WASP-156 b, and WASP-166 b, which were either unpublished yet or published for the purpose of atmospheric characterization. This yields a total of 26 datasets, summarized in Table~\ref{tab_log}. Properties used and derived in our analysis are reported in tables specific to each system in Appendix \ref{apn:sys_prop}.

The paper is structured as follows. Section \ref{sec:photom} presents the long-term and transit photometry that was used to refine host stars rotation and planetary ephemerides. Section \ref{sec:RV} presents the radial velocity (RV) data that were used to refine planetary orbital properties. In Sect.~\ref{sec:CCF_RM}, we describe the spectroscopic transit datasets for each planet and how we analyzed them to derive orbital architectures. Results from these analyses are reported and discussed for each system in Sect.~\ref{sec:pl_sample}. We conclude on this study in Sect.~\ref{sec:conclu}.

\begin{table*}
\tiny
	\caption{Log of RM observations.}
	\label{tab_log} 
	\centering  
%	\resizebox{\linewidth}{!}{
	
	\begin{tabular}{c  c  c c  c   c  c  c}          
		\hline \hline                       
		  Target & Instrument & Night & Program (PI) & S/N  & $N_{\rm data}$ & $t_{\rm exp}$ (s) & RM analysis    \\	
		\hline	
HAT-P-3 b  & HARPS-N  &  30 January 2020 &  OPT19B$\_$8 (V. Bourrier) & 28.3  & 24 & 900 & --  \\
HAT-P-11 b  & HARPS-N  &  13 September 2015 &  OPT15B$\_$19 (D. Ehrenreich) & 45.3  & 87 & 300 &  --  \\
             & HARPS-N  &  01 November 2015 &  OPT15B$\_$19 (D. Ehrenreich) & 40.4  & 52 & 300 & --   \\
             & CARMENES  &  07 August 2017 & OPT17B$\_$026 / 51 (R. Allart)  & 92.7  & 60 & 400 &  --  \\
              & CARMENES  &  12 August 2017 & OPT17B$\_$026 / 51 (R. Allart)  & 102.6  &  55 &  400 &  -- \\
HAT-P-33 b  & HARPS-N & 04 December 2019 & OPT19B$\_$8 (V. Bourrier)   &  20.4  & 56 & 400 & --  \\
HAT-P-49 b  & HARPS-N & 30 July 2020 & OPT20A$\_$8 (V. Bourrier)   & 23.1  & 126 & 180$^{\dagger}$ & --  \\
HD 89345 b  & HARPS-N & 02 February 2020 & OPT19B$\_$8 (V. Bourrier)   &  23.0 & 140 & 180 & --  \\
HD 106315 c  & HARPS & 09 March 2017 & 098.C-0304 (D. Ehrenreich)  & 62.1  & 74 & 350 & \citet{Zhou2018}   \\
            & HARPS & 30 March 2017 & 098.C-0304 (D. Ehrenreich)  & 65.0  & 47 &  600 & \citet{Zhou2018}   \\
            & HARPS & 23 March 2018 & 0100.C-0750 (D. Ehrenreich)  &  82.7  & 47  & 600  & --    \\
K2-105 b  & HARPS-N & 18 January 2020 & OPT19B$\_$8 (V. Bourrier)   & 19.4  & 35 & 600 & --  \\
Kepler-25 c & HARPS-N &  14 June 2019 & OPT19A$\_$8 (V. Bourrier)  &  36.5 & 47  & 500 & --   \\
Kepler-63 b  & HARPS-N & 13 May 2020 &  OPT20A$\_$8 (V. Bourrier)  & 14.9  & 13 & 900 & -- \\
Kepler-68 b  & HARPS-N & 03 August 2019 & OPT19A$\_$8 (V. Bourrier)  &  30.7 & 72 & 300 & -- \\
WASP-47 d  & HARPS-N & 30 July 2021 &  OPT20A$\_$8 (V. Bourrier)  & 20.2  & 19 & 900 & --   \\
WASP-107 b &   HARPS  &    06 April 2014 & 093.C-0474 (A.~H.~M.~J. Triaud)  & 13.8  & 25 & 750$^{\ddagger}$ & --   \\
         &    &     01 February 2018 &  0100.C-0750 (D. Ehrenreich)  & 26.8  & 24 & 800 & --    \\
          &    &   13 March 2018  & 0100.C-0750 (D. Ehrenreich)    & 26.2  & 34 & 800 & --   \\
          &   CARMENES  &    24 February 2018 & DDT.S18.188 (R. Allart)  & 37.7  & 19 & 950 & --   \\ 
WASP-156 b &    CARMENES  &    28 September 2019 & OPT19B$\_$032 / 53 (R. Allart)   & 52.6  & 17 & 1200 & --  \\
         &    &   25 October 2019 & OPT19B$\_$032 / 52 (R. Allart)   & 58.7  & 19 & 1200 & --   \\
          &    &   10 December 2019 & OPT19B$\_$032 / 52 (R. Allart)    &  49.8 & 20 & 890  & --    \\
WASP-166 b &    HARPS  &    14 January 2017 & 098.C-0304 (D. Ehrenreich)   &  52.6  & 71 & 300  &  \citet{Hellier2019}   \\
         &    &  04 March 2017 & 098.C-0304 (D. Ehrenreich)   &  58.7 & 52 & 300$^{\ddagger\dagger}$ & \citet{Hellier2019}    \\
          &    &   15 March 2017   & 098.C-0304 (D. Ehrenreich)   &  49.8 & 65 & 350 & \citet{Hellier2019}    \\
		\hline \hline 
	\end{tabular}
%	}
	\tablefoot{
		 The signal-to-noise ratio (S/N), number of data points $N_{\rm data}$, and exposure time $t_{\rm exp}$ relate to the exposures kept in our analysis. 
		 Time-averaged S/N are given at 550\,nm for HARPS and HARPS-N, and at 785\,nm for CARMENES.
		 For each planet we indicate whether datasets were already published for RM purposes.

		 $^{\dagger}$ The first seven exposures were obtained with $t_{\rm exp} = 360$ s.

		 $^{\ddagger}$ The last four exposures were obtained with $t_{\rm exp} = 900$ s. 

		 $^{\ddagger\dagger}$ The first two exposures were obtained with $t_{\rm exp} = 400$ s. 		 
	}
\end{table*}

%--------------------------------------------------------------------

\section{Photometry}
\label{sec:photom}

\subsection{Stellar rotation}
\label{sec:APT_photom}

\begin{figure}[tbh!]
\includegraphics[trim=0cm 0cm 0cm 0cm,clip=true,width=\columnwidth]{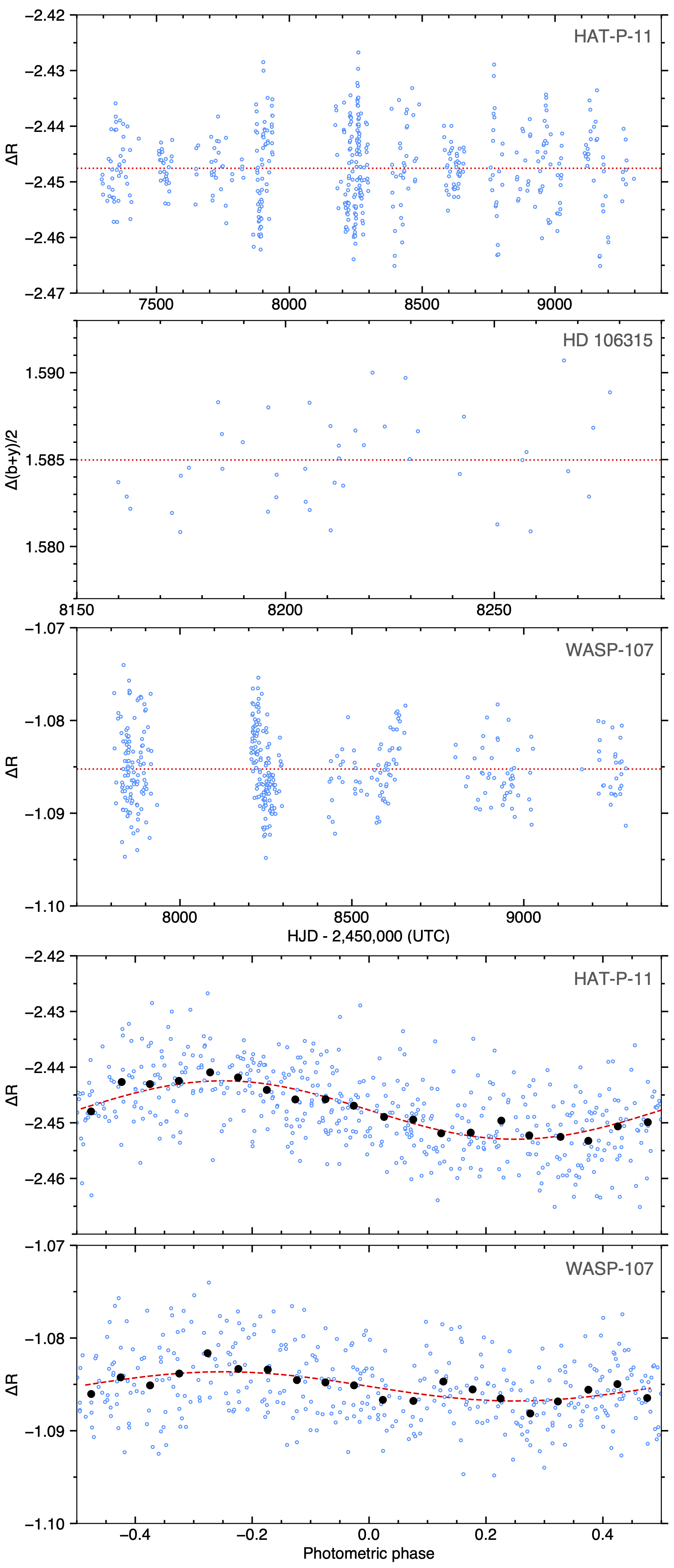}
\centering
\caption[]{\textit{Top panels}: Differential magnitudes of HAT-P-11, HD\,106315, and WASP-107 as a function of Heliocentric Julian Date minus 2,450,000. Red lines show the mean values over the sequence. \textit{Bottom panels}: HAT-P-11 and WASP-107 data phase-folded with the measured rotational modulation periods of 29.5 and 8.7\,d. Red curves show sinusoidal fits to the data, binned into black points.}
\label{fig:Longterm_photom}
\end{figure}

We acquired ground-based photometry of HAT-P-11, HD\,106315, and WASP-107 (Fig.~\ref{fig:Longterm_photom}) to search for starspot brightness modulation that would allow a direct measurement of the stellar rotation periods. We describe below the datasets obtained for each star.

We obtained a total of 43 good photometric observations of HD\,106315 between 2018 February 9 and 2018 June 7 with the T12 0.80~m Automatic Photoelectric Telescope (APT) at Fairborn Observatory. The T12 APT is essentially identical in construction and operation to the T8 0.80~m APT described in \citet{Henry_1999}. Differential magnitudes were computed as the brightness of HD\,106315 minus the mean brightness of the three comparison stars HD\,105374, HD\,105589, and HD\,106965. Like the T8 APT, T12 is equipped with a two-channel photometer that simultaneously measures each star in the Str\"omgren $b$ and $y$ passbands. Typical precision of a single nightly observation is $\approx 0.0015$~mag on good nights. To increase the data precision of our HD~106315 observations, we averaged the brightness in the $b$ and $y$ bands together into a $(b+y)/2$ ``passband".

HAT-P-11 and WASP-107 were observed with the Celestron 14~inch (C14) Automated Imaging Telescope (AIT) at Fairborn. We obtained 497 measurements of HAT-P-11 in seven observing seasons from 2015 to 2021 and 406 measurements of WASP-107 over five seasons from 2017 to 2021. The AIT is equipped with an SBIG STL-1001E CCD camera and uses a Cousins $R$ filter. Differential magnitudes are computed as the brightness of the target star minus the mean brightness of several nearby, constant comparison stars in the same field of view. The typical precision of a single nightly observation with the C14 is $\approx 0.0025$~mag \citep{Fu2021}. Instrumental changes made after the 2017 observing season resulted in brightness shifts of 0.2\% for HAT-P-11 and 0.3\% for WASP-107. Therefore, we normalized the observing seasons of each star to have the same mean. This does not affect our search for rotational modulation with periods of several days or weeks.

We used two approaches to search for rotation periods in our normalized photometric data sets. The method of \citet{Vanicek1971}, based on least-square fitting of sinusoids, measures the reduction in variance of the data over a range of trial frequencies in search of periodic variability \citep[e.g.,][]{Henry2022}. The second method of \citet{Hara2021} consists in searching for quasi-periodic, wavelet-like signals. Results are provided in the subsections relative to each system (Sect.~\ref{sec:pl_sample}).

%-----------------------
\subsection{Transits}

Rossiter--McLaughlin (RM) analyses require a precise ephemeris to identify which exposures are in and out of transit. An imprecise transit window can lead to the contamination of the in-transit signal by the stellar baseline (and conversely), and to positioning the fitted RM model at incorrect orbital phases. This biases the spin--orbit angle measurement \citep[e.g.,][]{CasasayasBarris2021} and can even lead to a nondetection \citep[e.g.,][]{Bourrier2021}. For each studied planet, we thus selected the published ephemeris that provides the most precise mid-transit times $T_0$ propagated to the epochs of our observations. When uncertainties remained too large we conducted ground-based observations of their transit with the STELLA (Sect.~\ref{sec:STELLA}, Table~\ref{tab_overview_STELLA}) and EulerCam (Sect.~\ref{sec:ecam}) photometers, or we reanalyzed TESS and K2 photometry, to derive more precise timings (Sect.~\ref{sec:K2+TESS}). Results of these analysis are given in the planet-specific subsections.   

Eventually, we revise the ephemerides of nine planets (HAT-P-3 b, HAT-P-33 b, HAT-P-49 b, HD\,89345 b, K2-105 b, Kepler-25 b, Kepler-63 b, WASP-47 d, and WASP-156 b) and improve their precision except for HAT-P-3 b and Kepler-63 b. Our final set of ephemerides yields precisions of $\approxinf1$~min on the mid-transit times at the epoch of the RM observations for ten of our targets, and precisions between about $2 - 5$~min for the rest.

%-----------------------
\subsubsection{STELLA}
\label{sec:STELLA}

We used the robotic 1.2 m twin telescope STELLA with its wide-field imager WiFSIP \citep{Strassmeier2004} in the $g$ passband to observe six of our targets over 13 transits (Table~\ref{tab_overview_STELLA}). The observational data were reduced following the description in \citet{Mallonn2015}. In short, the imaging frames were bias- and flat-field-corrected by the STELLA data reduction pipeline. Subsequently, we performed aperture photometry using Source Extractor \citep{Bertin1996}. The photometry software was run for a range of aperture sizes, and we finally employed the aperture size that minimized the photometric scatter in the light curve. We selected the ensemble of photometric reference stars which again minimized the scatter in the light curve.

The photometric uncertainties were enlarged in two steps as detailed in \citet{Mallonn2019b}. First, the uncertainties were adjusted to a reduced $\chi^2$ of unity after an initial transit fit. Second, we employed the time-binning method to calculate the $\beta$ factor \citep{Winn2008,Winn2008err} and applied it as a common factor to all photometric uncertainties of a light curve.

\begin{figure*}
\begin{minipage}[tbh!]{\textwidth}
\includegraphics[trim=0cm 0cm 0cm 0cm,clip=true,width=\columnwidth]{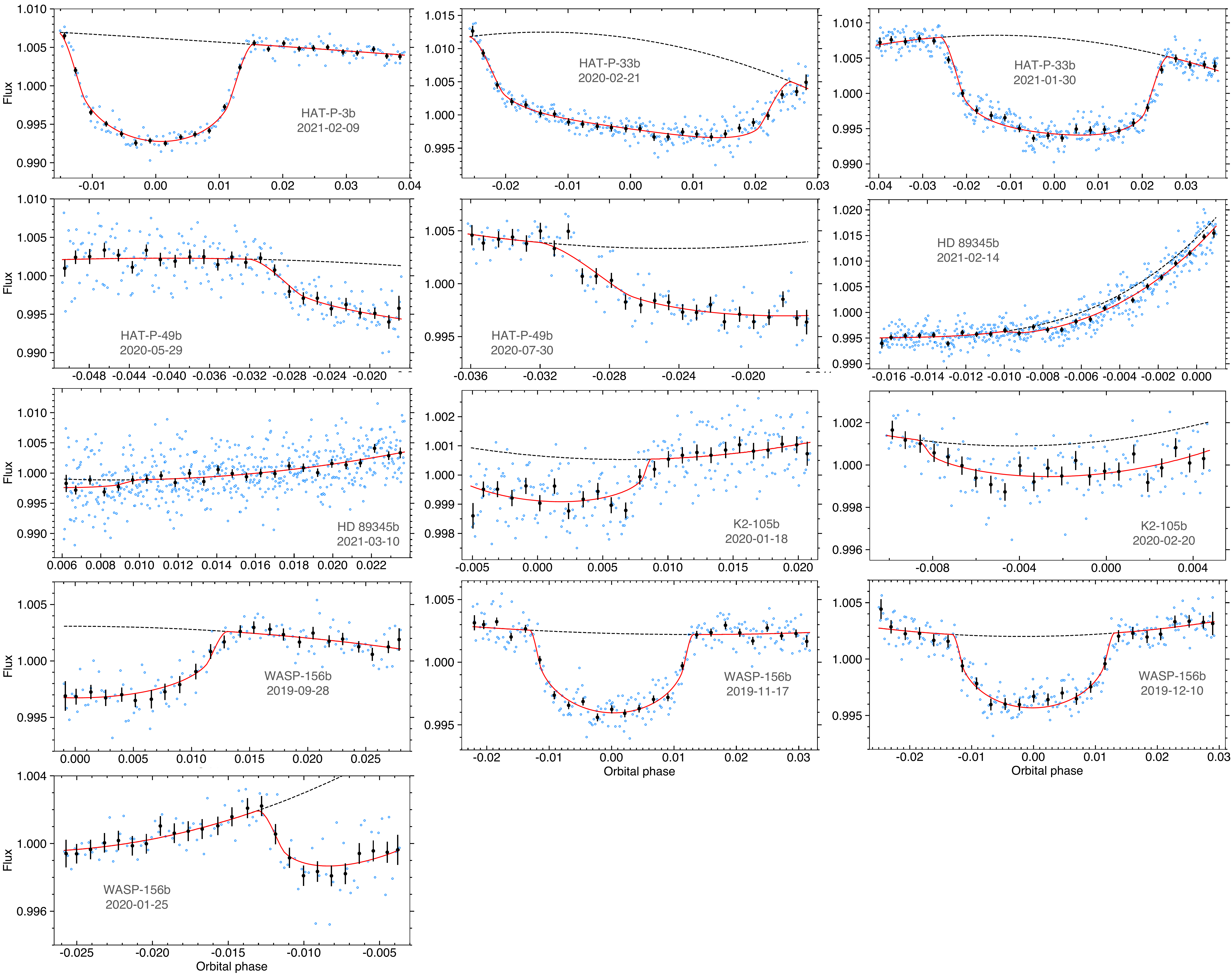}
\centering
\end{minipage}
\caption[]{STELLA lightcurves. Measurements, shown as blue points, were fit with a combined model (red curve) of the transit light curve and detrending polynomials (dashed black curves).}
\label{fig:STELLA_LC}
\end{figure*}

We derived the new ephemerides following the procedure described in \citet{Mallonn2019b}. The STELLA light curves (Fig.~\ref{fig:STELLA_LC}) were fit with the software JKTEBOP \citep{Southworth2005,Southworth2011}. The fit parameters $a/R_\star$ (with $a$ being the semi-major axis of the planet orbit and $R_\star$ the stellar radius), the planetary orbital inclination $i_{\rm p}$, and the planet-to-star radius ratio $k = R_{\rm p}/R_\star$ were held fixed to the values in the planet discovery papers. The limb darkening coefficients of a quadratic limb darkening law were held fixed to tabulated values of \citet{Claret2012,Claret2013}. To account for a smooth trend in the light curves common to ground-based observations, we simultaneously fit a second order polynomial over time together with the transit model for each data set. In the fit, we included the zero epoch transit time of the discovery paper. Free-to-fit values for each light curve were $P$ and $T_0$ of a new ephemeris and the detrending coefficients $c_0$, $c_1$, $c_2$. The epoch of $T_0$ was chosen to minimize the covariance between $T_0$ and $P$. The final ephemerides resulting from this analysis are reported in the tables of each observed planet.

\begin{table*}
\small
\begin{minipage}[h!]{\textwidth}
\caption{STELLA observations. The columns provide the observing date, the exposure time $t_{\rm exp}$, the number of observed individual data points $N_{\rm obs}$, the dispersion of the data points as root-mean-square (rms) of the observations after subtracting a transit model and a detrending function, the $\beta$ factor \citep[see][]{Winn2008,Winn2008err}, and the fitted mid-transit time $T_0$.}
\label{tab_overview_STELLA}
\begin{center}
\begin{tabular}{lcccccl}
\hline
\hline
\noalign{\smallskip}
Planet & Date  &  $t_{\mathrm{exp}}$ (s) & $N_{\mathrm{data}}$ &  rms (mmag) &  $\beta$ & $T_0$ (BJD$_{\rm TDB}$) \\
\hline
\noalign{\smallskip}
HAT-P-3 b    &   09 February 2021     &  60 & 157   & 1.01 & 1.03  &   59255.60337 $\pm$ 0.00038 \\
\hline
\noalign{\smallskip}
HAT-P-33 b   &  21 February 2020   &  30 & 287   & 1.55 & 1.46  &   58901.58213 $\pm$ 0.00097 \\ 
  &  30 January 2021    &  30 & 405   & 1.46 & 1.64  &   59245.55509 $\pm$ 0.00063 \\ 
\hline
\noalign{\smallskip}
HAT-P-49 b   &  29 May 2020    &  10 & 219   & 2.68 & 1.00  &   58999.66484 $\pm$ 0.00165 \\
   & 30 July 2020   &  10 & 123   & 1.67 & 1.06  &   59061.57234 $\pm$ 0.00172 \\
\hline
\noalign{\smallskip}
HD\,89345 b    &  14 February 2021   &   8 & 512   & 2.18 & 1.29  &   59260.64543 $\pm$ 0.00828 \\
    &  10 March 2021  &   5 & 568   & 3.71 & 1.06  &   59284.27525 $\pm$ 0.00971 \\
\hline
\noalign{\smallskip}
K2-105 b     &  18 January 2020   &  50 & 243   & 0.89 & 1.06  &   58867.53425 $\pm$ 0.00368 \\
          & 20 February 2020    &  50 & 139   & 1.14 & 1.00  &   58900.60588 $\pm$ 0.00509 \\
\hline
\noalign{\smallskip}
WASP-156 b   & 28 September 2019   &  50 & 123   & 1.44 & 1.04  &   58755.55641 $\pm$ 0.00140 \\
            &  17 November 2019     &  50 & 233   & 1.07 & 1.00  &   58805.42313 $\pm$ 0.00052 \\
          &  10 December 2019    &  50 & 235   & 1.19 & 1.52  &   58828.44049 $\pm$ 0.00127 \\
           & 25 January 2020     &  50 &  97   & 1.26 & 1.00  &   58874.47570 $\pm$ 0.00104 \\
\hline
\noalign{\smallskip}
\hline                                                                                                     
\end{tabular}
\end{center}
\end{minipage}
\end{table*}

%---------------------------------------------

\subsubsection{EulerCam photometry}
\label{sec:ecam}
We used EulerCam, the CCD imager installed at the 1.2\,m Euler telescope located at La Silla observatory to perform photometric transit observations. The instrument and the associated data reduction procedures are described in detail by \citet{Lendl2012}. In short, relative aperture photometry is performed on the target using an iteratively chosen set of bright nearby references, with the extraction aperture and reference star selection optimized to achieve the minimal light curve rms. EulerCam data are analyzed using a Markov Chain Monte Carlo (MCMC) approach as implemented in CONAN \citep{Lendl2020b}, which allows fitting the system parameters via the jump parameters $R_{\rm p}/R_\star$, $b$ (impact parameter), $T_{14}$ (transit duration), $T_0$, $P$, as well as $\sqrt{e}\sin{\omega}$ and $\sqrt{e}\cos{\omega}$, with $e$ the eccentricity and $\omega$ the argument of periastron, and quadratic limb darkening parameters $u_1$ and $u_2$. The latter are derived with LDCU\footnote{\url{https://github.com/delinea/LDCU}} \citep{Deline2022}.

%---------------------------------------------

\subsubsection{K2 + TESS}
\label{sec:K2+TESS}

Transit observations performed by the Kepler Space Telescope and the Transiting Exoplanet Survey Satellite (TESS) were used to refine the ephemerides of the targets HD\,89345 b and K2-105 b. The data were drawn from the Mikulski Archive for Space Telescopes (MAST) and analyzed with the \texttt{Python} tool \texttt{allesfitter} \citep{Guenther2021}. Data by TESS is provided at 2-min cadence, while data acquired by Kepler is provided at either 1-min cadence (short cadence) or 30-min cadence (long cadence). For targets with Kepler data only available at long cadence, the fitter was set up to interpolate to a 2-min sampling rate when evaluating the transit-model. To account for the effects of limb-darkening, a quadratic limb-darkening law was assumed. The corresponding coefficients were fit jointly with the transit parameters but priored using Gaussian priors to estimates calculated using the \texttt{Python} tool \texttt{limb darkening} \citep{Espinoza2019}. To remove long-term time trends from the data, a spline function was fit to the TESS data, while a Gaussian Process (GP) using a Matern-3/2-Kernel was fit to the Kepler data. The corresponding detrending parameters were fit jointly with the transit parameters. The transit parameters themselves were only constrained by uniform priors. The fit was performed using the nested-sampling option of the \texttt{allesfitter} tool.

%---------------------------------------------

\section{Radial velocities}
\label{sec:RV}

A Keplerian model is required to align cross-correlation functions (CCFs) in the stellar rest frame, where the RM signal is modeled. For each system, we evaluated whether additional existing RV data could be used to refine the planetary orbital properties. We identified three systems for which a reanalysis was warranted: HAT-P-33, HAT-P-49, and HD\,89345.

\subsection{Observations}

Altogether, we used RV measurements obtained with SOPHIE, HIRES, TRES, HARPS, HARPS-N, FIES and APF. We retrieved available data from the DACE platform\footnote{Data \& Analysis Center for Exoplanets (DACE), see \url{https://dace.unige.ch}} and, when relevant, complemented it with data retrieved from the literature (HIRES data from \citealt{hartman2011} for HAT-P-33, HIRES and APF data from \citealt{VanEylen2018} and \citealt{Yu2018} for HD\,89345). 

We also included in our analysis unpublished SOPHIE data for the three systems. SOPHIE is a stabilized \'echelle spectrograph dedicated to high-precision RV measurements on the 193-cm Telescope at Observatoire de Haute-Provence, France \citep{Perruchot2008}. Data were obtained in either one of the two possible observation modes: HR (high resolution) with a resolving power $R=75\,000$ and HE (high efficiency) with $R=40\,000$. All the SOPHIE RVs used here were extracted with its standard pipeline using CCFs \citep{Bouchy2009b} and including CCD charge transfer inefficiency correction \citep{Bouchy2013}. We checked that none of the observations were significantly affected by moonlight pollution.

The discovery paper of HAT-P-33 by \citet{hartman2011} referred to preliminary SOPHIE data but did not include them. Several additional observations of that target were acquired with SOPHIE thereafter. We used that dataset, which includes a total of 20 observations made in HE mode between December 2006 and December 2008. Their median exposure time is 1200~s and their typical $\rm{S/N} = 45$ per pixel at 550\,nm corresponds to a median accuracy of $\pm17$\,m/s.

The discovery paper of HAT-P-49 by \citet{Bieryla2014} included six SOPHIE observations secured in 2012. Six new SOPHIE observations of that target were obtained in $2015 - 2017$ to search for a possible outer companion in the system. We used that full 12-measurement SOPHIE dataset, reextracted in an homogeneous way. It was secured in HR mode with typical exposure times of 1400~s, S/N of 50, and accuracy of $\pm12$\,m/s.

Finally, we used four new SOPHIE observations of HD\,89343 secured in HR mode in Feb--May 2018 as part of the K2 follow-up. Their exposure times range between 1200 and 1800~s and their S/N of about 60 corresponds to $\pm2$\,m/s, except for one of them acquired in bad weather conditions ($\rm{S/N} = 10$).

None of those systems presented significant RV drifts in addition to the planetary signals.

\subsection{RV model and parameterization}
\label{sec:keplerian_model_and_parameterization}

The first element of our RV analysis is to determine an independent zero point for each data set to compensate for the expected offsets in the measurements between the different instruments. The RV time series becomes $RV\:-\:\braket{RV}_{\textsf{SET}}$, where \textsf{SET} represents each spectrograph. Secondly, our model includes a linear background component that accounts for the RV trend induced by a possible long-period companion within the system. The parameters describing a Keplerian orbit are the time of inferior conjunction $T_{0}$, orbital period $P$, eccentricity $e$, argument of periastron $\omega$, and RV semi-amplitude of the stellar reflex motion $K$. 

\citet{Fulton2018b} showed that reparameterizing the fit parameters as $T_{0}$, $\ln P$, $\ln K$, $\sqrt{e}\,\sin\omega$, and $\sqrt{e}\,\cos\omega$ forces $P$ and $K>0$, avoids biasing $K$, prevents the numerical overestimation of $e$, and helps to speed up the MCMC convergence. Our three targets are known to show distinct signs of stellar activity. White noise parameters $\sigma_{\textsf{SET}}$ are thus added to the RV model for each spectrograph to account for the jitter and instrumental noise \citep[e.g.,][]{Gregory2005,Baluev2009}.

Following \citet{Konig2022}, the modeling and fitting method were coded in the \texttt{exoplanet}\footnote{\texttt{\href{https://docs.exoplanet.codes/en/latest/}{https://docs.exoplanet.codes}}} \citep{Foreman2019} and \texttt{PyMC3}\footnote{\texttt{\href{https://docs.pymc.io}{https://docs.pymc.io}}} \citep{Salvatier2016} \texttt{Python} toolkits for Bayesian statistical modeling that focuses on advanced MCMC and variational fitting algorithms.

\subsection{Prior distributions and posterior sampling}
\label{sec:prior_distributions_and_posterior_sampling}

The priors chosen for the 15 parameters of the model and parameterization are presented in Appendix \ref{apn:sys_prop} and Table~\ref{tab:prior_RV}.
For most parameters we use uninformative priors with large bounds. The normal distributions for the priors on $K$, $T_{0}$ and $P$ are centered on the literature values and the respective instrumental median values of the signal for $\braket{RV}_{\textsf{SET}}$. Nonrestrictive uniform prior distributions were chosen for the remaining parameters.

The posterior distribution was sampled using an MCMC algorithm implemented in \texttt{PyMC3} \citep{Salvatier2016}. We ran the \texttt{PyMC3} algorithm with 16 walkers through 5000 iterations. We discarded the first 1000 steps, considering them as tuning draws. The walkers mixed homogeneously and converged before the end of the chains in the same region of the parameter space, around a maximum of the posterior density. % (see trace plots in Fig.~\ref{fig:trace}). 
This indicates that the algorithm has converged properly, and the corresponding corner plots reveal no clear correlations between the model parameters.%(Fig.~\ref{fig:corner_HAT-P-33}, \ref{fig:corner_HAT-P-49}, and \ref{fig:corner_HD89345}) 

\subsection{Updated Keplerian solution}
\label{sec:updated_keplerian_solution}

\begin{figure}[hbtp]
    \centering
    \includegraphics[trim=0.0cm 0.cm 0.0cm 0.2cm, clip=true, width=0.49\textwidth]{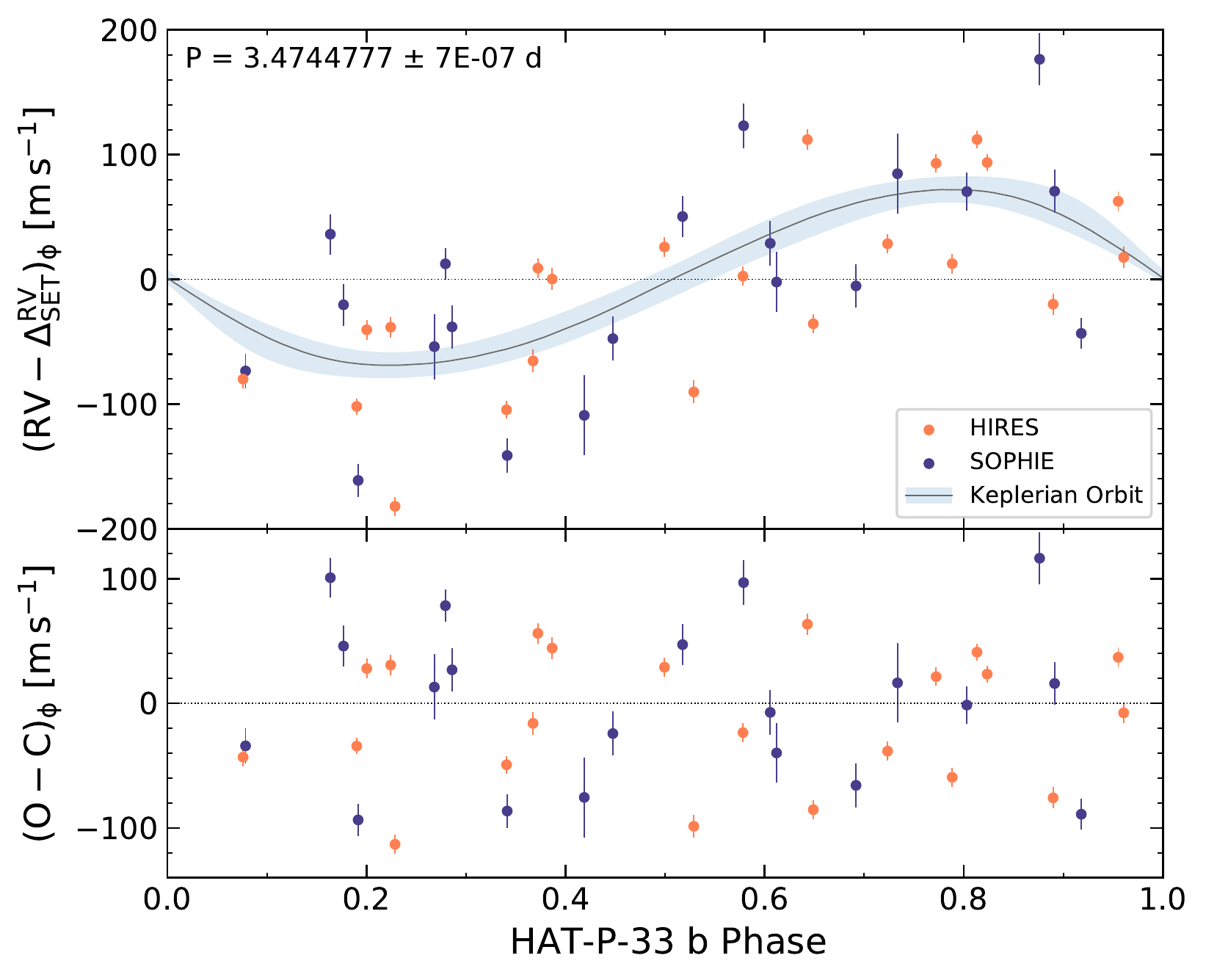}
    \includegraphics[trim=0.0cm 0.cm 0.0cm 0.2cm, clip=true, width=0.49\textwidth]{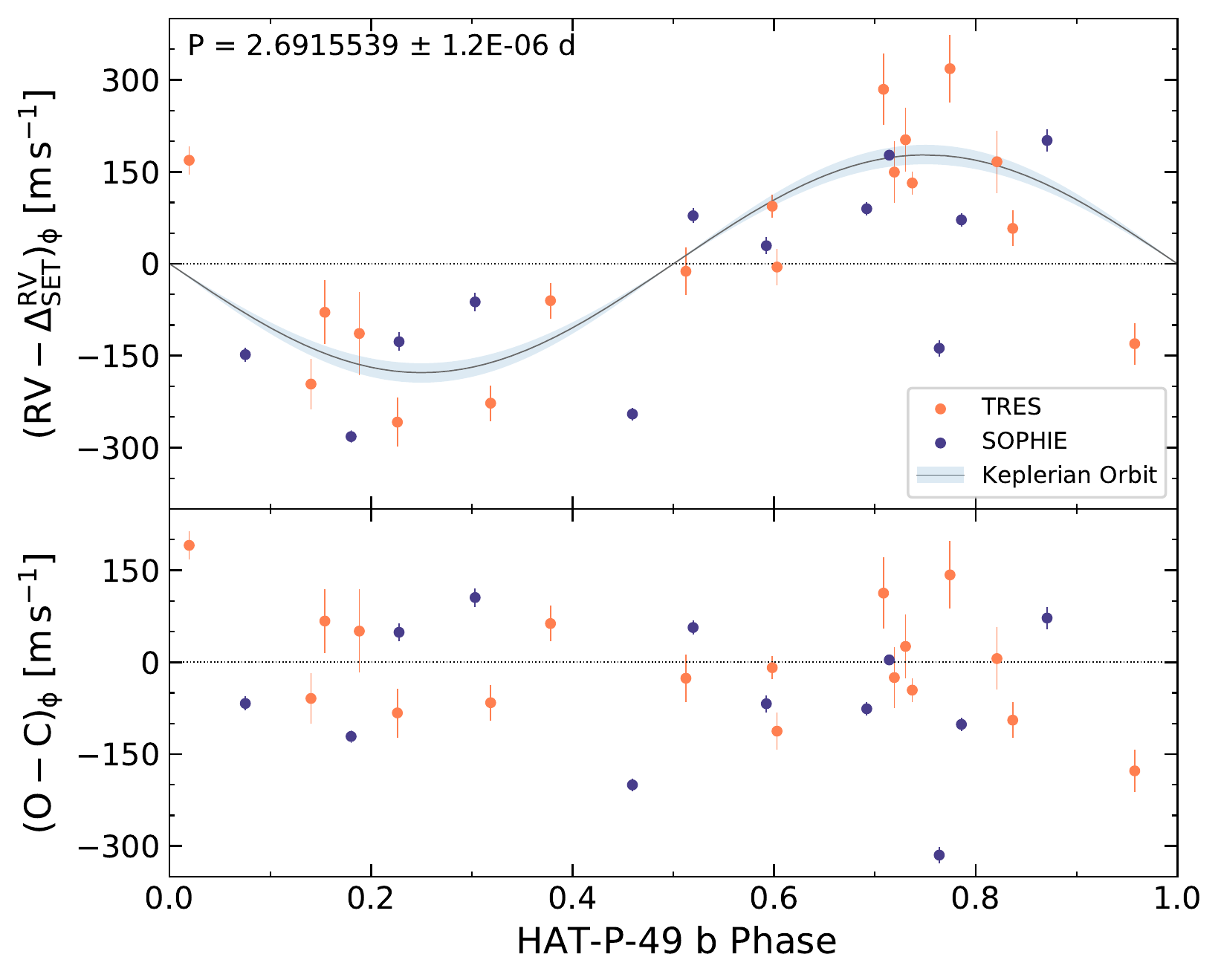}
    \includegraphics[trim=0.0cm 0.cm 0.0cm 0.2cm, clip=true, width=0.49\textwidth]{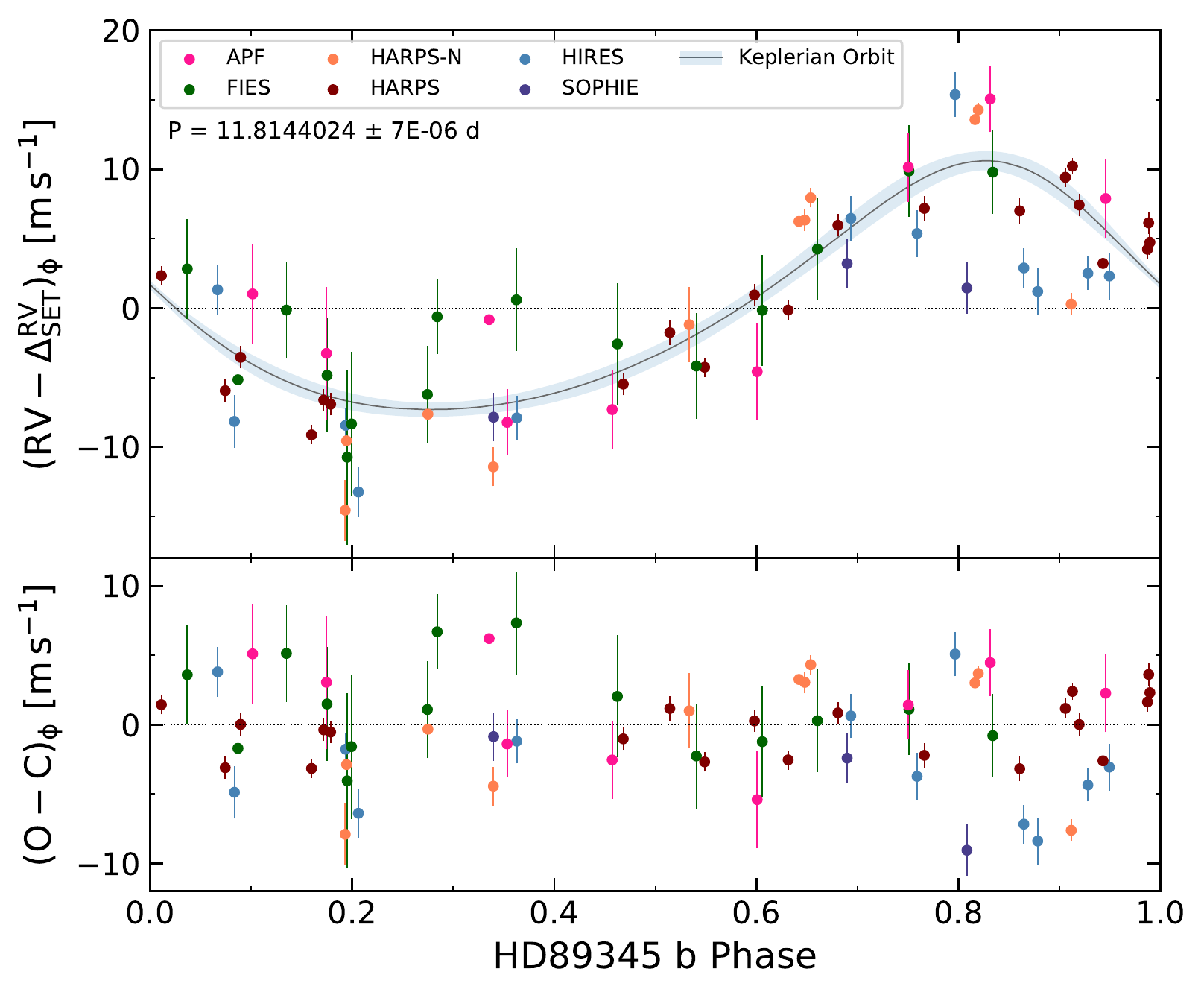}
    \caption{Phase-folded RV evolution of HAT-P-33, HAT-P-49, and HD\,89345. On each figure, the top panel shows the RV variation relative to a fitted offset and linear trend, the solid line and its blue overlay correspond to the orbital solution presented in this study (Sect.~\ref{sec:updated_keplerian_solution}) with its relative 1$\sigma$ uncertainty, respectively, and used to compute the residuals in the bottom panel. The dispersion of the measurements around the Keplerian solutions is larger than the individual estimated error bars due to the stellar jitter, which is not included in the plotted uncertainties.}\label{fig:RV_phase}
\end{figure}

The results of our analysis are discussed in the sections specific to each of the three systems. We adopt as final parameters the median of the probability density function (PDF) from the MCMC samples and set their 1$\sigma$ uncertainties to the $\pm$34.1\% quantiles. The values are reported in Tables~\ref{tab:HATP33} (HAT-P-33), ~\ref{tab:HATP49} (HAT-P-49), and ~\ref{tab:HD89345} (HD\,89345). Our best-fit RV models are shown in Fig.~\ref{fig:RV_phase} along with the corresponding RV residuals, whose quality estimates are given in Table~\ref{tab:quality_RV}.

\begin{table}[h!]
\tiny
\centering
\caption{Quality of RV residuals.}\label{tab:quality_RV}
\begin{tabular}{llccc}
\hline
\hline
\noalign{\smallskip}
    Planet & Instrument &   Dispersion   &   Median error  &  White noise \\
    		   &  		    &  $\mathrm{rms}^{\rm res}$ (m\,s$^{-1}$)  & $\widetilde{\sigma}$ (m\,s$^{-1}$) & $\sigma$ (m\,s$^{-1}$) \\
    \hline
    \noalign{\smallskip}
	HAT-P-33 b & HIRES  &  53.3 &  7.8 & 55$\pm$9  \\
			 & SOPHIE & 64.2 &  17.3 & 64$\pm$11   \\
    \hline
    \noalign{\smallskip}
	HAT-P-49 b &  TRES &  92  &  39.5  &  93$\pm$21   \\
			 & SOPHIE & 129  &  11.9  &  140$\pm$36 \\
    \hline
    \noalign{\smallskip}
	HD\,89345 b &   SOPHIE   &  5.4  &   7.5$\pm$8.8 \\
			& HIRES   &	4.8  & 1.8  &    5.0$\pm$1.7 \\
			& HARPS   &  2.1  &  1.7  &   2.2$\pm$0.5 \\
			& HARPS-N   &  4.4 &  0.8  &   4.7$\pm$1.5 \\
			& FIES   &  3.4  &  0.8  &   2.0$\pm$1.4 \\
			& APF &  3.9  &  3.7  &   3.1$\pm$2.1 \\
\hline
\hline                                                                                                     
\end{tabular}
\end{table}

% FIGURE - RV Phase plots

%---------------------------------------------

\section{CCF transit series}
\label{sec:CCF_RM}

\subsection{CCF reduction}
\label{sec:CCF_reduc}

HARPS \citep{Mayor2003}, HARPS-N \citep{Cosentino2012}, and CARMENES \citep{Quirrenbach2016,Quirrenbach2018,Quirrenbach2020} are fiber-fed spectrographs installed at the ESO La Silla 3.6m\,telescope (Chile), at the 3.6\,m Telescopio Nazionale Galileo (La Palma, Spain), and at the 3.5\,m telescope at the Calar Alto Observatory (Almer\'ia, Spain). Their light is dispersed on 71, 69, and 61 spectral orders over the ranges $383 - 690$\,nm (HARPS and HARPS-N), and $520 - 960$\,nm (only the visible arm of CARMENES is used in this study). The spectral resolution of HARPS and HARPS-N is $\sim$ 2.6\,km\,s$^{-1}$, and that of CARMENES about 3.2\,km\,s$^{-1}$ in the visible. Exposure times were adjusted for each target based on the expected S/N for the stellar flux and RM signal. Exposure indices are counted from 0 throughout the paper.

HARPS and HARPS-N spectra were extracted from the detector images, corrected and calibrated using versions 3.5 and 3.7 of the Data Reduction Softwares (DRS), adapted from the ESPRESSO pipeline \citep{Dumusque2021}. The so-called ``color correction", compensating for the variability of extinction induced by Earth’s atmosphere \citep[e.g.,][]{bourrier2014b,Wehbe2020}, is automatically applied by the DRS using standard spectrum templates of stars with spectral types closest to that of the target. Spectra were then passed through weighted cross-correlation \citep{baranne1996,pepe2002} with numerical masks to compute CCFs. We used a step of 0.82\,km\,s$^{-1}$ to match the HARPS and HARPS-N pixel size and thus limit correlations.

The CARMENES data were reduced with our custom pipeline, ANTARESS, which will be described in detail in a forthcoming publication. We summarize here the main steps. We exclude from the reduction the first spectral order of the WASP-107 dataset, too noisy to be processed, and orders between indices 57 and 60 for all datasets, too strongly contaminated by telluric lines. 2D spectra are first scaled back from flux to count values to avoid amplifying CCF errors due to low count levels at the edges of spectral orders. Polynomial functions, fit to the ratio between each exposure spectrum and the master out-of-transit spectrum, are used to correct for the color effect and low-frequency instrumental variations. ANTARESS then applies a cosmic correction and mask persistent features (bad pixels and telluric emission lines). CCFs are finally calculated through cross-correlation of the corrected 2D spectra with numerical masks, using a step of 1.1 \,km\,s$^{-1}$ to match CARMENES pixel size.   

The CCF$_\mathrm{DI}$ (for ``disk-integrated") produced by the pipelines correspond to the light coming from the entire star. CCF$_\mathrm{DI}$ of individual exposures are aligned by correcting their velocity table for the Keplerian motion of the star, accounting for all planets in the system that induce a measurable drift over the duration of the visit (properties of planetary companions are listed in the tables specific to each system). CCF$_\mathrm{DI}$ outside of the transits are then coadded to build master-out CCFs representative of the unocculted star, which are fit to measure the systemic velocity of the star and align all CCF$_\mathrm{DI}$ into the common stellar rest frame. We emphasize the importance of measuring the systemic velocity in each visit, as it can vary by a few m\,s$^{-1}$ due, for example, to instrumental variations or different color corrections. We used either Gaussian or Voigt models to fit the CCF$_\mathrm{DI}$ and analyze their properties, which are the contrast, full width at half maximum (FWHM), and RV residuals to the Keplerian model.

%---------------------------------------------------

\subsection{CCF corrections}
\label{sec:CCF_corr}

For all datasets obtained with HARPS and HARPS-N, we used the DRS to compute sky-corrected CCFs, exploiting the monitoring of the sky with the second instrument fiber. The only exception is the first transit of WASP-107 b, as the current version of the HARPS DRS does not yet correct for sky contamination in datasets older than 2015. We also note that the CARMENES pipeline performs a different reduction for data obtained with its two fibers, making it difficult to correct one for the other. Using sky-corrected CCFs is a trade-off between an increased white noise and the possible correction of systematics (mainly due to moonlight contamination). To make this decision, we assessed whether sky-correction decreased the dispersion of the out-of-transit CCF properties (Table~\ref{tab:disp_full}). 

\begin{figure}
\includegraphics[trim=0cm 0cm 0cm 0cm,clip=true,width=\columnwidth]{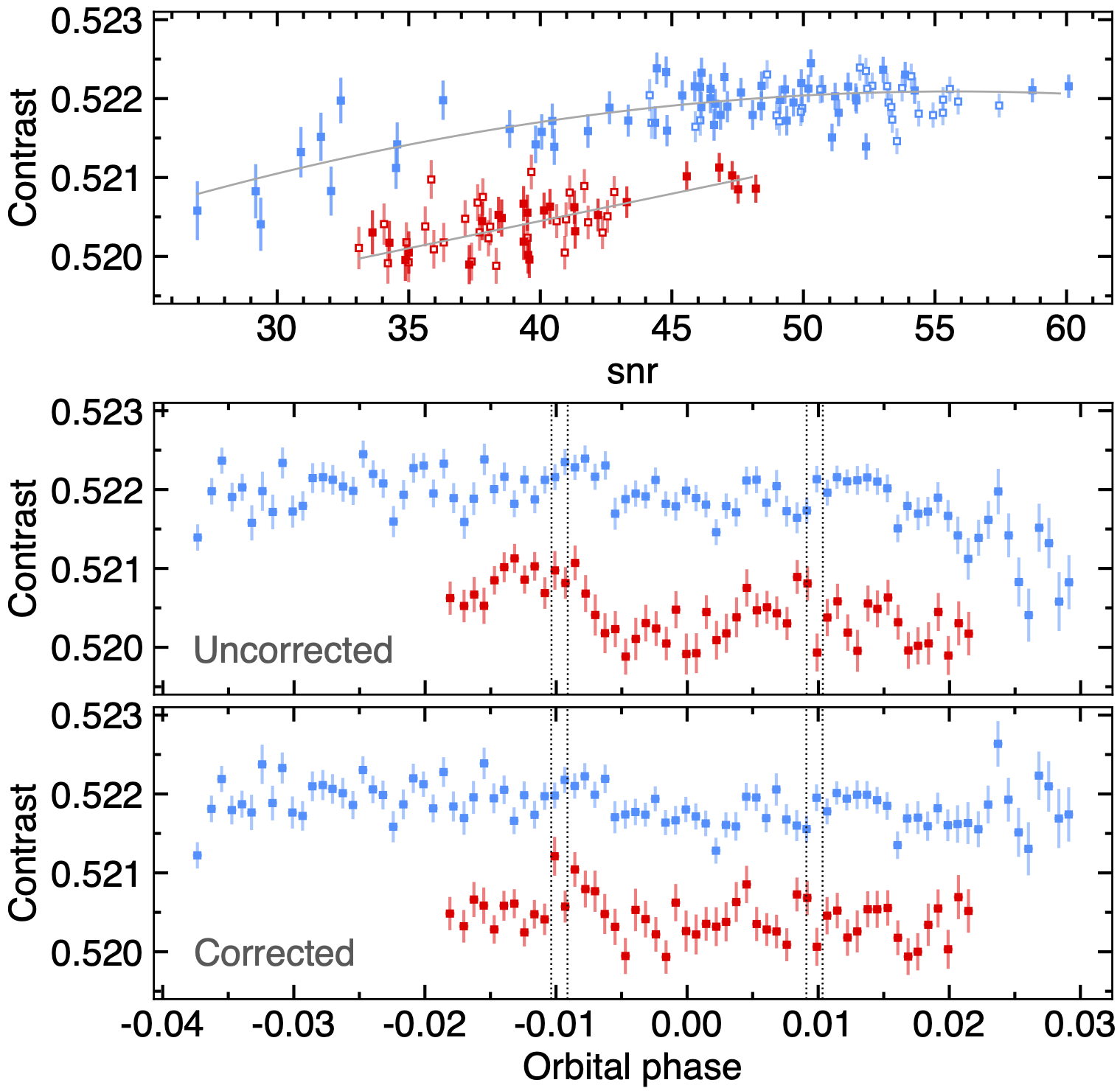}
\centering
\caption[]{Contrast of the HAT-P-11 CCF$_\mathrm{DI}$ in HARPS-N Visit 1 (blue) and 2 (red). \textit{Top panel}: correlation between the contrast and S/N, fitted on the out-of-transit measurements (gray line). In-transit measurements are plotted with empty symbols. \textit{Middle and bottom panels}: contrast as function of orbital phase before and after correction using the best-fit model from the top panel.}
\label{fig:HAT_P11_C_S/N}
\end{figure}

Then, for each dataset, we searched for correlations between the out-of-transit CCF properties and time or S/N ratio. Correlations are identified and corrected following the same approach as in \citet{Bourrier2022}. All corrections are summarized in Table~\ref{tab:disp_full}, and an example is given in Fig.~\ref{fig:HAT_P11_C_S/N}. The origin of S/N correlations is unclear, as they are observed in a broad range of S/N regimes, environmental conditions, and across different instruments. Temporal correlations can be linked to short-term stellar activity, such as the HAT-P-49 dataset (Sect.~\ref{sec:HAT_P49}). However, we note that all CARMENES CCF series had to be corrected for some correlations of their properties with time, which could be due to the impossibility to correct for Moon contamination, to instrumental systematics not accounted for by the pipeline, or to a systematic issue in the standard reduction of the CARMENES 2D spectra.

Within the precision of our data we found the average stellar lines to be symmetrical, so that correcting for variations in one of their properties does not affect the others. Corrections of the RV series allow better aligning CCFs in the stellar rest frame and preventing the creation of P-Cygni like profiles in the planet-occulted CCFs. Corrections of the contrast and FWHM series allow making in- and out-transit CCF profiles more comparable and prevent distortions in the shape of the extracted planet-occulted CCFs. The need for these corrections however depends on the slant of the planet-occulted lines and its depth. For example, the contrast correction shown in Fig.~\ref{fig:HAT_P11_C_S/N} slightly improves the precision of $\lambda$ for HAT-P-11 b but does not change its value, while without a similar correction for HAT-P-33 b, $\lambda$ is changed by $\sim$3$^{\circ}$. We also emphasize the importance of measuring the unocculted stellar flux both before and after the transit to derive a correlation model for the entire visit and correct the in-transit data.

For each HARPS and HARPS-N dataset, we compared the CCF series calculated with two types of masks. First, a set of ``improved" masks (\citealt{Bourrier2021}) with better line selection and weighting (instead of the relative line depth, weights are computed based on the Doppler content of the stellar lines, \citealt{Bouchy2001}), which are now used by default in the ESPRESSO, HARPS, and HARPS-N DRS. The mask closest in spectral type to the target star is automatically chosen by the DRS among a representative set of F9, G2, G8, G9, K2, and K6 stars (as well as M dwarfs, unused here). Second, ``custom" masks that are built like the improved masks but directly using the spectrum of the target star to define the mask lines and their weights (\citealt{Cretignier2020a}). We then compared the dispersion of the out-of-transit CCF properties to determine which mask yields the most stable and precise CCFs. Our goal is both to select the optimal CCF mask for each dataset and to assess the suitability of the new standard DRS masks to any type of star. Standard CCF masks are not available for CARMENES data other than M dwarfs, so we built custom masks based on the CARMENES spectra.

Using masks specific to the target star typically decreases the uncertainties on all CCF properties by $10 - 20$\%. In terms of dispersion, we see a clear difference between spectral types. Custom masks increase the depth and width of CCFs for our F-type targets and can improve (typically by $\sim 10 - 30$\%) or degrade the precision of their properties. For all our G- and K-type targets, custom masks decrease the depth and width of the CCFs and substantially improve the stability of their properties. The diminution in dispersion is noticeably stronger for K-type stars ($\sim 20 - 50$\%, up to 80\% for the FWHM) compared to G-type stars ($\sim 10 - 30$\%). When considering contrast, FWHM, and RV together, we find that it is worth using custom masks for all our targets except Kepler-25. A custom mask was selected by default for Kepler-63, as there are not enough out-of-transit exposures to measure dispersions. Our comparison suggests that a finer sampling of the DRS masks as a function of subspectral types and possibly other stellar properties (age, metallicity, etc.) is needed for at least G- and K-type stars. A similar investigation of the CCF time series of M-type stars is required to assess whether they can be improved as well. 
We note that, in several cases, the custom masks decrease the dispersion of the CCF properties by reducing its correlation with the S/N. A possible origin for this correlation may thus be found in the (dis)agreement between the mask lines and the actual stellar lines. We highlight that one of the advantages of using CCFs rather than template matching is the possibility to identify and correct for such variations in the average line properties.

%---------------------------------------------

\subsection{Rossiter-McLaughlin Revolutions analysis}

\begin{figure*}
\begin{minipage}[tbh!]{\textwidth}
\includegraphics[trim=0cm 0cm 0cm 0cm,clip=true,width=\columnwidth]{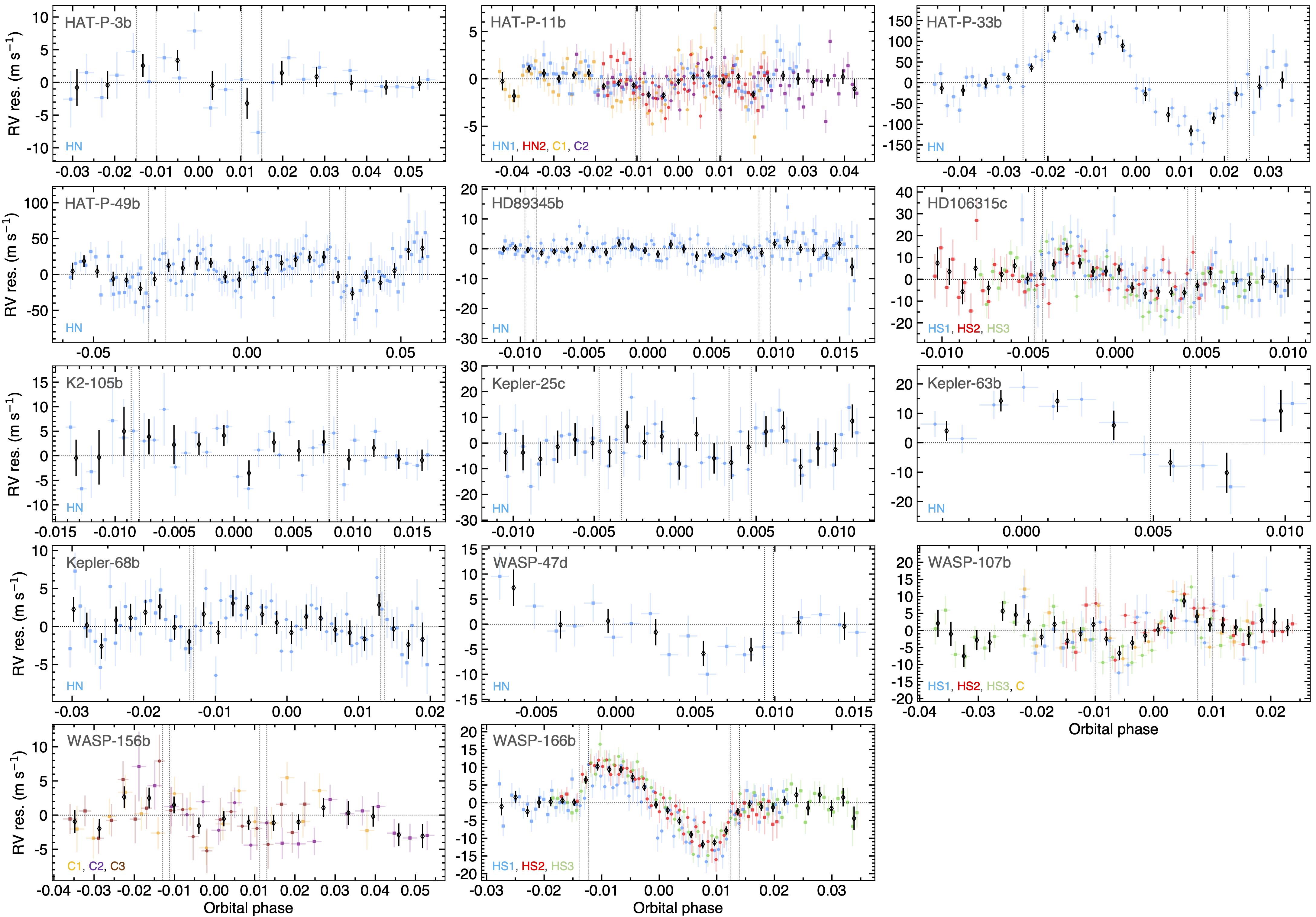}
\centering
\end{minipage}
\caption[]{RV residuals from the stellar Keplerian motion, phase-folded over the period of the transiting planet. Vertical lines mark transit contacts. In- and out-of-transit points are respectively plotted as disks and squares, with different colors for different transits and instruments (see definitions in Table~\ref{tab:quality_RV}). Measurements were binned as black points to enhance the visibility of the RV signal.}
\label{fig:class_RM}
\end{figure*}

Figure \ref{fig:class_RM} shows the final RV time series derived from the CCF$_\mathrm{DI}$. A visual inspection shows that the RV anomaly (due to the occultation of the rotating stellar photosphere by the planet, and the resulting distortion of the stellar lines) is barely detectable in many datasets, and that several of them show instrumental or stellar RV jitter, which is detrimental to the analysis of the disk-integrated RVs. Besides its limited precision, the classical RV technique can be subject to biases associated with the shape of the occulted stellar line profile and its variations along the transit chord \citep[e.g.,][]{Cegla2016,Bourrier2017_WASP8,Bourrier2022}, or with spurious features undetectable in the CCF$_\mathrm{DI}$ (Grouffal et al., under review). We thus analyze all datasets using the Rossiter-McLaughlin Revolutions (RMR) method, which avoids these biases through the direct analysis of the planet-occulted starlight rather than the disk-integrated starlight. The complete description of this approach can be found in \citet{Bourrier2021}.

\subsubsection{Extraction of planet-occulted CCFs}

\begin{figure*}
\begin{minipage}[tbh!]{\textwidth}
\includegraphics[trim=0cm 0cm 0cm 0cm,clip=true,width=0.95\columnwidth]{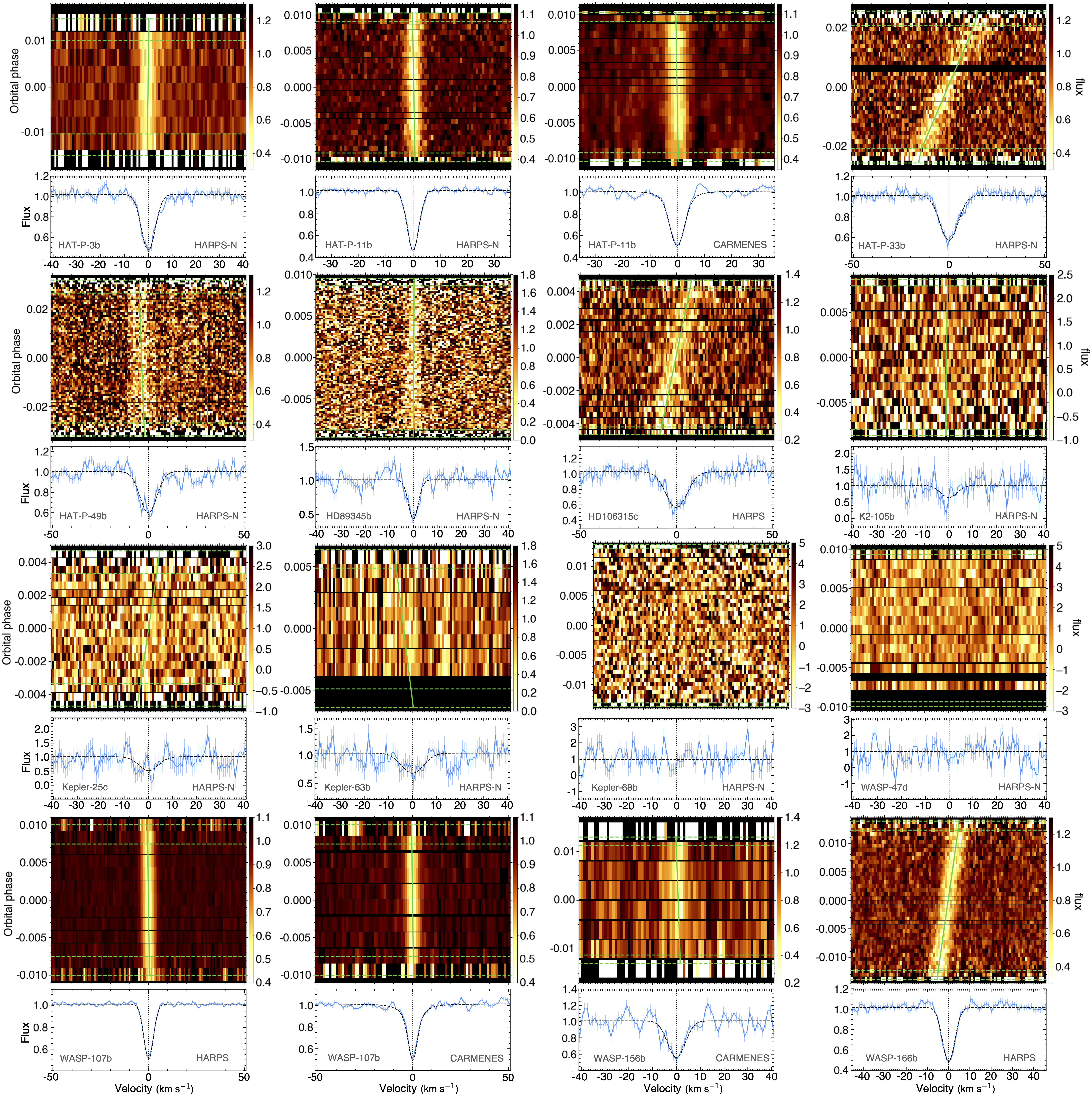}
\centering
\end{minipage}
\caption[]{CCF$_\mathrm{intr}$ for all transiting planets. When several datasets are available for a given instrument, they were binned together for clarity. \textit{Top subpanels}: Maps of the CCF$_\mathrm{intr}$, colored as a function of the flux (right axis), and plotted as a function of RV in the star rest frame (in abscissa) and orbital phase (in ordinate). Missing or out-of-transit data is plotted in black. When the S/N is high enough, the core of the stellar line from the planet-occulted regions can be seen as a bright streak. Transit contacts are shown as green dashed lines. The green solid line shows the stellar surface RV model from the RMR best fit, when detected. \textit{Bottom subpanels}: Master CCF$_\mathrm{intr}$, calculated as the weighted average of the fitted in-transit exposures after they were aligned in a common rest frame using the model surface RVs. The occulted lines of WASP-107 vary in contrast with $\mu$ (see text), and we show here their best-fit model at disk center. }
\label{fig:RMR_maps}
\end{figure*}

CCF$_\mathrm{DI}$ are scaled to the expected flux during the observed transits using light curves computed with the \texttt{batman} package \citep{Kreidberg2015}. For a homogeneous analysis, we used quadratic limb-darkening coefficients derived using the EXOFAST calculator\footnote{\url{http://astroutils.astronomy.ohio-state.edu/exofast/limbdark.shtml}} \citep{Eastman2013}. Exposures are considered as being in-transit if they overlap with the transit window. CCFs from the regions of the stellar photosphere that are occulted by the planet are retrieved by subtracting the scaled CCF$_\mathrm{DI}$ from the master-out CCF in each visit. These planet-occulted CCFs are reset to a common flux level to yield intrinsic CCFs, called CCF$_\mathrm{intr}$, which are directly comparable and trace the variations in the local stellar line profiles. Time series of CCF$_\mathrm{intr}$ for each planet are displayed in Fig.~\ref{fig:RMR_maps}.

For all RMR analyses, the posterior PDFs of free parameters describing fitted models to the data are sampled using \texttt{emcee} \citep{Foreman2013}, as detailed in \citet{Bourrier2021}. The number of walkers is set based on the number of free parameters, and the number of steps and the burn-in phase are adjusted to ensure well-mixed, converged chains. Best-fit values for the model parameters are set to the median of their PDFs, and their 1$\sigma$ uncertainty ranges are defined using highest density intervals. 

%---------------------------------------

\subsubsection{Individual and joined fits to planet-occulted CCFs}
\label{sec:CCF_fits}

In a first step, the CCF$_\mathrm{intr}$ are independently fit to assess their quality, using a simple Gaussian profile and broad uniform priors on its RV centroid $\mathcal{U}$($-3 \times v_{\rm eq} \sin i_\star$, $3 \times v_{\rm eq} \sin i_\star$), $v_{\rm eq} \sin i_\star$ being the equatorial projected stellar rotation velocity, its FWHM $\mathcal{U}$(0, 0.3 $\times$ FWHM$_{\rm DI}$), considering that the local line is always narrower than the disk-integrated line, and its contrast $\mathcal{U}$(0, 1). We found that the S/N is too low to detect the planet-occulted stellar line in all or most individual CCF$_\mathrm{intr}$ for eight planets: HAT-P-49 b, HD\,89345 b, HD 106315 c, K2-105 b, Kepler-25 c, Kepler-63 b, Kepler-68 b, WASP-47 d. The fits result in broad PDFs for the line properties, preventing in particular the derivation of their RV centroids and the interpretation of the surface RVs along the transit chord with the reloaded RM approach \citep{Cegla2016}. While the occulted stellar line track could be revealed by binning exposures in some cases, it would degrade the temporal sampling and limit our ability to probe the stellar properties. This highlights the interest of the RMR technique to fully exploit the signal from small planets across faint stars and to allow improving the spatial sampling of the transit chord for larger planets by using shorter exposure times (as is the case, e.g., for HAT-P-49 b and HD\,893455 b).

Indeed, the main novelty of the RMR technique consists in exploiting the full temporal and spectral information contained in the transit data by directly fitting a model of the stellar line to all CCF$_\mathrm{intr}$ simultaneously \citep[see details in][]{Bourrier2021}.
Several cases were tested for each planet, using the Bayesian information criterion \citep[BIC,][]{Schwarz1978,Kass1995,Liddle2007} to determine which model of the local stellar line is most justified. First, we tried whether the planet-occulted stellar lines are better modeled with a Gaussian or with a Voigt profile and found that the latter was justified for the CARMENES datasets. In that case, the damping coefficient, which relates the width of the Lorenzian and Gaussian components of the Voigt profile as $a = \sqrt{\ln(2)} {\rm FWHM}_{\rm Lor}/{\rm FWHM}_{\rm Gauss}$, is included as a free parameter. We then explored the possibility for variations in the local line contrast and FWHM along the transit chord, modeled as polynomial functions of the center-to-limb coordinate $\mu$. When several datasets are available, we allowed for a common modulation of the line properties $x(\mu) = x_0 (1 + \sum_{i\geq1}c_i \mu^i)$, with x representing the contrast or the FWHM, and $x_0$ specific to the instrument and/or the epoch. Only WASP-107 showed variations of the line shape along the transit chord (see Sect.~\ref{sec:WASP107}), and for all other datasets we assumed a constant profile and only varied $x_0$. By default, the centroids of the theoretical stellar lines are set by a RV model of the stellar surface along the transit chord, assumed to rotate as a solid body and constrained by the sky-projected spin--orbit angle $\lambda$ and stellar rotational velocity $v_{\rm eq} \sin i_\star$. We also investigated the possibility for differential rotation and convective blueshift as additional RV contributions \citep[see][for details]{Cegla2016}. While we did not find evidence for differential rotation in any of our target stars, the datasets of HAT-P-33 b and HD\,106315 c revealed a hint of convective blueshift (Sects.~\ref{sec:HAT_P33} and \ref{sec:HD106315}). As a final test, we allowed the semi-major axis and orbital inclination to vary with priors set from the literature values, as these properties define the theoretical chord transited by the planet and can be degenerate with $\lambda$ and $v_{\rm eq} \sin i_\star$ \citep{Bourrier2020_HEARTSIII}. We found that no dataset has the precision to constrain those properties better than our current knowledge, and they were thus kept fixed in our analysis. We did not attempt to let free the mid-transit time and transit depth, given the high precision on the final set of transit properties we use.

The planet-occulted stellar regions are discretized with a Cartesian grid to calculate their brightness-averaged properties, and each theoretical exposure is oversampled to account for the blur induced by the planet motion \citep{Cegla2016,Bourrier2017_WASP8}. The grid resolution and oversampling factor are adjusted to each planet. The time series of theoretical stellar lines is fit to the CCF$_\mathrm{intr}$ map of each visit, after being convoluted with a Gaussian profile of equivalent width to the corresponding spectrograph resolving power. Our pipeline thus allows to jointly fit transit datasets obtained with several instruments, using a common unconvolved stellar line profile when relevant. We exclude nonconstraining exposures that display uniform RV PDFs, and/or contrast and FWHM PDFs consistent with null values in the individual fits, typically at the edges of the transit where the planet only partially occults the limb-darkened star. Uncertainties on the CCF$_\mathrm{intr}$ were scaled with a constant factor to ensure a reduced unity $\chi^{2}$ for the best fit. MCMC jump parameters are the coefficients describing the line properties along the transit chord, $\lambda$, and $v_{\rm eq} \sin i_\star$. Uniform priors are set on the local FWHM $\mathcal{U}$(0, 0.3 $\times$ FWHM$_{\rm DI}$), the local contrast $\mathcal{U}$(0, 1), and $\lambda$ $\mathcal{U}$(-180, 180)$^{\circ}$. We performed preliminary fits with broad uniform priors on $v_{\rm eq} \sin i_\star$, and when relevant set normal priors based on literature values (see Sect.~\ref{sec:pl_sample} for details).

The analysis and results of the RMR fits are discussed in detail in the specific subsections of Sect.~\ref{sec:pl_sample}. The best-fit surface RVs along the transit chord are overplotted to the CCF$_\mathrm{intr}$ maps in Fig.~\ref{fig:RMR_maps}, where we further show the best-fit local line models overplotted to the master CCF$_\mathrm{intr}$. Multiepoch CCF$_\mathrm{intr}$ series from a common instrument were binned together for the sake of clarity, but we emphasize they were analyzed without binning. The agreement between the theoretical and measured CCF$_\mathrm{intr}$ time series in each epoch can be assessed by inspecting their residuals in Fig.~\ref{fig:RMR_resA} and \ref{fig:RMR_resB}. We make a distinction between the detectability of the RM signal, which we evaluate through the detectability of the occulted stellar line (i.e., the PDFs of C$_0$ and FWHM$_0$ excluding zero, and a significant BIC difference between the best-fit RMR model and the null hypothesis of a constant CCF$_\mathrm{intr}$), and our ability to constrain the planet transit chord (i.e., nonuniform PDFs for $\lambda$ and $v_{\rm eq} \sin i_\star$). In the end, we detect the RM signal confidently for all our targets except for Kepler-68 b and WASP-47 d. Out of these 12 detections, the transit chord is poorly constrained for K2-105 b and constrained thanks to priors on $v_{\rm eq} \sin i_\star$ for Kepler-25 c and WASP-156 b.

The sky-projected spin--orbit angle and stellar rotational velocity derived from the RMR fits are reported in the tables for each planet in Appendix \ref{apn:sys_prop}. Throughout the paper, we use the term spectroscopic for the values of $v_{\rm eq} \sin i_\star$ derived from line-broadening in the literature. In some cases, the stellar inclination is known from asteroseismology, and we combine its distribution with those for $\lambda$ and $i_\mathrm{p}$ to sample the 3D spin--orbit angle:
\begin{equation}
\label{eq:cospsi}
    \psi = \arccos\left( \sin i_\star \sin i_\mathrm{p} \cos \lambda + \cos i_\star \cos i_\mathrm{p}\right).
\end{equation}
In other cases, knowledge of the stellar radius $R_\star$ and the equatorial rotation period $P_\mathrm{eq}$ can be used to estimate $i_\star$. In this context, we highlight the warning raised by \citet{Masuda2020} that the distributions for $v_\mathrm{eq}$ = $2 \pi R_\star/P_\mathrm{eq}$ and $v_\mathrm{eq} \sin i_\star$ should not simply be combined, due to their interdependency. In those cases, we ran again our final MCMC fit using the independent variables $R_\star$, $P_\mathrm{eq}$ and $\cos i_\star$ as jump parameters instead of $v_\mathrm{eq} \sin i_\star$, and we set uniform priors on $\cos i_\star$ (assuming an isotropic stellar inclination distribution) and priors from measured values on $R_\star$ and $P_\mathrm{eq}$. We then derive from the results the PDF on the stellar inclination, which we use to derive $\psi$ using Eq.~(\ref{eq:cospsi}). Except when the degeneracy on $i_\star$ can be broken, we provide a ``northern" ($\psi_{\rm N}$) and ``southern" ($\psi_{\rm S}$) value for the two degenerate configurations corresponding respectively to $i_\star$ and $\pi - i_\star$, as well as the value of $\psi$ resulting from their combined distributions, assumed to be equiprobable.

%--------------------------------------------------------------------
\section{Orbital architecture orrery}
\label{sec:pl_sample}

In this section we provide some context on each of the studied systems, present our revision of their properties (listed in appendix \ref{apn:sys_prop}), and discuss how it changes or improves our understanding of their evolution. 

%-----------------------------

\subsection{HAT-P-3}
\label{sec:HAT_P3}

\subsubsection{Background}

HAT-P-3 b \citep{Torres2007} is a hot Jupiter with no known companion on a circular orbit around a metal-rich, early K-dwarf star. From a classical RM analysis of a single HARPS-N transit, \citet{Mancini2018} concluded that HAT-P-3 b is on a moderately misaligned orbit with $\lambda$ = $21.2\pm8.7^{\circ}$. Little is known about its atmosphere \citep{Todorov2013}, although its small radius implies a massive core \citep{Torres2007} or a metal-enriched composition \citep{Chan2011} that raises questions about its migration pathway and the influence it may have had on its atmospheric evolution and evaporation.

\subsubsection{Update}

We observed two spectroscopic transits of HAT-P-3 b with HARPS-N on 15 April 2019 and 30 January 2020. The first visit was excluded from our analysis since observations had to be stopped just before the transit due to deteriorating weather conditions, and could only restart right at the end of the transit (Sect.~\ref{sec:CCF_reduc}). The second visit was graciously granted as a compensation by the TNG director. Photometric observations of the 2020 transit with STELLA failed, but we observed it successfully in February 2021. The derived ephemeris is consistent with that of \citet{Baluev2019}, which yields $T_0$ with a precision of 20\,s for the second visit. We thus used the \citet{Baluev2019} ephemeris for the RM analysis.

The planet-occulted stellar line is clearly detected, but the transit chord is poorly constrained with $\lambda$ = $-25.3^{+29.4}_{-22.8}$\,$^{\circ}$ and $v_\mathrm{eq} \sin i_\star$ = 0.46$\stackrel{+0.22}{_{-0.25}}$\,km\,s$^{-1}$. These values are marginally different from those derived by \citealt{Mancini2018}, even though we both analyzed a single HARPS-N transit and used the same star and planet properties. We also derive $\lambda$ with a lower precision, probably due to the lower quality of our dataset. Our analysis however does not change the overall conclusion that HAT-P-3 b has a small sky-projected spin--orbit angle. Interestingly, the comparison between $v_\mathrm{eq} \sin i_\star$ and the stellar rotation period, derived by \citet{Mancini2018} from the stellar activity level, suggests that HAT-P-3 is seen nearly pole-on ($i_\star\sim16^{\circ}$, or $\sim$34$^{\circ}$ using the \citealt{Mancini2018} values). The resulting 3D spin--orbit angle, $\psi\sim$76$^{\circ}$, implies that HAT-P-3 b is on a polar orbit. \citet{Mancini2018} estimated that the orientation of HAT-P-3 b's orbital plane has not been significantly affected by tides during the main-sequence stellar evolution. If our result of a polar orbit is confirmed with additional RM measurements and a direct estimate of the stellar rotation period from photometry, detailed dynamical simulations will be required to determine whether the present-day architecture traces a disruptive dynamical history or a primordial misalignment between the protoplanetary disk and the star. The former scenario would be particularly interesting, considering that the massive core or high metallicity of HAT-P-3b could both point toward the partial evaporation of its volatile content.

%-----------------------------

\subsection{HAT-P-11}

\subsubsection{Background}

HAT-P-11 b \citep{Bakos2010} is a close-in, Neptune-size planet on an eccentric, highly misaligned orbit \citep{winn2010b,Hirano2011c} around a K-dwarf. A long period ($\sim$ 10 yr) companion in the system, likely mutually inclined with HAT-P-11 b \citep{Yee2018,Xuan2020} could be partly responsible for its migration. Strong general relativistic precession, nodal precession \citep{Yee2018}, disk-driven resonance \citep{Petrovich2020} or strong scattering \citep{Pu2021} rather than high-eccentricity migration have been proposed as possible migration pathways. Helium \citep{Allart2018,Mansfield2018}, hydrogen, and carbon \citep{BenJaffel2021} have been detected evaporating from the atmosphere of HAT-P-11 b. Like GJ 436 b, HAT-P-11 b is thus part of the group of warm Neptunes at the edge of the desert whose present atmospheric and orbital state is linked to a disruptive dynamical history \citep{Bourrier_2018_Nat,Attia2021}, making it a prime target for secular evolutionary simulations constrained by precise measurements of its spin--orbit angle.  

\subsubsection{Update}

We exploited two spectroscopic transits of HAT-P-11 b observed with HARPS-N on 13 September 2015 and 01 November 2015, and two transits observed with CARMENES on 07 August 2017 and 12 August 2017 (published by \citealt{Allart2018} for transmission spectroscopy). We used the ephemeris derived by \citet{Huber2017}, which yields a precision on our mid-transit times below 2\,s, and are closer to our observing epochs than those from \citet{Chachan2019}. We excluded exposures at indices 1, 2, and 57 (outliers, likely due to a low S/N) from the CARMENES visit on 12 August 2017. Furthermore, the CARMENES CCF$_\mathrm{Intr}$ time series shows residual streaks, likely due to telluric lines that could not be corrected during our reduction process. Indeed, the strength of the streaks correlates with airmass, which strongly increases toward the end of the visits. Calculating the master-out CCF$_\mathrm{DI}$ with pretransit exposures in the first visit, and excluding exposures after phase 0.02 in the second visit, removes most of the in-transit contamination.

The planet-occulted stellar track is nonetheless well-defined in all datasets (Fig.~\ref{fig:RMR_maps}). Even a visual inspection shows that it shifts from positive to negative surface RVs during the transit, indicating that the orbit is not exactly polar and is defined by $\lambda > 90^{\circ}$. Indeed, we derive $\lambda$ = $133.9^{+7.1}_{-8.3}$ $^{\circ}$, which is consistent within $1 - 2 \sigma$ with the literature but more precise thanks to the analysis of four combined RM datasets with the RMR technique. \citet{winn2010b} measured $\lambda = 103^{+26}_{-10}$ $^{\circ}$; \citet{Hirano2011c} measured $\lambda = 103^{+22}_{-18}$ $^{\circ}$; \citealt{SanchisOjeda2011} measured $\lambda = 90\pm28$ or $83^{+77}_{-65}$ $^{\circ}$ from spot-crossing anomalies, and $106^{+15}_{-12}$ or $121^{+24}_{-21}$ $^{\circ}$ when using the \citet{winn2010b} values as priors (the two values correspond to the star seen edge-on and pole-on, respectively). Our analysis favors a common line profile between the two visits of each instrument, suggesting that the stellar photosphere did not substantially evolve over the few weeks separating the observing epochs.

Our analysis of HAT-P-11 long-term photometry (Sect.~\ref{sec:APT_photom}, Fig.~\ref{fig:Longterm_photom}) yields an unambiguous rotation signal at 29.6 days, with a peak-to-peak amplitude of $\sim 0.01050$~mag, consistent with the period of 30.5$^{+4.1}_{-3.2}$ days determined from Kepler data by \citet{SanchisOjeda2011}. Combining this period with our precise measurement for $v_\mathrm{eq} \sin i_\star$ yields two possible inclinations of $i_\star = 33.3^{+6.3}_{-7.6}$ $^{\circ}$ and $146.8^{+7.6}_{-6.2}$ $^{\circ}$ for the stellar spin axis. Meanwhile, the spot-crossing anomalies analyzed by \citet{SanchisOjeda2011} can be explained by an edge-on ($i_\star = 80^{+5}_{-3}$ $^{\circ}$) or pole-on ($i_\star$ = $160^{+9}_{-19}$ $^{\circ}$) configuration for the star. These independent constraints allow us to break the degeneracy between the different architectures, favoring the configuration where the stellar south pole is visible ($i_\star$ = $160^{+9}_{-19}$ $^{\circ}$) and leading to a 3D spin--orbit angle $\psi = 104.9^{+8.6}_{-9.1}$ $^{\circ}$. 

HAT-P-11 is an active star \citep{Deming2011,Morris2017}. Photometric analyses revealed the presence of two active latitudes, where long-lived spots are repeatedly occulted by the planet and possibly phased with its revolution \citep{SanchisOjeda2011,Beky2014}. While our disk-integrated RVs show systematic variability possibly due to stellar activity (Fig.~\ref{fig:class_RM}), the CCF$_\mathrm{Intr}$ (Fig.~\ref{fig:RMR_maps}) and their residuals (Fig.~\ref{fig:RMR_resA}) show no evidence that the planet crossed spots during our observations. Given the high frequency of HAT-P-11 b spot-crossings \citep{SanchisOjeda2011}, it is more likely that spots were occulted during our observations but have spectral line profiles similar to the rest of the transit chord at our current precision level, or that our scaling with an nonspotted light curve erased the spot signatures. The very good agreement between our results and those from the literature, obtained with different instruments and techniques (RM spectroscopy with Subaru/HDS, Keck/HIRES, HARPS-N, and CARMENES; spot-crossing anomalies with Kepler) yet suggests that any spot effects were smoothed out in our analysis. Our refined value for the 3D orbital architecture of the HAT-P-11 system will be useful to inform more detailed simulations of its past evolution.

%-----------------------------

\subsection{HAT-P-33}
\label{sec:HAT_P33}

\subsubsection{Background}
This highly inflated hot Jupiter orbits at a very close distance ($a/R_\star = 5.7$) to a late, fast-rotating F star \citep{hartman2011,Wang2017} and is accordingly the second most irradiated planet in our sample. \citet{Turner2017} measured an excess transit depth in the $R$-band, which contains the H\,$\alpha$ transition, suggesting that HAT-P-33 b may be undergoing hydrodynamical escape.

\subsubsection{Update}

We observed one spectroscopic transit of HAT-P-33 b with HARPS-N on 04 December 2019. Observations were performed with strong wind and variable seeing. The last exposure only reached a S/N of 6 and was excluded from the analysis.

We observed two photometric transits with STELLA, allowing us to reach a better precision on the mid-transit time propagated to the time of our observations (42\,s) than using the ephemeris from \citet{Wang_Y2017}. Interestingly, the times predicted by the two ephemerides differ by 3.5$\pm$1.3\,min, which could motivate further transit monitoring of the planet to search for tidal decay, even though we checked that there is no evidence for it in the mid-transit times reported by \citet{Wang_Y2017}. We used the STELLA ephemeris for the RM analysis. 

The RV residuals from our Keplerian fit (Sect.~\ref{sec:RV}) show rms values larger than the median RV errors (Table~\ref{tab:quality_RV}). This is likely due to the known RV jitter for this active star, with an amplitude we estimate at $\sim$50-70\,m/s. The corresponding values for our white noise parameters support this observation (Table~\ref{tab:quality_RV}). \citet{hartman2011} cautioned about the difficulty of constraining the planetary orbit of HAT-P-33 b due the stellar jitter, and indeed our RV analysis does not constrain the orbital shape as precisely as the joint photometry and velocimetry fit from \citet{Wang_Y2017}, with an upper limit on $e\leqslant0.26$ at the $1\text{}\sigma$ level and accordingly loose constraints on the argument of periastron $\omega=62^{\circ}\pm85^{\circ}$. We do however refine the Keplerian semi-amplitude ($K=74.4\pm8.5~\mathrm{m\,s^{-1}}$ to be compared with $78\pm12~\mathrm{m\,s^{-1}}$, \citealt{Wang_Y2017}; $82.8\pm12.0~\mathrm{m\,s^{-1}}$, \citealt{hartman2011}; $72^{+19}_{-16}~\mathrm{m\,s^{-1}}$, \citealt{knutson2014_HJ}) and used this value in our RM analysis. The best-fit linear background trend yields a RV drift of $-2.1\pm9.0~\mathrm{m\,s^{-1}\,yr^{-1}}$, which sets an upper limit of $148~M_{\rm \oplus}$ for a potential outer planet on a circular orbit with minimal orbital period of $P_{\mathrm{min}}=2911~\mathrm{d}\simeq8~\mathrm{yr}$.

The HAT-P-33 b dataset yields the clearest RM anomaly of our sample, due to the fast stellar rotation and an aligned orbit. The RMR analysis provides $v_\mathrm{eq} \sin i_\star$ = $15.57\pm0.31$\,km s$^{-1}$, marginally larger than the spectroscopic value of $13.9\pm0.5$\,km s$^{-1}$ derived by \citet{hartman2011}, and the first measurement of HAT-P-33 b projected spin--orbit angle with $\lambda$ = $-5.9\pm4.1^{\circ}$. 

While the planet-occulted stellar track is clear and well-fit (Figs.~\ref{fig:RMR_maps}, \ref{fig:RMR_resA}), there is a hint of redshift from the solid-body rotation model toward the end of the transit. Differential rotation does not explain this feature but a model with convective blueshift varying linearly with $\mu$ yields the same BIC as the solid-body model. The strong RV jitter of HAT-P-33 is thought to be caused by convective inhomogeneities in the host star, possibly due to time-varying photospheric magnetic fields locally suppressing convection \citep{hartman2011}. An interesting alternative is that HAT-P-33 b occulted a region of the stellar surface with redshifted spectral lines due to the suppression of convective blueshift. The potential of constraining convection effects at the surface of HAT-P-33 makes this system a target of choice for spectroscopic follow-up. 

The revision performed by \citet{Wang2017} appears to confirm the eccentricity of the orbit. In that context, a truly aligned orbit for HAT-P-33 b would be surprising. Indeed, the F-type host star, with its shallow convective envelope, cannot realign the planetary orbit (\citealt{winn2010a}), and the tidal damping timescale is estimated to be much shorter than the age of the system. The mechanism that shaped HAT-P-33 b's present-day orbital architecture would thus need to have kept its primordial alignment, while exciting the eccentricity of its orbit and inducing its migration in recent times. This rather suggests that the system is actually misaligned due to the inclination of the host star, and that HAT-P-33 b underwent a high-eccentricity migration. A measurement of the stellar inclination is needed to determine the 3D spin--orbit angle and assess our prediction.  \\

%-----------------------------

\subsection{HAT-P-49}
\label{sec:HAT_P49}

\subsubsection{Background}

With an equilibrium temperature in excess of 2000\,K, due to its close-in orbit around an evolved F-star, HAT-P-49 b (\citealt{Bieryla2014}) belongs to the category of ultra-hot Jupiters.

\subsubsection{Update}

We observed one spectroscopic transit of HAT-P-49 b with HARPS-N on 31 July 2020. Exposure time was reduced after about half of the pretransit baseline to benefit from improved seeing.

We observed two photometric transits with STELLA. The resulting ephemeris is consistent within 1$\sigma$ with the values published by \citet{Bieryla2014}, but more precise (1.5\,min), and they were thus used for our RM analysis.

As with HAT-P-33, the RV residuals from our Keplerian fit (Sect.~\ref{sec:RV}) show rms values larger than the corresponding median RV errors, and the white noise parameters are significantly larger than zero (Table~\ref{tab:quality_RV}). In the case of HAT-P-49 this might be linked with the rapid oscillations characteristic of stellar pulsations observed in photometry (\citealt{Ivshina2022}), which induce here an estimated RV jitter amplitude of $\sim$90-130\,m/s. We chose to fix a circular orbit to avoid biasing $P$, determined precisely through photometric observations. Our revision of the orbital solution is consistent with the analysis of \citet{Bieryla2014}. The best-fit linear background trend yields a RV drift of $+11.0\pm8.6~\mathrm{m\,s^{-1}\,yr^{-1}}$, which sets an upper limit of $3.7~{\rm M}_{\rm Jup}$ for a potential outer planet on a circular orbit with minimal orbital period of $P_{\mathrm{min}}=3941~\mathrm{d}\simeq11.8~\mathrm{yr}$.

Our Keplerian model strongly deviates from the out-of-transit RVs of the HARPS-N visit. We noticed that the contrast and FWHM also show linear trends with time and that the line properties do not correlate with the S/N. This, along with the known stellar pulsations,  suggests that the stellar line variations are induced by short-term stellar activity. After correcting for these linear trends, there remained spurious features in the CCF$_{\rm Intr}$ profiles, which we could attribute to residual RV offsets between different groups of exposures (see Fig.~\ref{fig:class_RM}). The first seven measurements are abnormally redshifted, possibly because of the change in exposure time, while the post-transit measurements increase with time, possible again due to stellar activity. We minimized the spurious features by excluding these two groups of exposures from the master-out CCF$_{\rm DI}$, preventing the various RV shifts to blurr its profile and to offset it with respect to in-transit CCF$_{\rm DI}$.

The planet-occulted track is well detected and modeled (Fig.~\ref{fig:RMR_maps}). Some residual features are still visible (Fig.~\ref{fig:RMR_resA}), likely because our corrections could not completely remove the impact of the disk-integrated RV jitter. We derive $v_\mathrm{eq} \sin i_\star$ = $10.68^{+0.46}_{-0.47}$\,km s$^{-1}$, significantly lower than the spectroscopic value of $16.00\pm0.50$\,km\,s$^{-1}$ derived by \citet{Bieryla2014}. We surmise that this discrepancy might arise from their use of a synthetic stellar spectra library that may not be representative of HAT-P-49 (\citealt{Buchhave2012}). We measure for the first time the spin--orbit angle of HAT-P-49 b with $\lambda$ = $-97.7\pm1.8$ $^{\circ}$. With $\lambda$ and $i$ close to 90$^{\circ}$, the planet is likely truly on a polar orbit, supporting a disruptive dynamical formation or evolution for the system whose architecture remained unaltered by tidal interactions with the shallow convective envelope of the F-type host star.

%-----------------------------

\subsection{HD\,89345}
\label{sec:HD89345}

\subsubsection{Background}

HD 89345 b is a warm sub-Saturn (\citealt{VanEylen2018,Yu2018}) on an eccentric orbit around a slightly evolved and oscillating star. The star appears to have recently left the main sequence, moving toward the giant branch.

\subsubsection{Update}

We observed one spectroscopic transit of HD 89345 b with HARPS-N on 03 February 2020.

\begin{figure}[tbh!]
\includegraphics[trim=0cm 0cm 0cm 0cm,clip=true,width=\columnwidth]{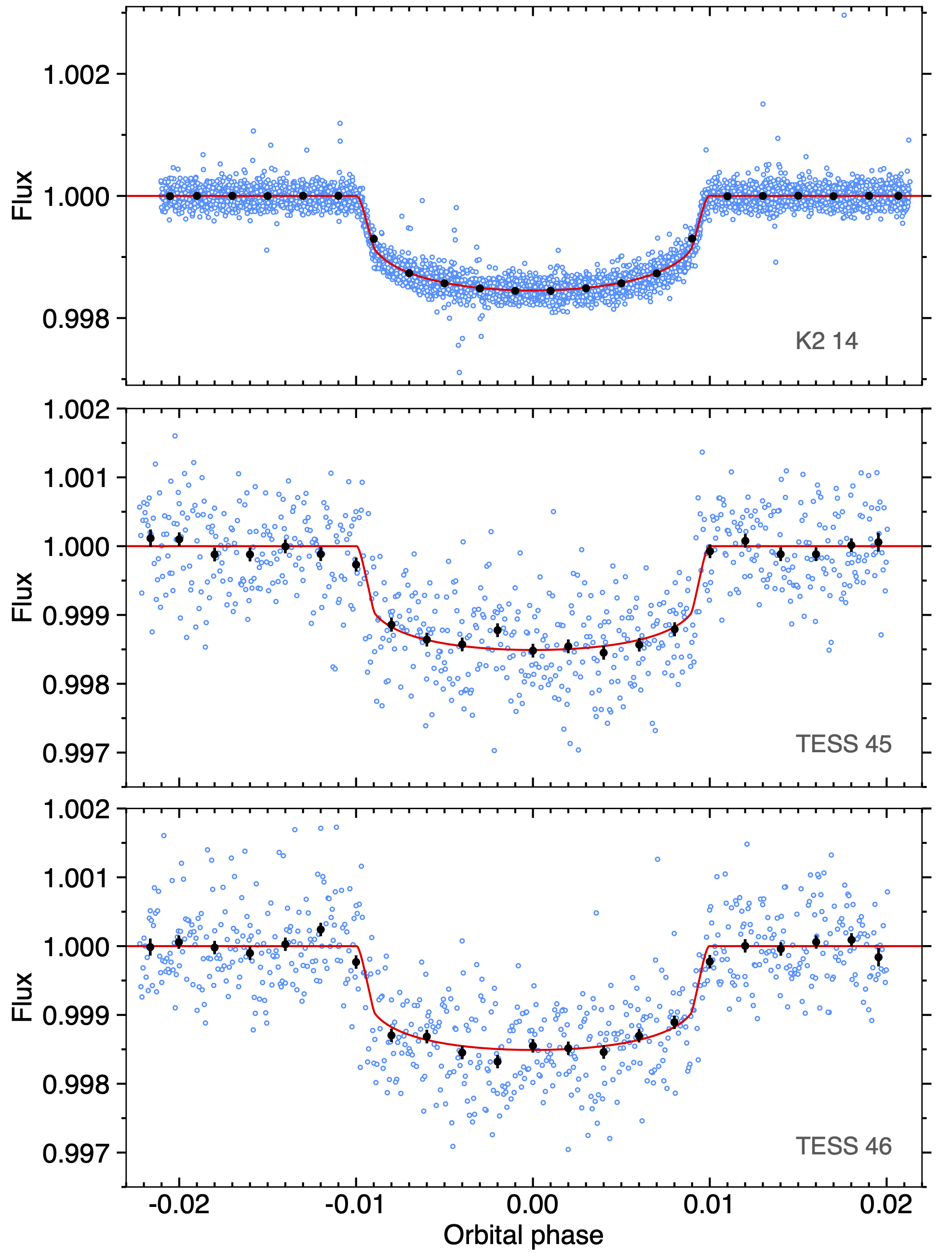}
\centering
\caption[]{Kepler and TESS lightcurves of HD 89345 b. Measurements corrected for the detrending models are shown as blue points, and binned into the black points. The red curve shows the best-fit transit model.}
\label{fig:Light_curves_HD89345}
\end{figure}

We observed two partial photometric transits with STELLA. A model including the transit combined with individual time-correlated detrending polynomials yields a BIC only marginally lower (by 4.7) compared with a model of detrending-only using the aforementioned polynomials. As a second test for significance, we allowed the transit depth to vary while still fixing $a/R_s$ and $i$ to literature values. This fit yields $(R_{\rm p}/R_\star)^2\,=1940 \pm 770$~ppm, which is in 1$\sigma$ agreement with the 1513~ppm value from \citealt{Yu2018}, but does not rule out a nondetection of the transit depth by more than $3\,\sigma$. We thus consider our detection of the transit with STELLA as tentative and analyzed TESS+K2 data to refine the planet ephemeris, orbital properties, and transit depth. 
HD 89345 b was observed by the Kepler Space Telescope as part of campaign 14 of the K2 mission from  31 March 2017 to 19 August 2017 at short cadence (1-min). Additionally, it was observed by TESS in sectors 45 and 46 from the 7 November 2021 to 30 December 2021. Kepler observed a total of seven transits during campaign 14, while TESS observed two transits in each sector, totaling 11 transits observed by both space telescopes. This data was used to perform a joint fit of the transit parameters according to the methods described in section \ref{sec:K2+TESS}. The resulting parameters are listed in table \ref{tab:HD89345}, and the detrended, phase-folded light curves are displayed in Fig.~\ref{fig:Light_curves_HD89345}. The resulting mid-transit time propagated at the epoch of the RM transit has a precision of 39\,s.

As with HAT-P-33 and HAT-P-49, the RV residuals from our Keplerian fit (Sect.~\ref{sec:RV}) show rms values ($3.7~\mathrm{m\,s^{-1}}$ over all instruments) larger than the corresponding median RV errors ($2.8~\mathrm{m\,s^{-1}}$), and the white noise parameters are significantly larger than 0 (Table~\ref{tab:quality_RV}). In the case of HD\,89345 this might be linked with solar-like oscillations (\citealt{VanEylen2018}), which induce here an estimated RV jitter amplitude of $\sim$5\,m/s. Our analysis yields an eccentric orbit ($e=0.208\pm0.039$;  $\omega=21.7\pm19.1^{\circ}$) consistent, but more precise, than the results of \citealt{VanEylen2018}. The instrumental jitter terms for FIES, HARPS-N, and HARPS reach about the same amplitude levels as in the latter authors' analysis. The best-fit linear background trend yields a RV drift of $-6.0\pm6.6~\mathrm{m\,s^{-1}\,yr^{-1}}$, which sets an upper limit of $15.8~{\rm M}_{\rm \oplus}$ for a potential outer planet on a circular orbit with minimal orbital period of $P_{\mathrm{min}}=244~\mathrm{d}$.

The planet-occulted track is clearly detected (Fig.~\ref{fig:RMR_maps}), although the CCF$_{\rm Intr}$ show some residual features possibly due to short-term stellar activity (Fig.~\ref{fig:RMR_resA}). We find that HD 89345 b is likely on a highly misaligned orbit with $\lambda$ = $74.2^{+33.6}_{-32.5}$\,$^{\circ}$. Although our derived $v_\mathrm{eq} \sin i_\star$ = $0.58\pm0.28$\,km\,s$^{-1}$ is inconsistent with the spectroscopic value from \citet{VanEylen2018} (2.6$\pm$0.5\,km\,s$^{-1}$), our results are consistent with their tentative detection of the RM signal ($v_\mathrm{eq} \sin i_\star$ = $1.4^{+1.1}_{-0.8}$\,km\,s$^{-1}$, $\lambda$ = $2^{+54}_{-30}$\,$^{\circ}$).

By modeling the rotational splitting of stellar oscillation frequencies, \citet{VanEylen2018} constrained the stellar inclination and excluded a pole-on configuration. We reproduced their PDF on $i_\star$ to derive $\psi$ = 80.1$\stackrel{+22.3}{_{-23.1}}$\,$^{\circ}$. Even though it depends on the internal structure of HD 89345 b, the time scale for tidal circularization is likely quite long (between $\sim$1 and 20\,Gyr (\citealt{VanEylen2018}). The cool host star would be efficient at circularizing and realigning the orbital plane of a hot Jupiter, but the lower mass and larger orbit of the warm sub-Saturn HD 89345 b reduce the strength of tidal effects, so that we are likely measuring the orbital architecture of the system unaltered by interactions with the star. If the misaligned orbit traces the primordial formation of the system, it could arise from the tilt of the early star or protoplanetary disk. Alternatively HD 89345 b could have followed a similar evolution as GJ 436 b (\citealt{Bourrier_2018_Nat}), migrating in recent times after exiting a Kozai resonance with an outer companion, which would have excited the eccentricity and inclination of its orbit. Further RV and imaging campaigns are required to search for the companion that would be responsible for this migration, as yet undetected (\citealt{VanEylen2018,Yu2018}). A late migration for HD 89345 b could imply that it arrived near the star at the end of its main-sequence lifetime, changing our view of its irradiative history and our interpretation of its inflation (\citealt{Yu2018}).

%-----------------------------

\subsection{HD\,106315}
\label{sec:HD106315}

\subsubsection{Background}

The F-star HD\,106315 is orbited by an inner super-Earth, planet b, and a warm Neptune, planet c (\citealt{Crossfield2017,Rodriguez2017,Barros2017}). No third planet (\citealt{Barros2017}) or stellar companion (\citealt{Crossfield2017,Rodriguez2017,Barros2017,Kosiarek2021}) were detected. Global modeling of the system and dynamical stability arguments support circular and coplanar orbits for HD 106315 b and c (\citealt{Barros2017,Rodriguez2017}), likely well-aligned with the star given the low spin--orbit angle of HD 106315 c (\citealt{Zhou2018}).

\subsubsection{Update}

We exploited three spectroscopic transits of HD 106315 c observed with HARPS on 09 March 2017, 30 March 2017, and 23 March 2018. We excluded exposures at index 73 from the visit on 09 March 2017 (low S/N and outlying properties), and the last two exposures from the visit on 30 March 2017 (high noise and spurious CCF features). We used the ephemeris from \citet{Kosiarek2021}, which yields precisions on the mid-transit times of the RM observations of 2.4, 2.5, and 5.2\,min. 

Our limited, single-season photometric data of HD\,106315 (Sect.~\ref{sec:APT_photom}, Fig.~\ref{fig:Longterm_photom}) were 
analyzed by \citet{Kosiarek2021}, who found no significant periodicities between 1 and 100 days. The standard deviation of the 43 nightly observations from the seasonal mean is 0.00259~mag, somewhat larger than the precision of a single observation with the T12 APT, implying possible low-level brightness variability in HD~106315. \citet{Kosiarek2021} analyzed $Kepler$ and K2 observations and found a weak periodogram peak at 4.8 days, which they attribute to the rotation period, and a slightly larger peak at 9.6 days, which would be the second harmonic of the period.

The planet-occulted track is well detected and modeled (FigS.~\ref{fig:RMR_maps}, \ref{fig:RMR_resA}). There is a hint that the local stellar line and thus photosopheric properties varied during the second epoch, but BIC comparison favors a common line profile for all three visits. Our RMR fit also hints at a convective blueshift on HD\,106315 decreasing linearly with $\mu$. The linear coefficient $c_{1}$ = -6.0$\pm$1.8\,km\,s$^{-1}$ differs from zero by more than 3$\sigma$, and this model yields a BIC similar to the pure solid-body rotation model. This makes follow-up RM observations of HD\,106315 interesting for stellar characterization.
The first two visits were included in a Doppler tomographic analysis of four transits with the MIKE, TRES, and HARPS facilities by \citet{Zhou2018}, who reported $\lambda$ = $-10.9^{+3.6}_{-3.8}$\,$^{\circ}$ and $v_\mathrm{eq} \sin i_\star$ = $13.00\pm0.28$\,km\,s$^{-1}$.
Our rotational velocity ($v_\mathrm{eq} \sin i_\star$ = $9.66^{+0.64}_{-0.65}$\,km s$^{-1}$) is significantly lower than the value from \citet{Zhou2018}. However their result seems mainly constrained by the strong prior ($13.08\pm0.28$\,km\,s$^{-1}$) they derived from a fit to the disk-integrated line, including solid-body rotation and mactroturbulence, which may be biased by the strong correlation between these two broadening velocities. 
We derive a smaller spin--orbit angle ($\lambda$ = $-2.68^{+2.7}_{-2.6}$\,$^{\circ}$) than \citet{Zhou2018}, but our results are consistent within 2$\sigma$ and point toward HD 106315 c being on a well-aligned orbit if the stellar inclination is low. However, if the stellar equatorial rotation period is indeed about 4.8 days, our value for $v_\mathrm{eq} \sin i_\star$ yields $i_\star\sim$46$^{\circ}$ and a 3D spin--orbit angle $\psi\sim$43$^{\circ}$. A precise measurement of the stellar rotation period would thus be of particular interest.

The probable low mutual inclination with HD 106315 b support disk-driven migration for both planets. HD 106315 c has a radius almost twice as large as HD 106315 b but a similar mass (\citealt{Barros2017}). They could thus have formed in different regions of the protoplanetary disk, or they could have formed with the same core and envelope mass but planet b migrated close enough to the star that a substantial fraction of its envelope photo-evaporated. If HD 106315 c is truly misaligned with the star, it would suggest a primordial tilt of the star or protoplanetary disk.

%-----------------------------

\subsection{K2-105}

\subsubsection{Background}

K2-105 b (\citealt{Narita2017}) is a warm planet orbiting a G-dwarf. It stands at the transition between the mini-Neptune and Neptune populations and is a good candidate to understand the processes behind the formation of ice giants and the possible resilience of sub-Neptunes to atmospheric escape (\citealt{Owen2018}). Additional observations are required to determine the presence of planetary companions, but no stellar companions have been detected through direct imaging. \\

\subsubsection{Update}

\begin{figure}[tbh!]
\includegraphics[trim=0cm 0cm 0cm 0cm,clip=true,width=\columnwidth]{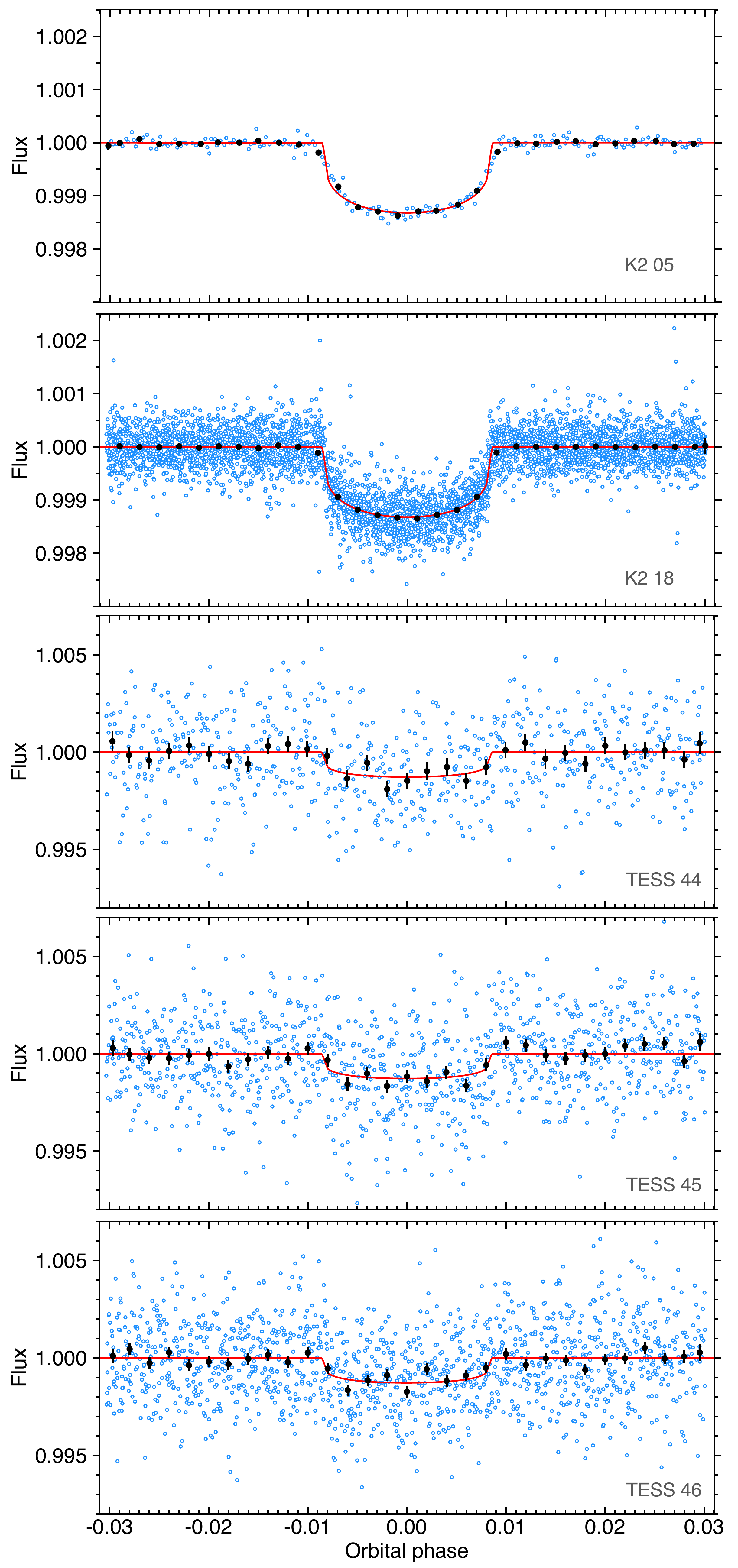}
\centering
\caption[]{Kepler and TESS lightcurves of K2-105 b. Measurements corrected for the detrending models are shown as blue points, and binned into the black points. The red curve shows the best-fit transit model.}
\label{fig:Light_curves_K2-105}
\end{figure}

We observed one spectroscopic transit of K2-105 b with HARPS-N on 19 January 2020. The fourth exposure of the visit (lowest S/N and outlying CCF properties) was excluded from our analysis. We observed two photometric transits with STELLA, yielding a precision on the mid-transit time at the epoch of the RM transit of 4.2\,min. K2-105 b was observed by the Kepler Space Telescope as part of the campaigns 05 (ten transits between 24 April 2015 and 11 July 2015) and 18 (six transits between 12 May 2018 and 02 July 2018) of the K2 mission. Noteworthy, K2-105 b was chosen as one of 42 already confirmed exoplanets for which short cadence (1-min) data was acquired during campaign 18. Additionally, it was observed by TESS in sectors 44 (two transits), 45 (three transits) and 46 (four transits) from 12 October 2021 to 30 December 2021. Data from these 25 transits were used to perform a joint fit of the transit parameters (Sect.~\ref{sec:K2+TESS}). Results are listed in Table \ref{tab:K2-105}, and the detrended, phase-folded light curves are displayed in Fig.~\ref{fig:Light_curves_K2-105}. The timing precision was reduced to 1\,min at the epoch of the RM transit. This analysis also improved substantially the uncertainties on the planet transit depth and orbital properties, which we used for the RM analysis.

The RMR fit yields a detection for the planet-occulted stellar line (Fig.~\ref{fig:RMR_maps}) but no constraints on the stellar rotation. The PDF for $\lambda$ has a well-defined peak at -81$^{\circ}$ with a 1$\sigma$ confidence interval in [-128 , -31]$^{\circ}$, but it displays broad wings that cover the full parameter space within 3$\sigma$. We perform the final fit with a prior on $v_\mathrm{eq} \sin i_\star$ set to the spectroscopic value of $1.76\pm0.86$\,km s$^{-1}$ from \citet{Narita2017}, but it does not change the result on $\lambda$. Thus, even though the best-fit RMR model is favored over the null hypothesis with a BIC difference of 11, additional observations are required to confirm our measurement.

If confirmed, a misaligned orbit for K2-105 b could support a disruptive dynamical past and the late arrival of the planet on its close-in orbit. Alternatively the presence of additional planets could support a primordial tilt of the star or protoplanetary disk, as this low-mass planet is likely far enough from its G-type star that its spin--orbit angle was not much influenced by tidal interactions. Further monitoring of the system and a refinement of the spin--orbit angle are necessary to investigate these scenarios.

%-----------------------------

\subsection{Kepler-25}

\subsubsection{Background}

The late F-type star Kepler-25 hosts three known planets (\citealt{Steffen2012b}, \citealt{Marcy2014}). The hot super-Earth Kepler-25 b and the warm Neptune Kepler-25 c are transiting, in contrast to the long-period giant Kepler-25 d. There is no evidence for additional companions (\citealt{Marcy2014}, \citealt{Mills2019}). Properties of the system were revised by \citet{Mills2019}, using a photodynamical model to interpret velocimetry and photometry data, and to account for the TTVs observed between Kepler-25 b and c (\citealt{Steffen2012b}). The two planets lie close to but outside of the 2:1 mean-motion resonance (MMR; $P_{\rm b}=6.238$, $P_{\rm c}=12.721$ days). The results by \citet{Mills2019} favor low orbital eccentricities (see also \citealt{Lithwick2012}, \citealt{VanEylen2015}, \citealt{Hadden2017}), which support a near-resonant state between Kepler-25 b and c (\citealt{Migaszewski2018}). This observed configuration is a natural outcome of the capture in resonance during the migration within a disk followed by long-term tidal dissipation \citep{Lee2013,Delisle2014}. Disk-driven migration is also supported by the lack of evidence for high mutual inclination between planets Kepler-25 b and c (\citealt{Mills2019}), although measurements of the 3D spin--orbit angle spurred some debate as to the actual alignment of the system (\citealt{Albrecht2013,Benomar2014,Campante2016}).

\subsubsection{Update}

We observed one spectroscopic transit of Kepler-25 c with HARPS-N on 14 June 2019.
Observations of the photometric transit with STELLA failed. We thus used the mid-transit times derived for individual transits of Kepler-25 b and c by \citet{Battley2021} to perform a TTV analysis. For a set of orbital parameters and planetary masses, the transit timings of the planets are modeled using the {\ttfamily TTVfast} algorithm \citep{Deck2014} with the set-up described in \cite{Leleu2021}. We sample the posterior using the adaptive MCMC sampler \texttt{samsam}\footnote{\url{https://gitlab.unige.ch/Jean-Baptiste.Delisle/samsam}} \citep[see][]{Delisle2018}. The main period of the TTV is the super-period associated with the 2:1 MMR \citep{Lithwick2012}:
\begin{equation}
P_\text{TTV}=\frac{1}{1/6.238-2/12.721} \approx 325.48\, \text{d}.
\end{equation}
TTVs are indeed retrieved at the predicted period, with peak-to-peak amplitude of $\sim 8$ minutes for the inner planet and $\sim 2$ minutes for the outer one. We then propagate the solution to the chosen date for 1000 samples of the posterior in order to estimate the time of transit at the epoch of the RM observation and its uncertainty (about 50\,s).

Measuring the RM signal of Kepler-25 c is challenging due its small transit depth. A classical RM analysis by \citet{Albrecht2013} yielded ($\lambda$ = 7$\pm$13$^\circ$ ; $v_\mathrm{eq} \sin i_\star$ = 6.2$\pm$3\,km s$^{-1}$). A spectroscopic prior on $v_\mathrm{eq} \sin i_\star$ improved their results to ($\lambda$ = 5$\pm$8$^\circ$ ; $v_\mathrm{eq} \sin i_\star$ = 8.5$\pm$0.6\,km s$^{-1}$), while a tomographic analysis of the same dataset returned ($\lambda$ = -0.5$\pm$5.7$^\circ$ ; $v_\mathrm{eq} \sin i_\star$ = 8.2$\pm$0.2\,km s$^{-1}$). \citet{Benomar2014} then refined the orbital architecture of the system through a combined analysis of asteroseismology, transit lightcurve and the RM effect. They derived ($\lambda$ = 9.4$\pm$7.1$^{\circ}$ ; $v_\mathrm{eq} \sin i_\star$ = 9.34$_{-0.39}^{+0.37}$\,$^{\circ}$ ; $i_\star$ = 65.4$_{-6.4}^{+10.6}$\,$^{\circ}$ ; $\psi$ = 26.9$_{-9.2}^{+7.0}$\,$^{\circ}$) and concluded that Kepler-25 c is on a mildly misaligned orbit. This claim was disputed by \citet{Campante2016}, who concluded from an asteroseismology study that Kepler-25 c is on an aligned orbit with the star  ($\psi$ = 12.6$^{+6.7}_{-11.0}$\,$^{\circ}$), seen edge-on ($i_\star$ in [68.7, 90.0]$^{\circ}$ at 1\,$\sigma$). However we note that $\psi$ and $i_\star$ are poorly constrained by the Kepler data alone, and that the results from \citet{Campante2016} mainly derive from the priors they set on $\lambda$ and $i$ using \citet{Albrecht2013} values.

Our first RMR fit yielded a detection of the planet-occulted stellar line (Fig.~\ref{fig:RMR_maps}) with a BIC difference of 13 between the best-fit model and the null hypothesis. While the PDF for $\lambda$ favored values close to 0$^\circ$, the data is not sufficient to strongly constrain the stellar rotational velocity (consistent with 0\,km s$^{-1}$) and thus the orientation of the transit chord. We performed a second fit using the stellar projected rotational velocity and inclination derived by \citet{Benomar2014} from asteroseismology alone, as these values are independent from priors and from their classical analysis of the RM effect. Setting a prior on $v_\mathrm{eq} \sin i_\star$ to 9.13$^{+0.60}_{-0.69}$\,km s$^{-1}$ refined substantially our PDF for $\lambda$, yielding a value of -0.9$^{+7.7}_{-6.4}$\,$^{\circ}$. The stellar inclination from \citet{Benomar2014} then allowed us to derive $\psi$ = $24.1^{+9.2}_{-9.3}$\,$^{\circ}$.

Our results are fully consistent with those from \citealt{Albrecht2013}, \citealt{Benomar2014}, and \citealt{Campante2016}. The 3D spin--orbit angle we derive is significantly larger than 0$^{\circ}$, but smaller than the value of 30$^{\circ}$ usually taken as criterion for misalignment given the typical uncertainties on RM measurements. Rather than debating whether Kepler-25 c is aligned or not, we ask whether $\psi\sim24$\,$^{\circ}$ can trace a disruptive dynamical process or is within the variations expected from disk-driven migration. In both cases, the system architecture was likely not impacted by later tidal interactions with the star due to the large orbital distance of Kepler-25 c and shallow convective envelope of its F-type host star. A way to better constrain the dynamical past of the system would be to measure the spin--orbit angle of Kepler-25 b with a spectrograph on a larger telescope, to assess whether the two planets are truly coplanar or show some mutual misalignment (see the case of HD\,3167, \citealt{Bourrier2021}). 

We note that \citet{McQuillan2013} derived a rotation period for Kepler-25 (23.147$\pm$0.039\,days) using starspot modulation in Kepler data. This would imply a maximum value of 2.9\,km\,s$^{-1}$ for $v_\mathrm{eq} \sin i_\star$, which is inconsistent with all analysis of the system architecture. However, combining our results for the projected stellar rotational velocity with the stellar inclination from \citet{Benomar2014} yields an equatorial rotation period of $6.8^{+0.6}_{-0.7}$\,days, which is consistent with a third of 23.147 and thus suggest that \citet{McQuillan2013}'s value is an alias of the true rotation period.

%-----------------------------

\subsection{Kepler-63}

\subsubsection{Background}

Kepler-63 b is a giant planet intermediate in radius to Neptune and Saturn, orbiting a young and active solar-type star (\citealt{SanchisOjeda2013_K63b}). Due to the high-level of stellar activity, RVs only set a 3$\sigma$ upper limit of 0.4\,$M_{\rm Jup}$ on the planet. No planetary or stellar companion has been detected. \citet{SanchisOjeda2013_K63b} used RM measurements and spot-crossing anomalies to constrain the orbital architecture of the system, showing Kepler-63 b to be on a highly misaligned orbit around the star.

\subsubsection{Update}

We observed one spectroscopic transit of Kepler-63 b with HARPS-N on 13 May 2020. Pretransit baseline and ingress could not be observed due to high humidity, and the post-transit baseline was cut short due to increasing seeing and the loss of the autoguider. 
We observed three photometric transits with STELLA, which were published in \citet{Mallonn2022}. The derived ephemeris is consistent but less precise than those published by \citet{Gajdos2019}, which were thus used for our analysis. They yield a precision of 12\,s on the mid-transit time at the epoch of the RM observation.

Despite the low S/N of the CCF$_{\rm Intr}$, the planet-occulted line is detected with a BIC difference of 34 compared to the null hypthesis (Fig.~\ref{fig:RMR_maps}). The master CCF$_{\rm Intr}$ and residual map (Fig.~\ref{fig:RMR_resB}) however highlight some features that likely perturb the RMR fit. They result in a bimodal PDF for the FWHM of the intrinsic stellar line with a low-value mode consistent with the FWHM of the disk-integrated line and a nonphysical mode at larger values. The low-value mode only blends with the other one in its high-value wing but otherwise has a Gaussian profile. We thus set a Gaussian prior on the intrinsic FWHM, which we adjusted to match the unblended part of its low-value mode. This constraint provides a cleaner fit to the planet-occulted stellar line with little impact on the other properties. We thus conclude that $v \sin{i_\star}$ = 7.47$^{+2.6}_{-2.7}$\,km\,s$^{-1}$, consistent with the spectroscopic (5.4$\pm0.5$\,km\,s$^{-1}$) and classical RM (5.6$\pm0.8$\,km\,s$^{-1}$) values derived by \citet{SanchisOjeda2013_K63b}, and $\lambda$ = $-135^{+21.2}_{-26.8}\,^{\circ}$, consistent as well with their classical RM value ($-110^{+22}_{-14}\,^{\circ}$).

The analysis of spot-crossing anomalies by \citet{SanchisOjeda2013_K63b} allowed them to break the degeneracy on the stellar inclination and favor the configuration where the southern stellar pole is visible ($i_{\star}$ = $138\pm7$\,$^{\circ}$). Combined with our measurement for $\lambda$, this yields $\psi$ = $114.6^{+16.6}_{-12.5}$\,$^{\circ}$, which is in agreement with the value of $104^{+9}_{-14}$\,$^{\circ}$ derived by \citet{SanchisOjeda2013_K63b} from a combined analysis of the RM effect and spot-crossing events.

In contrast to \citet{SanchisOjeda2013_K63b} we lack a high-precision light curve contemporaneous with our RM transit to account for occulted spots in the scaling of the CCF$_{\rm DI}$. However, the quality of our data is likely not high enough to be sensitive to spot signatures in the planet-occulted lines, and our results are fully consistent with those of \citet{SanchisOjeda2013_K63b}. We thus confirm their result of a polar orbit for Kepler-63 b. Given the young age of the system, it is of high interest to study the primordial processes behind the misalignment of close-in giant planets.

%-----------------------------

\subsection{Kepler-68}

\subsubsection{Background}

The solar-type star Kepler-68 hosts three confirmed planets (\citealt{Gilliland2013}): an inner transiting pair formed of the sub-Neptune Kepler-68 b and the super Earth Kepler-68 c (\citealt{Borucki2011b,Batalha2013,Ofir2013,Huang2013}), and the long-period, nontransiting giant Kepler-68 d (\citealt{Marcy2014}). Transit photometry and RV analyzed by \citet{Mills2019} hint at a fourth planetary or stellar companion on a larger orbit. A marginal detection of a bound, distant stellar companion has been obtained from direct imaging by \citet{Ginski2016}. RV and imaging follow-up of the system are required to characterize these two candidates and investigate their relation to the overall dynamics of the Kepler-68 system.

Kepler-68 b has a low density that is consistent with a water-rich envelope (\citealt{Gilliland2013}) and could explain its resilience to the strong stellar irradiation (\citealt{Lopez2012,Zeng2014}) over the long life of the system (6.3\,Gyr, \citealt{Gilliland2013}).

Asteroseismology by \citet{Campante2016} exclude that the star is seen pole-on. From the duration of their transits, \citet{VanEylen2015} find Kepler-68 b to be consistent with a circular orbit and Kepler-68 c possibly to be on an eccentric orbit. Kepler-68 d is on an eccentric orbit (\citealt{Gilliland2013,Marcy2014}) and could have induced late orbital instability on the inner planets, trapping the pair near MMR and exciting the eccentricity of Kepler-68 c (\citealt{Pan2020}). Self-excitation could further have led to a high mutual inclination between Kepler-68 d and the inner system (\citealt{Becker2016,Kane2015,Read2017}).

The peculiar configuration of the Kepler-68 system highlights the interest of determining the alignment of the inner planets with the star and investigating how their dynamical history influenced their atmospheric evolution.

\subsubsection{Update}

We observed one spectroscopic transit of Kepler-68 b with HARPS-N on 03 August 2019. We used the ephemeris from \citealt{Gajdos2019}, which yields $T_0$ with a precision of 32\,s at the epoch of the RM transit.

Despite the good quality of the data (Fig.~\ref{fig:RMR_resB}), the RMR fit is unable to detect the planet-occulted line and thus brings no constraints on $\lambda$ and $v \sin{i_\star}$. Setting a prior on $v \sin{i_\star}$ based on the spectroscopic value (0.5$\pm$0.5\,km\,s$^{-1}$, \citealt{Gilliland2013}) does not change this result, and we thus conclude a nondetection. Follow-up RM observations with a larger telescope are warranted to determine this system architecture.

%-----------------------------

\subsection{WASP-47}
\label{sec:WASP47d}

\subsubsection{Background}

WASP-47 is a G-dwarf that hosts three transiting planets: an ultra-short period planet (WASP-47 e, \citealt{Becker2015}), a hot Jupiter (WASP-47 b, \citealt{Hellier2012}), and an outer Neptune-like planet with a dense core and lightweight envelope (WASP-47 d, \citealt{Becker2015,Vanderburg2017}). A fourth giant planet WASP-47 c has been detected on a wider eccentric orbit, with no transit observed for now (\citealt{Neveu2016}). 
No stellar companions have been detected (\citealt{Wollert2015,Becker2015}). Properties of the system were successively refined by \citet{Dai2015,Almenara2016,Sinukoff2017,Weiss2017, Vanderburg2017} and more recently by \citet{Bryant2022}, whose results we mainly use in our analysis.

The density of WASP-47 e is too low to be explained by a rocky Earth-like composition and requires the presence of a high-metallicity envelope, such as a steam-rich layer (\citealt{Dai2015,Vanderburg2017,Dorn2019,Bryant2022}). WASP-47 e could be the remnant core of a larger progenitor that photo-evaporated its gaseous envelope, which makes the dynamical evolution of the system even more interesting. Indeed the configuration of the system, with a hot Jupiter surrounded by two smaller planets, is rather singular and suggests that WASP-47 formed differently than multiplanet and single hot Jupiter systems (\citealt{Huang2016,Bryant2022}). WASP-47 b might stand at the transition between hot Jupiters, many of which may undergo high-eccentricity migration (\citealt{Mustill2015}), and warm Jupiters, which could form in situ (\citealt{Huang2020}). High-eccentricity migration is unlikely for WASP-47 b, as it would have disrupted the orbits of the smaller WASP-47 e and d. \citet{Huang2016} thus speculated that WASP-47 b might be at the tail end of the in situ warm Jupiter formation mechanism. Alternatively, \citet{Weiss2017} proposed a two-stages process with the giant planets forming first in a gas-rich disk and migrating to their present locations, and then the smaller high-density planets forming in situ in a gas-poor environment. In that scenario the eccentric orbit of WASP-47 c would need to have been excited after the dampening by the disk, possibly by another outer companion (\citealt{Weiss2017}). In any case the system likely underwent a quiescent dynamical evolution with the migrating planets remaining within the plane of the protoplanetary disk. This is supported by the combined spectroscopic and photometric analysis of \citet{SanchisOjeda2015}, which excludes highly misaligned and retrograde orbits for WASP-47 b, and by the dynamical analyses of \citet{Becker2017} and \citet{Vanderburg2017}, which suggest that WASP-47 c orbits close to the plane of the inner three planets. Observational constraints on the orbital architectures of all planets in the system would thus be helpful in understanding its peculiar origin.\\

\subsubsection{Update}

We observed one spectroscopic transit of WASP-47 d with HARPS-N on 05 August 2021. While most of the visit was carried out in good observing conditions, the sequence stopped for $\sim$15\,min after the first exposure due to a problem with writing its AG image, and the S/N decreased below 13 toward the end of the sequence due to cirrus clouds. As a result the last five exposures (indexes 19 to 23) show abnormal CCF properties and were excluded from our analysis.

\begin{figure}[tbh!]
\includegraphics[trim=0cm 0cm 0cm 0cm,clip=true,width=\columnwidth]{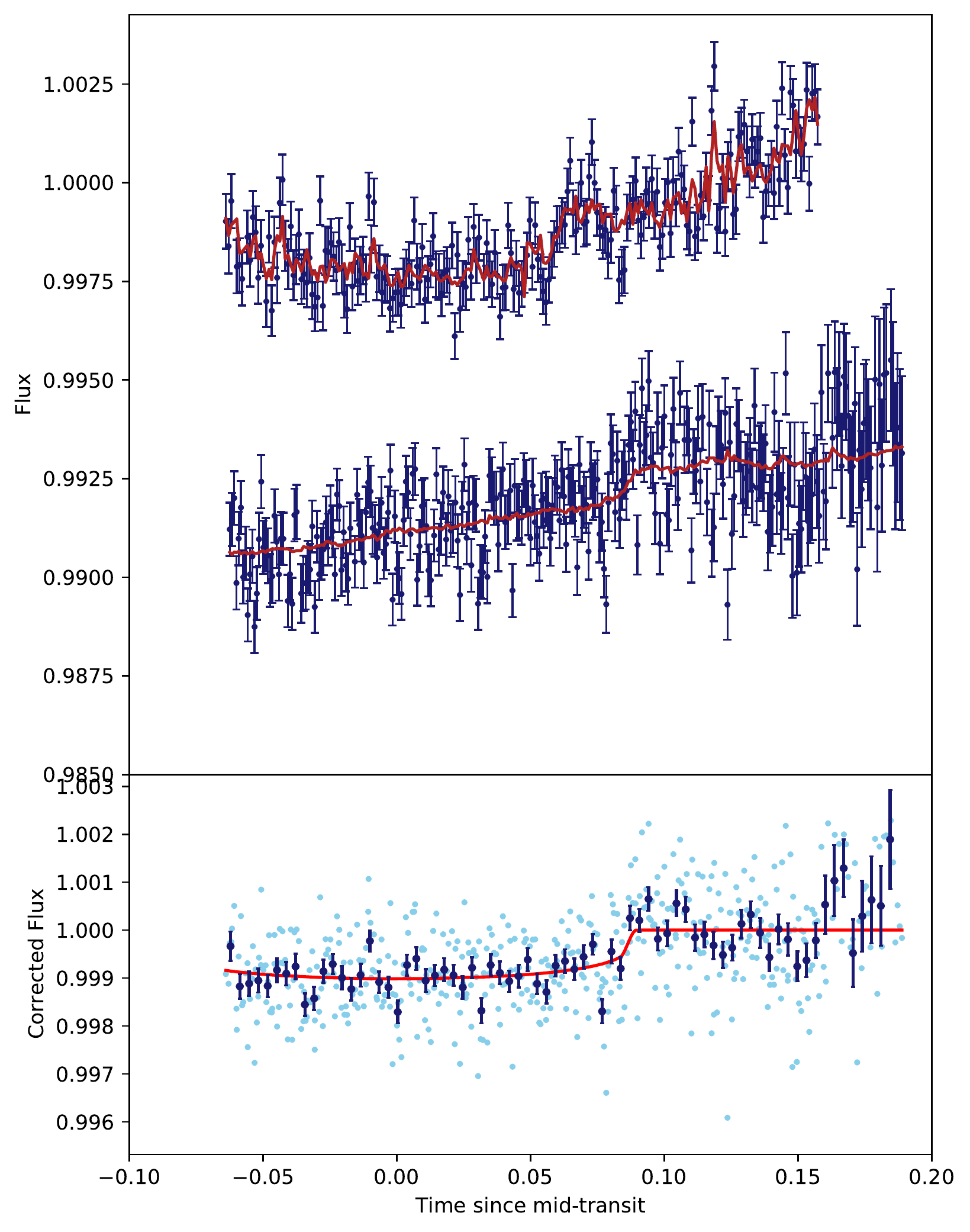}
\centering
    \caption{EulerCam light curves of Wasp-47\,d. \emph{Top panel:} raw data together with the transit and systematic model corresponding to the median posterior values. 
    \emph{Bottom panel:} Systematics-corrected, phase-folded data together with the transit model. The unbinned data points are shown in light blue, while the dark blue points show the data binned into 2-minute intervals.}
\label{fig:EULERcam}
\end{figure}

We used EulerCam to observe two transits of WASP-47\,d in 26 August 2021 and 04 September 2021. We used an \emph{r'-Gunn} filter and slightly defocused the telescope to optimize observation efficiency and PSF sampling, using exposure times of 75 and 60 s during the first and second night, respectively. Owing to the small transit depth and the exquisite K2 measurements available \citep{Becker2015}, we opted to fix $R_{\rm p} / R_\star$,  $b$, $T_{\mathrm{14}}$, $P$ to the values quoted by \citet{Vanderburg2017}, as well as the limb-darkening coefficients to those derived with LDCU \citep{Deline2022} ($u_1 = 0.462$, $u_2 = 0.197$), and assumed a zero eccentricity. The mid-transit time was allowed to vary within 1 hour of the predicted value. Correlated noise was fit using an approximate Mat\'ern-3/2 kernel implemented through \emph{celerite} \citep{FM2017}. For both light curves, we included an evident correlation between the residual flux and the stellar FWHM as a linear trend fit together with the transit model and GP. We allowed for additional white noise by inclusion of a jitter term for each light curve. We carried out a joint fit assuming a constant period and individual fits to both data sets allowing $T_0$ to vary. We note that both transits are partial due to ephemeris drift occurring since the K2 observations. The raw and phase-folded light curves are shown in Figure \ref{fig:EULERcam}. The derived mid-transit times are $2459453.6435_{-0.0031}^{+0.0041}$ for 26 August 2021 and $2459462.6676_{-0.0033}^{+0.0032} $ for 04 September 2021.

WASP-47 b and WASP-47 d exhibit significant TTVs (\citealt{Becker2015,Weiss2017,Bryant2022}). We used the WASP-47 b transit times from K2 (\citealt{Becker2015}) and TESS (\citealt{Bryant2022}), along with the WASP-47 d transit times from K2 (\citealt{Becker2015}) and Euler (our analysis), to model the TTVs of the two planets (\citealt{Lithwick2012}). We predict the TTV super-period (52.7360$\pm$0.0007\,days) and amplitudes (0.72$\pm$0.14\,min for WASP-47 b and 5.76$\pm$1.30\,min for WASP-47 d), which are consistent with the results of \citet{Becker2015}, and the mid-transit time of WASP-47 d at the epoch of our RM observation, with a precision of 4\,min.

\begin{figure*}
\begin{minipage}[tbh!]{\textwidth}
\includegraphics[trim=0cm 0cm 0cm 0cm,clip=true,width=\columnwidth]{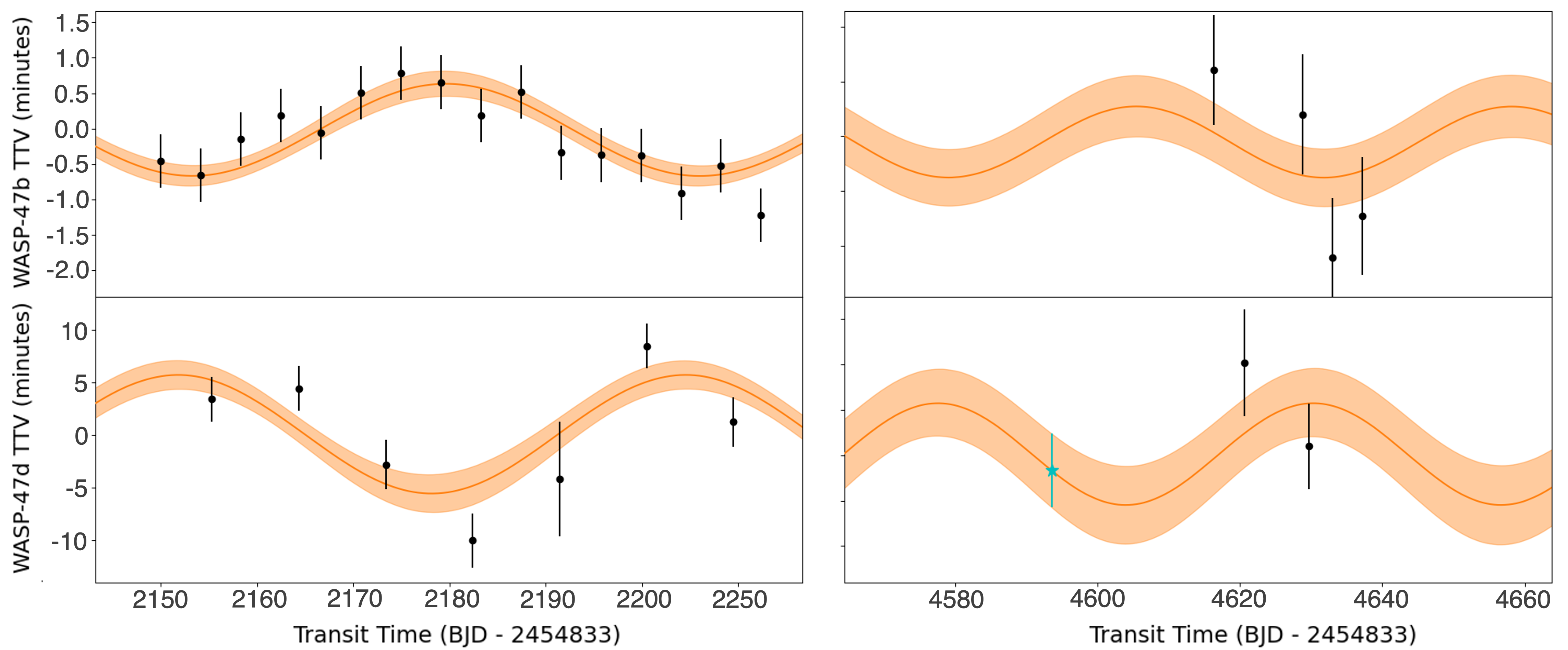}
\centering
\end{minipage}
    \caption[]{Measured TTVs for WASP-47 b (K2, \textit{upper left panel}; TESS, \textit{upper right panel}) and WASP-47 d (K2, \textit{lower left panel}; Euler, \textit{lower right panel}). The orange line shows our best-fit TTV model with the associated 1$\sigma$ envelope. The blue star shows the mid-transit time for WASP-47 d predicted at the time of our RM observation. }
\label{fig:TTV_WASP47}
\end{figure*}

By chance, WASP-47 e transited toward the end of the RM visit. The last and first contacts of WASP-47 d and e, respectively, occurred 3\,h 51\,min and 4\,h 44\,min after the start of the RM visit, that is during exposures at indexes 13 and 18. Only during the last exposure kept in our analysis may we thus have observed the transit of WASP-47 e. Given that the planet was then transiting the stellar limb, and that it yields a similar transit depth as Kepler-68 b (for which the RM signal could not be detected) across a star two orders of magnitude fainter, we can safely neglect its influence.

The occultation signal from WASP-47 d is too faint to be detected confidently in our data (Fig.~\ref{fig:RMR_maps}). The RMR fit hints at an aligned system ($\lambda$ = 4$\pm$53\,$^{\circ}$) but the model local line is consistent with a null detection within 2$\sigma$, and the projected stellar rotational velocity is not constrained. Follow-up RM observations with a larger telescope are needed to confirm this tentative measurement and determine whether WASP-47 d shares the same orbital plane as WASP-47 b.

We can still put constraints on the system architecture by combining the projected stellar rotational velocity and spin--orbit angle of WASP-47 b from \citet{SanchisOjeda2015} with the stellar rotation period recently derived by \citet{Bryant2022}. This yields $i_\star$ = 69.9$^{+10.9}_{-9.3}\,^{\circ}$ and $\psi$ = 29.2$^{+11.1}_{-13.3}\,^{\circ}$, showing that the star is seen nearly equator-on and that the giant planet is consistent with being aligned.

%-----------------------------

\subsection{WASP-107}
\label{sec:WASP107}

\subsubsection{Background}

With twice the mass of Neptune but a radius similar to Jupiter, WASP-107 b is one of the least-dense known exoplanets (\citealt{Anderson2017}). This requires an atmosphere dominated by hydrogen and helium, which has been observed to be evaporating under the strong irradiation from the active K dwarf host (\citealt{Spake2018,Allart2019,Kirk2020,Spake2021}). \citet{Piaulet2021} showed that the internal structure of WASP-107 b is consistent with an envelope mass fraction larger than 85\% and a core mass smaller than 4.6\,$M_\oplus$. These authors propose that accretion of primordial gas beyond 1\,au was stunted by migration to the inner disk, limiting WASP-107 b to a Neptune-mass planet, and that it only reached its current orbit in recent times because the core would not have been massive enough for the atmosphere to survive photoevaporation over the age of the system.

This makes the dynamical history of WASP-107 b particularly interesting. It cannot be constrained by the present orbital eccentricity, which is not well constrained and consistent with being circular, as expected from the short timescale for tidal circularization ($\sim$60\,Myr, \citealt{Piaulet2021}). The lack of recurring spot crossings during consecutive transits led \citet{Mocnik2017} and \citet{Dai2017} to conclude the orbit is highly misaligned, with a 3D spin--orbit angle between $40 - 140^{\circ}$. Indeed, because the star shows large persistent spots and its rotation period is about three times the planet orbital period, spot crossings would recur every three transits in an aligned system. A classical analysis of the RM effect confirmed these results, concluding it has a polar, retrograde orbit (\citealt{Rubenzahl2021}). This points toward dynamical scenarios in which WASP-107 c, a massive nontransiting companion on a wide eccentric orbit (\citealt{Piaulet2021}), played a significant role. The orbital architecture of the system could be explained by nodal precession, disk dispersal-driven tilting, or Kozai--Lidov resonance, provided that WASP-107 b and c had some degree of mutual inclination (\citealt{Piaulet2021}, \citealt{Rubenzahl2021}). The latter scenario was discarded because Kozai–Lidov cycles would be suppressed by general relativity precession unless the true mass of WASP-107 c is ten times larger than its projected mass, which requires that it is on a near face-on orbit. Yet this configuration is actually quite likely, as WASP-107c would orbit within the stellar equatorial plane if it remained within the protoplanetary disk, and the star is seen nearly pole-on (\citealt{Rubenzahl2021}). Gaia astrometric observations of the orbit of WASP-107c and precise measurements of WASP-107 b orbital architecture are needed to constrain detailed secular simulations of the system and further investigate these scenarios. 

\subsubsection{Update}

We exploited three spectroscopic transits of WASP-107 b observed with HARPS on 06 April 2014, 01 February 2018 and 13 March 2018, and one transit observed with CARMENES on 24 February 2018. The last exposures were excluded from the CARMENES visit and from the HARPS visits on 06 April 2014 and 13 March 2018 (outlying CCF properties). The first HARPS visit was obtained with a S/N of about half that of the other visits, and we had to apply a contrast-vs-S/N correction possibly linked with uncorrected Moon contamination (see Sect.~\ref{sec:CCF_corr}), so that the CCF for this visit are of lower quality. We use the ephemeris from \citet{Dai2017}, which yields precisions between 37 and 54\,s on the mid-transit times at the epoch of the RM observations. 

Our analysis of WASP-107's long-term photometry (Sect.~\ref{sec:APT_photom}, Fig.~\ref{fig:Longterm_photom}) yields a dominant signal at about 8.7 days (consistent with the analysis of the first observing season by \citealt{Spake2018}) and a secondary signal at 15.5 days. A signal close, but not exactly on the first harmonic of the rotational period, can dominate the photometric modulation (see the case of RV data from the Sun, \citealt{Hara2021}). Thus we consider the signal at 15.5 days to trace the true rotation period, as it is consistent with the values derived from WASP data over 2009-2010 (17.1$\pm1$\,days, \citealt{Anderson2017}) and from K2 data of 2016 (17.5$\pm$1.4\,days, \citealt{Mocnik2017,Dai2017}).

\begin{figure}[hbtp]
\centering
\includegraphics[trim=0cm 0cm 0cm 0cm, clip=true, width=\columnwidth]{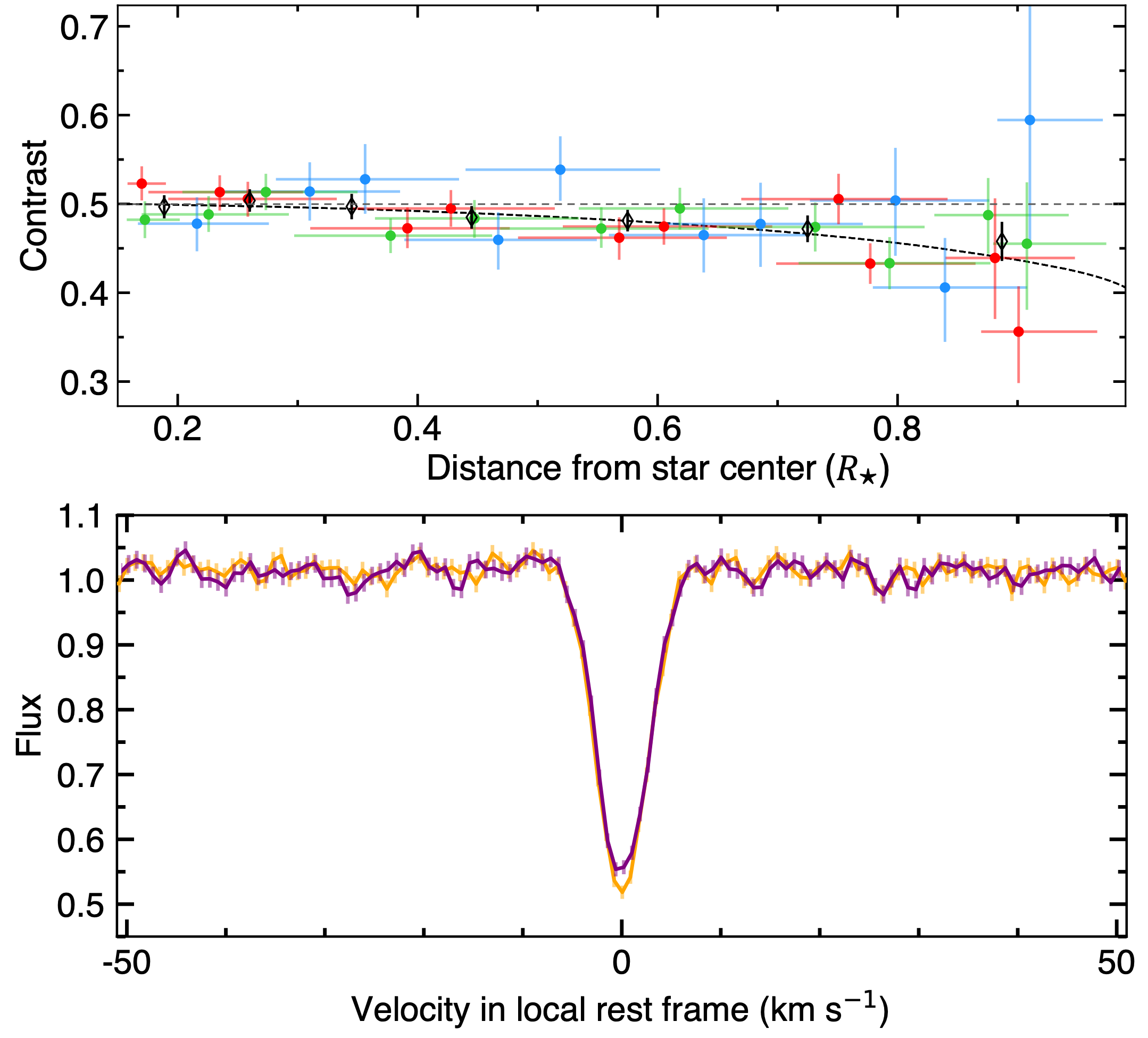}
\caption{Variations in WASP-107 line contrast measured with HARPS. \textit{Upper panel:} Contrast as a function of the projected distance to the star center $r = \sqrt{1-\mu^2}$ (for a given exposure time the range covered by the planet in $r$ is closer to being constant than in $\mu$). Measurements, derived from the fit to individual exposures (Sect.~\ref{sec:CCF_fits}) in each visit (1 = blue, 2 = red, 3 = green), are binned into the black points. The dotted black line is the best-fit model from the joint RMR fit, to be compared with a constant contrast of 50\% (horizontal gray line). \textit{Lower panel:} Average CCF$_\mathrm{Intr}$ at r = 0.34 (orange, from profiles over r = 0--0.5) and at r = 0.72 (purple, from profiles over r = 0.5--1). The stellar line keeps its Gaussian profile but is markedly deeper toward the limbs.}
\label{fig:WASP107_var}
\end{figure}

The RMR model provides a very good fit to the CCF$_\mathrm{Intr}$ in all four epochs (Fig.~\ref{fig:RMR_maps}, ~\ref{fig:RMR_resB}). We detect a clear center-to-limb variation in the contrast of the intrinsic stellar line, with a linear dependence in $\mu$ at a rate of 0.28$\pm$0.06. This model is strongly preferred over a constant line profile, with a common rate for all epochs and instruments and a trend consistent with predictions from 3D MHD
simulations of K dwarfs (\citealt{Cegla2016}), supporting a variation of stellar origin. The S/N of the combined HARPS data is high enough that the contrast difference between the disk center and its limbs can be seen by eye (Fig.~\ref{fig:WASP107_var}). This makes WASP-107 b particularly interesting for follow-up transit spectroscopy aimed at characterizing the stellar surface.

We derive $v_\mathrm{eq} \sin i_\star$ = $0.507^{+0.072}_{-0.086}$\,km s$^{-1}$ and $\lambda$ = $-158.0^{+15.2}_{-18.5}$\,$^{\circ}$, consistent with the results of \citet{Rubenzahl2021}. We note however that they adopted transit parameters from \citet{Dai2017}, which come from the sole analysis of K2 photometry and are less precise than the properties we used. In particular their impact parameter is consistent with 0, which results in a degeneracy over |$\lambda$| = 118$^{+38}_{-19}$\,$^{\circ}$ and a PDF for the projected rotational velocity consistent with 0\,km\,s$^{-1}$. 

Priors on the stellar rotation period and radius, combined with the precise constraint on $v_\mathrm{eq} \sin i_\star$ brought by the fitted dataset, allow us to derive a low stellar inclination $i_\star$ = $15.1^{+2.1}_{-2.6}\,^{\circ}$ and a 3D spin--orbit angle $\psi$ = 103.5$\stackrel{+1.7}{_{-1.8}}\,^{\circ}$. We thus confirm that WASP-107 b is on a retrograde and, due to a low impact parameter and a star seen pole-on, polar orbit. As an evaporating, low-density giant planet on a polar orbit at the edge of the Neptunian desert, WASP-107 b is a target of choice to investigate the processes behind its formation.

%-----------------------------

\subsection{WASP-156}

\subsubsection{Background}

With half the radius of Jupiter, a mass of 0.128\,$M_{\rm Jup}$, and a heavy element mass fraction of $\sim$90\% similar to that of Uranus and Neptune, WASP-156 b (\citealt{Demangeon2018}) stands at the transition between ice and gas giants. On a short orbit around a K star, this hot super-Neptune lies within the Neptunian desert. \citet{Demangeon2018} propose that the discrepancy between its gyrochronological/isochronal age hints at high-eccentricity migration, in relation with the transfer of angular momentum from the planet to the star during tidal circularization. WASP-156 b may thus have migrated recently into the desert and would be losing its gaseous envelope in a short-lived evolutionary phase. Determining the orbital architecture of the system is of high interest to investigate this scenario.

\subsubsection{Update}

We exploited three spectroscopic transits of WASP-156 b observed with CARMENES on 28 September 2019, 25 October 2019, and 10 December 2019. We excluded from the visit of 28 September 2019 the exposures at indexes 0 (used to adjust the exposure time), 17 (star guidance lost) and 18 (interrupted). 

We observed four photometric transits of the planet with STELLA. The derived ephemeris is consistent within 1$\sigma$ with the literature (\citealt{Demangeon2018}) and were used for our analysis. They yield a precision of about 40\,s on the mid-transit times of the RM observations.

The planet-occulted line is detected in the three visits. However, analysis of the residual CCFs revealed spurious features in all visits, correlating with airmass and likely caused by tellurics.

In the visit on 28 September 2019 the strongest features are found in the pretransit and in the last exposures, all of which were obtained at airmass $>$1.5. We found that excluding those exposures from the master-out CCF$_\mathrm{DI}$ yields much cleaner residual CCFs in the remaining exposures, in particular during transit (Fig.~\ref{fig:RMR_maps}). A preliminary RMR fit returned a broad PDF for $v_\mathrm{eq} \sin i_\star$, peaking at 0\,km s$^{-1}$ and $<$6\,km s$^{-1}$ at 3$\sigma$ and a PDF for $\lambda$ with a well-defined peak at about 100\,$^{\circ}$ but shallow wings covering the entire parameter space. Setting a prior on $v_\mathrm{eq} \sin i_\star$ from the spectroscopic value of \citet{Demangeon2018} ($3.80\pm0.91$\,km s$^{-1}$) removes these wings and yields $\lambda$ = $105.7^{+14.0}_{-14.4}$\,$^{\circ}$.

The visit on 25 October 2019 shows even stronger telluric features, as airmass decreases below 1.5 only at mid-transit. Even with the master-out CCF$_\mathrm{DI}$ calculated from post-transit exposures alone, the CCF$_\mathrm{Intr}$ obtained during the second half of the transit remain contaminated. Fitting them with the same prior on $v_\mathrm{eq} \sin i_\star$ returns a model stellar line that is too deep and narrow compared to the disk-integrated line and a bi-modal PDF for $\lambda$ with the main mode peaking at about -60\,$^{\circ}$. Although the second mode corresponds to the PDF for $\lambda$ derived in the visit on 28 September 2019, it might still be biased by the residual contamination, and we opt to exclude the visit on 25 October 2019 from our analysis.

In the visit on 10 December 2019 the airmass only increases beyond 1.5 after egress. However the contrast of the CCF$_{\rm DI}$ shows abnormal variations over the entire visit, which we could not fully correct because they only partly correlate with time and S/N. The residual CCFs show spurious features even during transit, which contaminate the planet-occulted line. We thus also excluded this visit from our analysis.

Our final results are thus derived from the visit on 28 September 2019 and, given that $\lambda$ is close to 90\,$^{\circ}$, suggest a polar orbit for WASP-156 b. This would support the high-eccentricity migration scenario proposed by \citet{Demangeon2018}, although the short orbital distance of WASP-156 b makes it possible that its spin--orbit angle was further altered by tidal interactions with its cool host star. Given the high level of contamination of the processed CARMENES datasets, follow-up RM observations are encouraged to confirm our measurement.

%-----------------------------

\subsection{WASP-166}
\label{sec:WASP166}

\subsubsection{Background}

WASP-166 b is a bloated, low-density super-Neptune on a close and aligned orbit around an F-star (\citealt{Hellier2019}). Sodium was detected at high velocities and up to high altitudes in the planetary atmosphere, hinting at its hydrodynamical escape (\citealt{Seidel2020b,Seidel2022}). 

\subsubsection{Update}

We exploited three spectroscopic transits of WASP-166 b observed with HARPS on 14 January 2017, 04 March 2017, and 15 March 2017. We excluded exposures at indexes 0, 1, 2, 5 from the visit on 14 January 2017 (used to adjust the exposure time, or outliers to the CCF property series), and at index 32 from the visit on 15 March 2017 (lowest S/N). The ephemeris from \citet{Bryant2020} yields precisions of $\sim$1.2\,min on the mid-transit times at the epochs of our RM observations.

The planet-occulted lines are well-detected and modeled in the three epochs (Fig.~\ref{fig:RMR_maps},~\ref{fig:RMR_resB}). There is a hint that the local stellar line started evolving in the last epoch, but BIC comparison favors a common line profile for all three visits. Analyzing two ESPRESSO transits with the reloaded RM method (\citealt{Cegla2016}), \citet{Doyle2022} noticed that the FWHM of the local line profile appears to increase linearly with $\mu$ toward the stellar limb, with a slope of about 2.4\,km\,s$^{-1}$. Interestingly the HARPS data appears to confirm this hint, with a slope of 2.3$\pm$0.8\,km\,s$^{-1}$ when allowing the width of our model line profile to vary linearly with $\mu$. The corresponding BIC is however the same than for a constant line profile, and the values for $v_\mathrm{eq} \sin i_\star$ and $\lambda$ are unchanged, so that we adopt this simpler model. 

\citet{Doyle2022} further conclude that the ESPRESSO data is best modeled by solid-body rotation plus quadratic convective RV motions when fit at its original temporal resolution, or by solid-body rotation plus cubic convective RV motions when fit at a downsampled resolution of 10\,min to mitigate p-mode oscillations. They further claim a constraint on differential rotation when modeled together with linear convective RV motions. To compare with the results from \citet{Doyle2022} we fit these various models to the HARPS data at its original temporal resolution or binned by two exposures to reach a 10\,min resolution. We find that the HARPS data, at their original or binned resolution, are not sensitive enough to constrain convective motions and/or differential rotation with coefficients consistent with zero and BIC values significantly larger than for a pure solid-body model. We derive consistent values for the original ($v_\mathrm{eq} \sin i_\star$ = $5.4\pm0.14$\,km\,s$^{-1}$, $\lambda$ = $-0.7\pm1.6$\,$^{\circ}$) and binned ($v_\mathrm{eq} \sin i_\star$ = $5.4\pm0.2$\,km\,s$^{-1}$, $\lambda$ = $-1.0\pm1.8$\,$^{\circ}$) data, showing that our analyis is not impacted by p-mode oscillations. These results are consistent with those from \citet{Hellier2019} ($v_\mathrm{eq} \sin i_\star$ = $5.1\pm0.3$\,km\,s$^{-1}$, $\lambda$ = $3\pm5$\,$^{\circ}$), derived at lower precision from a classical RM analysis of the same HARPS datasets. Our RM Revolutions analysis of these three HARPS transits yields a comparable precision on $\lambda$ than \citet{Doyle2022}'s Reloaded RM analysis of two ESPRESSO transits. Our value for $\lambda$ is consistent with their solid-body fit to the original ($-4.49\pm1.74$\,$^{\circ}$) and binned ($-5.93\pm2.00$\,$^{\circ}$) ESPRESSO data, but their values for $v_\mathrm{eq} \sin i_\star$ (original, $4.89\pm0.08$\,km\,s$^{-1}$; binned, $4.77\pm0.09$\,km\,s$^{-1}$) are significantly lower than ours. We do not know the origin of this discrepancy, but we note that our results are more consistent with those derived by \citet{Doyle2022} when they fit the RV centroids from the line core only ($\lambda$ = $1.01\pm1.70$\,$^{\circ}$, $v_\mathrm{eq} \sin i_\star$ = $5.09\pm0.08$\,km\,s$^{-1}$). We adopt the solid-body fit to the original HARPS data as our final results.

\citet{Hellier2019} measured the stellar rotation period to be $12.3\pm1.9$\,days from a Gaussian-process analysis of the RV residuals to the Keplerian motion. If this value, currently estimated over only part of the putative stellar rotation, is confirmed, it would point toward a rare occurrence. Our fit with priors on the stellar rotation period and radius indeed results in the star being seen edge-on ($i_\star$ = $87.9^{+22.9}_{-19.3}$\,$^{\circ}$) and a system likely truly aligned ($\psi<$22.0\,$^{\circ}$ at 1$\sigma$, $<$56.2\,$^{\circ}$ at 3$\sigma$).

An aligned orbit would suggest an in situ formation or early disk-driven migration, considering the small probability that a high-eccentricity migration ends with an aligned orbit. In those scenarios, however, the planetary atmosphere must have survived evaporation for 2\,Gyr despite its location at the rim of the Neptune desert. This is surprising considering its strong present-day irradiation, which would have been even larger during the stellar saturation phase. An alternative would be that the planet underwent high-eccentricity migration recently and was fast realigned, but the low convective mass of its F-type host star and subsequent weak tidal interactions belies that hypothesis. A more precise measurement of the system 3D spin--orbit angle might help determine if there is still a substantial misalignment that could trace high-eccentricity migration without the need for realignment, or if the nature of WASP-166 b makes it resilient to photo-evaporation.

%--------------------------------------------------------------------

%-----------------------------------------------------------------

\section{Discussion and conclusion}
\label{sec:conclu}

Planets around the Neptunian desert and savanna are targets of choice to investigate the processes behind the formation of these features, which is the main objective of the SPICE DUNE project. In particular, acquiring knowledge on the orbital architecture of close-in exoplanets is critical to our understanding of their dynamical history. This is the aim of this first paper in the DREAM series, where we used the Rossiter--McLaughlin Revolutions technique to perform a homogeneous analysis of 26 transit spectroscopy datasets on 14 close-in exoplanets. 

We first refined the properties of the studied planets and their host stars using RVs, long-term photometry and transit photometry. We revised the Keplerian properties of the HAT-P-33, HAT-P-49, and HD\,89345 systems, and improved the ephemeris precision for several planets in our sample, which should prove useful for follow-up transit observations. Transit spectroscopy datasets were reduced into CCFs using standard and custom pipelines, and corrected for various systematics to improve their quality. We exploited our large dataset to determine the optimal CCF masks as a function of stellar spectral type. We found that masks customized to a specific target star, rather than representative of a spectral type proxy, are not justified for F-type stars but substantially improve the quality of CCF time series for G-type and, particularly, K-type stars. Beyond the interest to study the RM signal and stellar activity over short timescales (e.g., granulation), we highlight the possible benefits of using custom masks for a Keplerian analysis if they similarly decrease the dispersion of RV series over longer timescales.

\begin{table*}
\tiny
	\caption{Summary of orbital architecture results.}
	\label{tab_RM_results} 
	\centering  
%	\resizebox{\linewidth}{!}{
	
	\begin{tabular}{c | cc  | cc | cc  }          
		\hline \hline             
		  Target &  \multicolumn{2}{c}{Stellar spin inclination} & \multicolumn{2}{c}{Sky-projected spin-orbit angle} & \multicolumn{2}{c}{3D spin-orbit angle}   \\	
		\hline	
		   &  \multicolumn{2}{c}{$i_\star$ (deg)} & \multicolumn{2}{c}{$\lambda$ (deg)} & \multicolumn{2}{c}{$\psi$ (deg)}   \\	
		\hline		
		   & Literature & This work & Literature & This work & Literature & This work   \\	
		\hline		
HAT-P-3 b  &  -- & 15.8$\stackrel{+6.2}{_{-7.3}}$  &  $21.2\pm8.7$ & $-25.3\stackrel{+29.4}{_{-22.8}}$ & -- & 75.7$\stackrel{+8.5}{_{-7.9}}$  \\
HAT-P-11 b  & $160^{+9}_{-19}$  &  $146.8^{+7.6}_{-6.2}$ &  $121^{+24}_{-21}$ & $133.9^{+7.1}_{-8.3}$  & $97^{+8}_{-4}$ &  $104.9^{+8.6}_{-9.1}$  \\
HAT-P-33 b  & -- & -- & --  & $-5.9\pm4.1$ & -- & --   \\
HAT-P-49 b  & -- & -- & --  & $-97.7\pm1.8$ & -- & --  \\
HD 89345 b  &  $>44$ & -- & --   &  $74.2^{+33.6}_{-32.5}$ & -- &    80.1$\stackrel{+22.3}{_{-23.1}}$  \\
HD 106315 c  & -- & -- &  $-10.9^{+3.6}_{-3.8}$  & $-2.68^{+2.7}_{-2.6}$ & -- & -- \\
K2-105 b  & -- & -- & --  &  $-81^{+50}_{-47}$  & -- & --  \\
Kepler-25 c & $66.7^{+12.1}_{-7.4}$ &  -- & $9.4 \pm 7.1$ &  $-0.9\stackrel{+7.7}{_{-6.4}}$ &  $26.9^{+7.0}_{-9.2}$   & $24.1\stackrel{+9.2}{_{-9.3}}$ \\
Kepler-63 b  & $138\pm7$ & -- &  $-110^{+22}_{-14}$ & $-135^{+21.2}_{-26.8}$  & $104^{+9}_{-14}$ & $114.6^{+16.6}_{-12.5}$  \\
Kepler-68 b  & -- & -- & -- & -- & -- & -- \\
WASP-47 b  & -- &  $69.9^{+10.9}_{-9.3}$ &  0$\pm$24  & --  & -- &  29.2$\stackrel{+11.1}{_{-13.3}}$  \\
WASP-47 d  &  &                        &  --  & --  & -- & --  \\
WASP-107 b &  --  &  $15.1^{+2.1}_{-2.6}$ &  $\pm118.1^{+37.8}_{-19.1}$  & $-158.0^{+15.2}_{-18.5}$ & $109.81^{+28.17}_{-13.64}$ &  103.5$\stackrel{+1.7}{_{-1.8}}$  \\
WASP-156 b & --  &  -- & -- & $105.7^{+14.0}_{-14.4}$  & -- & -- \\
WASP-166 b & --  & $87.9^{+22.9}_{-19.3}$ & $-4.5\pm1.7$  &  $-0.7\pm1.6$  & -- & $<$22.0 (1$\sigma$) ; $<$56.2 (3$\sigma$)   \\

		\hline \hline 
	\end{tabular}
%	}
	\tablefoot{
	References for the literature measurements can be found in the tables specific to each system (Appendix \ref{apn:sys_prop}). We only report here $\psi$ from the combined northern and southern stellar orientations, or from the preferred one (see text). 
	}
\end{table*}

Our RM analysis yielded a detection of the planet-occulted stellar line for 12 of our targets and a nondetection for the two smallest planets Kepler-68 b and WASP-47 d. We detected variations in the stellar line shape along the transit chord of WASP-107 b and found hints of stellar convective blueshift along the transit chords of HAT-P-33 b and HD\,106315 c, motivating transit follow-up of these targets. The orientation of the transit chord could be constrained for the 12 targets, albeit tentatively for the small K2-105 b, so that we measured the sky-projected spin--orbit angle of five targets for the first time and refined its value for six other targets. Four host stars have known stellar inclination, and it could be constrained for three more. This allowed us to derive the 3D spin--orbit angle in seven systems, three of which (HAT-P-3 b, HD\,89345 b, WASP-156 b) had no such measurement previously. Our measurements of $\lambda$ and $\psi$ (Table~\ref{tab_RM_results} ) thus represent a useful addition to the sample of known spin--orbit angles, which previously consisted in 191 sky-projected and 39 3D values (31 August 2022, based on the TEPCat catalog \citep{Southworth2011}). We caution about the over-interpretation of the sky-projected spin-orbit angle, which strongly differs from the 3D spin-orbit angle for several of our systems. We will delve more deeply into this matter in DREAM II and recommend the combination of RM observations with measurements of the stellar inclination or rotation period  whenever possible.

While a preponderance of polar orbits is expected from the recent analysis of \citet{Albrecht2021}, it is noteworthy that nine out of twelve planets in our constrained sample are on highly misaligned orbits. In fact, only WASP-166 b, Kepler-25 c, HD106315 c have aligned or moderately misaligned orbits. Remarkably, the latter two planets are the only ones in our sample with close planetary companions (alongside WASP-47 d, for which the data also hinted at an aligned orbit), supporting the idea that planets in compact systems migrate together within the disk. We can further surmise that primordial tilting only results in small misalignments between the star and the protoplanetary disk, in contrast to disruptive dynamical processes placing single close-in planets on highly misaligned orbits. 

Our nine planets on misaligned orbits range from mini-Neptune to Jupiter-size and orbit F-, G-, and K-type stars. This shows that high-misalignment processes are not restricted to a specific type of system, at least for gas-rich planets. The exact dynamical processes and their behavior, however, likely depend on the planet and star properties. High-eccentricity migration induced by an outer massive companion, for example, is favored for WASP-107 b, HAT-P-11 b, and WASP-156 b. It is particularly interesting that these three planets are warm Neptunes located at the border of the hot Neptunes desert, confirmed or suspected to evaporate. These features are similar to those of GJ\,436 b and GJ\,3470 b and strengthen the idea proposed by \citet{Bourrier_2018_Nat} that a fraction of planets at the desert rim are late Neptunian migrators, which survive atmospheric erosion because it was triggered only recently. In that context, the case of WASP-166 b is intriguing because it is similar to the above planets yet appears to be on an aligned orbit. Whether it survived evaporation because of a peculiar nature, or because it migrated late through a process that maintained or recreated its alignment, refining its 3D orbital architecture is of high interest.

Going further, one can wonder whether high-eccentricity migration leads Neptune-size planets to cover the breadth of the desert and savanna, or whether it ends preferentially within the desert. HD\,89345 b and Kepler-63 b, located on misaligned orbits right within the savanna, are prime targets for follow-up studies addressing this question. The weak star/planet tidal interactions of the former, and the young age of the latter, means that we can probe the primordial, unaltered processes that led to their current architecture. Extending spin-orbit angle measurements to a wider variety of systems across the savanna will be useful to investigate the relative roles of early smooth / late disruptive migrations and their impact of the atmopsheric evolution of close-in Neptunes.

%-----------------------------------------------------------------

\begin{acknowledgements}
We thank the referee for their concise review.
We thank the TNG and Calar Alto staff for their help in carrying out our observing programs.
We thank M. Battley for sharing the mid-transit times of Kepler-25 b and c.
We thank V. van Eylen for his opinion on Kepler-68 b ephemeris.
%We thank E. Bryant and D. Bayliss for discussions over the WASP-47 planet ephemeris. 

During the preparation of this paper we became aware that the HARPS datasets of WASP-107 b and WASP-166 b were searched for RM signal by Kunovac et al. (in prep). We conducted our respective studies independently.

This work has been carried out within the framework of the NCCR PlanetS supported by the Swiss National Science Foundation under grants 51NF40$_{}$182901 and 51NF40$_{}$205606. This project has received funding from the European Research Council (ERC) under the European Union's Horizon 2020 research and innovation programme (project {\sc Spice Dune}, grant agreement No 947634; grant agreement No 730890). This material reflects only the authors views and the Commission is not liable for any use that may be made of the information contained therein.

M.L. acknowledges support of the Swiss National Science Foundation under grant number PCEFP2\_194576. The contributions of M.L. and A.P. have been carried out within the framework of the NCCR PlanetS supported by the Swiss National Science Foundation.

R. A. is a Trottier Postdoctoral Fellow and acknowledges support from the Trottier Family Foundation. This work was supported in part through a grant from FRQNT.

G.W.H. acknowledges long-term support from NASA, NSF, Tennessee State University, and the State of Tennessee through its Centers of Excellence program. Astronomy at Tennessee State University is supported by Tennessee State  University and the State of Tennessee through its Centers of Excellence program.

HMC acknowledges funding from a UK Research and Innovation (UKRI) Future Leader Fellowship, grant number MR/S035214/1
\end{acknowledgements}

\bibliographystyle{aa} % style aa.bst
\bibliography{biblio} % your references Yourfile.bib

\begin{appendix}

\section{Selection of optimal reduction}
\label{apn:opt_reduc}

\begin{table*}
\centering
\rotatebox{90}{
\tiny
%\begin{adjustbox}{scale=0.95,center}
\begin{minipage}[t][][b]{1.5\textwidth}
\caption{Quality assessment of CCF$_\mathrm{DI}$}
\begin{tabular}{lc|l|cccccc|cccccc|cccc}
\hline
Star & Type   & Visit &  \multicolumn{6}{c}{Contrast}&   \multicolumn{6}{c}{FWHM} &   \multicolumn{4}{c}{RV residuals}  \\
\hline
 &  &  &   \multicolumn{3}{c}{Improved} & \multicolumn{3}{c}{Custom}  &  \multicolumn{3}{c}{Improved} & \multicolumn{3}{c}{Custom}   &  \multicolumn{2}{c}{Improved} & \multicolumn{2}{c}{Custom}   \\
\hline
&  &  & Mean &  e$^\mathrm{rel}$ & $\sigma^\mathrm{rel}$  & Mean   &  e$^\mathrm{rel}$& $\sigma^\mathrm{rel}$     & Mean  &   e$^\mathrm{rel}$  & $\sigma^\mathrm{rel}$ & Mean  &  e$^\mathrm{rel}$  & $\sigma^\mathrm{rel}$   &  e  & $\sigma$  &  e  & $\sigma$   \\ 
&  &  & \% &  \textpertenthousand &  \textpertenthousand & \%    &  \textpertenthousand& \textpertenthousand    & km\,s$^{-1}$ &   \textpertenthousand & \textpertenthousand  & km\,s$^{-1}$ &   \textpertenthousand  & \textpertenthousand   &  m\,s$^{-1}$ & m\,s$^{-1}$  &  m\,s$^{-1}$ &  m\,s$^{-1}$   \\   
\hline
\hline
HAT-P\,49     &  F3  & HN & 22 & 29.3  &  63.8 & 27   & 24.5 (-16\%)   & 56.5 (-11\%)   & 18.0 &  35.7  &  56.8 &  18.1  &  30.3  (-15\%)  & 46.0 (-19\%)   &  27.0  &  46.0 &  22.8 (-15\%)  &   41.6 (-10\%)      \\ 
              &    &  HN$^{+}$    &     &  29.1  &  33.1     &     &      \textbf{24.4 (-16\%)}     &       \textbf{25.0 (-24\%)} &     &  34.9  & 31.3 &   &  \textbf{29.8 (-15\%)} & \textbf{24.2 (-23\%)}   &  26.5  & 29.3   & \textbf{22.6 (-15\%)}  &  \textbf{29.0 (-1\%)}     \\ 
\hline 
HAT-P-33    &  F4  &   HN & 20 &  36.6  &  33.8 &  25  &  29.5 (-19\%)   &  39.4 (+17\%) &    19.9 &  44.0  &  40.6 &  20.3  & \textbf{36.1 (-18\%)}  &  \textbf{33.4 (-18\%)} &  36.7  &  37.4 &  \textbf{30.6 (-17\%)}  &  \textbf{38.3 (+2\%)}    \\ 
       &      & HN$^{\dagger}$  &    &    &   &  & \textbf{29.5} &  \textbf{24.0}  &    &   &   &    &   &     & &   &  &    \\ 
\hline 
HD\,106315    &  F5   & HS1$^{\leftmoon}$ & 20 & 11.0  &  9.1 &  29 & \textbf{9.3 (-15\%)}  &  \textbf{7.1 (-22\%)} & 17.9 & 13.4  &  11.7 & 18.3 & \textbf{11.6 (-13\%)}  &  \textbf{8.0 (-32\%)} &  10.0  & 9.6 & \textbf{ 8.8 (-12\%)}  &  \textbf{8.5 (-11\%)}    \\ 
  &     & HS2  & 20  &  10.0  &  11.1 & 29 &  \textbf{8.5 (-15\%)}  &  \textbf{8.9 (-20\%)} & 17.9 &  12.2  &  8.0 &  18.3 & \textbf{10.6 (-13\%)}  &  \textbf{11.3 (+40\%)} &  9.1  &  8.7 & \textbf{ 8.1 (-11\%)}  &  \textbf{9.1 (+5\%)}    \\ 
  &     & HS3  & 20 &  7.4  &  5.5 & 29 & \textbf{ 6.3 (-15\%)}  &  \textbf{5.4 (-2\%)} & 17.9 & 9.0  &  8.4 & 18.3 & \textbf{7.9 (-13\%)}  &  \textbf{5.9 (-30\%)} &  6.7  &  5.7 &  \textbf{5.9 (-11\%)}  &  \textbf{6.3 (+11\%)}    \\ 
\hline 
Kepler-25    &  F8   & HN       & 29 &  13.6  &   15.1 & 32 &  11.8 (-14\%)  &  13.0 (-14\%) &  13.9 &  \textbf{17.0}  &   \textbf{12.7} & 14.2 &  14.9 (-13\%)  &  13.2 (+4\%) &   \textbf{9.8}  &   \textbf{7.2} &  8.7 (-11\%)  &  8.3 (+14\%)    \\ 
        &       & HN$^{+}$ &    &  \textbf{13.6}  &   \textbf{10.8} &    &  11.8 (-14\%) &     9.3 (-14\%) &        &                 &                 &      &              &               &       &                 &                 &       \\     
\hline 
WASP-166         &  F9  & HS1$^{\leftmoon}$  & 50 &  6.1  &  7.9 &  42  &  \textbf{5.8 (-6\%)}  &  \textbf{6.7 (-16\%)} &     9.6 & 8.2  &  9.0 & 9.4 &  \textbf{7.5 (-8\%)}  &  \textbf{8.3 (-9\%)} &  3.2  &  3.1 &  \textbf{2.9 (-10\%)}  &  \textbf{3.0 (-3\%)}    \\ 
             &         & HS2  & 50 & 8.0  &  8.0 &  42 &  7.6 (-5\%)  &  9.9 (25\%) & 9.6 &  10.8  &  10.2 & 9.4 &  \textbf{10.0 (-7\%)}  &  \textbf{12.2 (+20\%)} &  4.3  &  3.6 &  \textbf{3.8 (-10\%)}  &  \textbf{2.7 (-26\%)}    \\ 
             &        &  HS2$^{+}$        &   &     &     &   & \textbf{7.6}  &     \textbf{9.0}  & &   &   &    &   &   &  &    &      \\ 
             &        & HS3  & 50 & 6.7  &  7.2 &  42 &  \textbf{6.5 (-4\%)}  &  \textbf{6.6 (-9\%)} & 9.6 & 9.2  &  8.2 &  9.4 & \textbf{8.5 (-7\%)}  &  \textbf{7.8 (-5\%)} &  3.6  &  3.7 &  \textbf{3.3 (-9\%)}  &  \textbf{2.7 (-27\%)}    \\ 

\hline
Kepler-68    &  G1   & HN &  58 &  7.9  &  11.3 &  51 &  7.5 (-6\%)  &  9.6 (-15\%)  & 7.2 &  10.9  &  11.4 &  7.2  &   \textbf{10.1 (-8\%)}  &   \textbf{11.0 (-4\%)} &  3.2  &  3.1 &   \textbf{3.0 (-7\%)}  &   \textbf{2.9 (-4\%)}    \\ 
            &        & HN$\dagger$   &      &  7.9  &  8.1 &   &  \textbf{7.5 (-6\%)}  &  \textbf{6.6 (-18\%)} &     &    &   &   &    &    &   &     &   &   \\ 
\hline    
HD\,89345    &  G5   & HN & 67 & 12.1  &  21.1 &  53 &  10.3 (-15\%)  &  14.1 (-33\%)  & 7.8 &  16.0  &  23.4 & 7.1  & 13.3 (-17\%)  &  16.9 (-28\%)  &  5.1  &  7.8 &  \textbf{3.9 (-23\%)}  &  \textbf{5.3 (-32\%)}    \\ 
           &    &  HN$^{\dagger}$    &   &   &      &  & \textbf{10.3}  &   \textbf{13.3} &    &   &   &   &  \textbf{13.3}    &  \textbf{13.7} &   &  &    &     \\ 
\hline 
WASP-47       &  G9 & HN  & 64 & 12.0  &  13.0 & 52 &  \textbf{11.0 (-8\%)}  &  \textbf{9.8 (-25\%)}  & 7.2 & 16.0  &  30.0 &  6.7 &  \textbf{14.4 (-10\%)}  &  \textbf{21.5 (-28\%)} &  4.7  &  3.5 &  \textbf{4.0 (-15\%)}  &  \textbf{1.5 (-58\%)}    \\ 
\hline 
HAT-P-3   &  K1  &  HN  &  69 & 7.9  &  13.7 & 55 & \textbf{6.9 (-12\%)}  &  \textbf{9.4 (-32\%)} & 6.8 &  11.0  &  16.4 & 6.3 &  \textbf{9.4 (-15\%)}  &  \textbf{7.8 (-52\%)} &  3.0  &  4.9 & \textbf{2.4 (-20\%)}  &  \textbf{1.7 (-65\%)}   \\ 
         &       &  HN$^{\dagger}$   &   & 7.9 &  10.9 &    &   &   &  &  11.0    &  9.6  &   &   &   &    &   &   &        \\ 
\hline
K2-105    &  K1 &  HN & 61 &  15.1  &  11.5 &  51  & \textbf{11.9 (-21\%) } & \textbf{ 10.8 (-6\%)} & 6.9 &  19.7  &  22.8 &  6.8  &  \textbf{15.4 (-22\%)}  &  \textbf{11.7 (-49\%)} &  5.6  &  8.6 & \textbf{ 4.3 (-23\%) } & \textbf{ 4.1 (-53\%) }   \\ 
\hline 
WASP-156      &  K3  & C1  & - &  -  &  - &  51 &   \textbf{6.9}  &  \textbf{8.6} & - &  -  &  - & 4.1 &  10.5  &  45.6 &  -  &  - &  \textbf{2.7}  &  \textbf{2.8}    \\ 
             &      &  C1$^{+}$      &  -  &  -  &  - &       &   &      &  - & -& -&   &  \textbf{10.5} &  \textbf{11.9}   & - &   - &   &     \\ 
 &   &     C2  & - &  -  &  - & 51 & \textbf{6.4} &  \textbf{6.8} & - & -  &  - &  4.0 &  9.8  &  75.4 &  -  &  - &  \textbf{2.5}  &  \textbf{3.6}    \\ 
              &    &   C2$^{+}$    &  -  &  -  &  - &       &   &     &  - & - & -&   &  \textbf{9.8} & \textbf{9.1}  & - &  - &   &      \\ 
\hline 
HAT-P\,11  &  K4  & HN1  &  70 & 3.8  &  15.9 & 52 & 4.0 (+2\%)  &  8.3 (-48\%) &  7.4 & 5.7  &  17.5 & 5.7 & \textbf{5.5 (-4\%)}  &  \textbf{5.9 (-67\%)} &  1.7  &  2.3 &  \textbf{1.3 (-24\%)}  &  \textbf{1.5 (-33\%)}    \\ 
           & &    HN1$^{\dagger}$    &         &   3.8 &   6.1   &  &    \textbf{4.0 (+2\%)} &  \textbf{5.2 (-16\%)} & &  5.7  &  7.4 &   &  &  &   &  &   &     \\ 
   &    & HN2  &  70 & 4.1  & 12.1 & 52 & 4.3 (+5\%)  &  6.7 (-44\%)  & 7.4 &  6.2  &  12.7 & 5.7 & \textbf{6.0 (-4\%)}   & \textbf{ 5.7 (-55\%)}  &  1.8  &  2.1 &  \textbf{1.4 (-24\%)}   &  \textbf{1.5 (-26\%)}     \\ 
          & &     HN2$^{\dagger}$   &         &   4.1  &   4.7   &   &  \textbf{4.3 (+5\%)}    &  \textbf{4.0 (-14\%)} &   &  6.2  &  8.0 &  &  &   &   &  &   &    \\ 
       & & C1  & - & -  &  - & 49 &  3.6  &  7.0 &   -  & -  & - & 7.4  & 5.7  &  22.7 &  -  & -  &  1.5  &  2.1    \\ 
     &    &     C1$^{+}$  &  - &   -  &  - &    & \textbf{3.3} &   \textbf{6.0} & - &  -  &  - &    &  \textbf{5.1} & \textbf{8.1} &    - &  - &  \textbf{1.3} &  \textbf{1.8}    \\ 
     &  & C2  & - & -  &  - &  49 & 3.2  &  11.0 &  - &  -  &  - & 7.4 & 5.0  &  47.7 &  -  &  - &  1.3  &  3.9    \\ 
     &    &  C2$^{+}$     & - &  -  &  - &  &  \textbf{2.9}  &  \textbf{3.5} &  - &  -  &  - &    &  \textbf{4.5} &  \textbf{5.8} &  -  &  - &   \textbf{1.2}  &  \textbf{1.6}    \\ 
\hline    
WASP-107  &  K6  & HS1  & 56 & 25.8  &  60.9 & 46 &  22.7 (-12\%)  &  49.6 (-18\%)  & 6.0 & 32.4  &  109.6 &  6.0 &  \textbf{28.2 (-13\%)}  &  \textbf{23.2 (-79\%)} &  8.1  &  9.4 &  \textbf{7.0 (-13\%)}  &  \textbf{7.6 (-20\%)}    \\ 
       &         &  HS1$^{\dagger}$ &   &  25.8  & 53.6  &    &  \textbf{22.6 (-12\%)} & \textbf{23.1 (-57\%)} & &    &   &  &  &   &  &    &  &   \\ 
         &       & HS2$^{\leftmoon}$  & 55 & 14.5  &  24.3 &  46    &  \textbf{12.7 (-13\%)}  &   \textbf{10.0 (-59\%)}  & 6.0 & 18.7  &  22.1 & 6.0 &  \textbf{16.1 (-14\%)}  & \textbf{19.1 (-14\%)} &  4.6  &  4.4 &  \textbf{4.0 (-13\%)}  &  \textbf{3.8 (-14\%)}    \\ 
       &         &  HS2$^{\leftmoon\dagger}$ &   & 14.5 & 11.1 &   &   &   &  &    &   &   & &    &   &    &   &   \\          
              &    & HS3$^{\leftmoon}$  & 55 &  12.8  &  16.1 &  45 &  \textbf{11.3 (-12\%)}  &  \textbf{13.2 (-18\%)}  & 6.0 &  16.7  &  19.9 & 6.0 & \textbf{14.5 (-13\%)}  &  \textbf{14.8 (-25\%)} &  4.1  &  5.6 &  \textbf{3.6 (-12\%)}  &  \textbf{4.5 (-20\%)}    \\ 
   &     & C & - &  -  &  - & 46 &  \textbf{12.6}  &  \textbf{10.4} & - &  -  &  - & 28.1 &  \textbf{19.3}  &  \textbf{29.0} &  -  &  - & \textbf{4.8}  &  \textbf{5.2}    \\ 
   &     & C$^{+}$ & - &  -  &  - &    &   & & - &  -  &  - &  &   \textbf{19.3}  &  \textbf{22.7} &  -  &  - & -  &  -    \\

\hline 

\end{tabular}
\begin{tablenotes}[para,flushleft]
Notes: Systems have been ordered by spectral type. HS, HN, and C indicate the HARPS, HARPS-N, and CARMENES instruments. $e$ indicates the mean error on out-of-transit values. $\sigma$ indicates standard deviations with respect to the mean out-of-transit value. The $^\mathrm{rel}$ properties have been normalized by this mean, to allow for a direct comparison between cases. Number in parentheses indicate the variation in error or dispersion going from the improved mask to the custom mask. $\leftmoon$ indicate datasets corrected for sky contamination. $^{\dagger}$ and $^{+}$ indicate datasets where the reported property was corrected for its correlation with the S/N and time, respectively (mean values are not reported as they are not modified by these corrections). Values in bold highlight the reduction that was retained for the final analysis of each target. 
\end{tablenotes}
\label{tab:disp_full}
\end{minipage}}
\end{table*}

\section{Orbital solutions}
\label{apn:orb_sol}

% TABLE - Prior Distributions
\begin{table*}[hbtp]
\centering
\caption{Prior parameter distributions$\mathrm{^{\text{\textasteriskcentered}}}$ of the global fit with the \texttt{PyMC3} and \texttt{exoplanet} packages (see Section~\ref{sec:prior_distributions_and_posterior_sampling}).}\label{tab:prior_RV}
\begin{tabular}{llccc}
    \toprule
    \multicolumn{2}{l}{Parameter} & HAT-P-33 & HAT-P-49 & HD\:89345 \\
    \hline
    \multicolumn{2}{l}{\footnotesize{\underline{Instrument Offsets}}} \rule{0pt}{2.6ex} \\ 
    $\braket{RV}_{\textsf{SOPHIE}}$ & $\mathrm{[m\,s^{-1}]}$ & $\mathcal{N}(23236.35,\:1.0)$ & $\mathcal{N}(14511.75,\:1.0)$ & $\mathcal{N}(2312.0,\:1.0)$ \\
    $\braket{RV}_{\textsf{HIRES}}$ & $\mathrm{[m\,s^{-1}]}$ & $\mathcal{N}(13.65,\:1.0)$ &  & $\mathcal{N}(1.35,\:1.0)$ \\
    $\braket{RV}_{\textsf{TRES}}$ & $\mathrm{[m\,s^{-1}]}$ &  & $\mathcal{N}(45.5,\:1.0)$ &  \\
    $\braket{RV}_{\textsf{HARPS}}$ & $\mathrm{[m\,s^{-1}]}$ &  &  & $\mathcal{N}(2357.5,\:1.0)$ \\
    $\braket{RV}_{\textsf{HARPS-N}}$ & $\mathrm{[m\,s^{-1}]}$ &  &  & $\mathcal{N}(2350.2,\:1.0)$ \\
    $\braket{RV}_{\textsf{FIES}}$ & $\mathrm{[m\,s^{-1}]}$ &  &  & $\mathcal{N}(-5.4,\:1.0)$ \\
    $\braket{RV}_{\textsf{APF}}$ & $\mathrm{[m\,s^{-1}]}$ &  &  & $\mathcal{N}(-2.0,\:1.0)$ \\
    
    \multicolumn{2}{l}{\footnotesize{\underline{Keplerian Parameters$\mathrm{^{\text{\textasteriskcentered\textasteriskcentered}}}$}}} & \rule{0pt}{2.6ex} \\ 
    $T_{0}$ & $\mathrm{[BTJD]}$    & $\mathcal{N}(6684.86508,\:0.00027)$ & $\mathcal{N}(6975.61736,\:0.00050)$ & $\mathcal{N}(8740.81147,\:0.00044)$ \\
    $\ln P$ &    & $\mathcal{N}(1.2454442,\:1.9\,10^{-7})$ & $\mathcal{N}(0.9901187,\:4.5\,10^{-7})$ & $\mathcal{N}(2.4693193,\:5.6\,10^{-7})$ \\
    $\ln K$ &    & $\mathcal{N}(4.3567,\:0.1538)$ & $\mathcal{N}(5.2402,\:0.1161)$ & $\mathcal{N}(2.2502,\:0.0885)$ \\
    $\sqrt{e}\,\sin\omega$ &  & $\mathcal{N}(0.42400562,\:0.13834369)$ &  & $\mathcal{N}(-0.11585252,\:0.16083117)$ \\
    $\sqrt{e}\,\cos\omega$ &  & $\mathcal{N}(0.0148066,\:0.25613424)$ &  & $\mathcal{N}(0.43540578,\:0.07368543)$ \\

    \multicolumn{2}{l}{\footnotesize{\underline{White noise}}} \rule{0pt}{2.6ex} \\ 
    $\sigma_{\textsf{SET}}$ & $\mathrm{[m\,s^{-1}]}$ & $\mathcal{N}^{+}(0,\:50.0)$ & $\mathcal{N}^{+}(0,\:50.0)$ & $\mathcal{N}^{+}(0,\:50.0)$ \\

    \multicolumn{2}{l}{\footnotesize{\underline{Linear trend} ($m\cdot t\,\mathrm{[BTJD]}+b$)}} \rule{0pt}{2.6ex} \\ 
    $m$ & $\mathrm{[m\,s^{-1}\,d^{-1}]}$ & $\mathcal{N}(0.0,\:0.1)$ & $\mathcal{N}(0.0,\:0.1)$ & $\mathcal{N}(0.0,\:0.1)$ \\
    $b$ & $\mathrm{[m\,s^{-1}]}$ & $\mathcal{N}(0.0,\:1.0)$ & $\mathcal{N}(0.0,\:1.0)$ & $\mathcal{N}(0.0,\:1.0)$ \\
    \bottomrule
\end{tabular}
\tablefoot{$\mathrm{^{\text{(\textasteriskcentered)}}}$~$\mathcal{N}(\mu,\:\sigma)$ indicates a normal distribution with mean $\mu$ and standard deviation~$\sigma$; and $\mathcal{N}^{+}(0,\:\sigma)$ a half-normal distribution with mean $0$ and standard deviation~$\sigma$. $\mathrm{^{\text{(\textasteriskcentered\textasteriskcentered)}}}$~The values for the prior distributions of parameters $K$, $\sqrt{e}\sin\omega$, and $\sqrt{e}\cos\omega$ are taken from \citet{Wang_Y2017}; for parameters $T_{0}$, $\ln P$, and $\ln K$, the values are taken from this work (STELLA, cf.~Sect.~\ref{sec:STELLA}).}
\end{table*}

\newpage

\section{System properties}
\label{apn:sys_prop}

%%%%%%%%%%%%%%%%%%%%%%%%%%%%%%%%%%%%%%%%%%%%%%%%%%%%%%%%%%%%%%%%%%%%%%%%%%%%%%%%
% HAT-P-3 b
%%%%%%%%%%%%%%%%%%%%%%%%%%%%%%%%%%%%%%%%%%%%%%%%%%%%%%%%%%%%%%%%%%%%%%%%%%%%%%%%

\begin{table*}  
\caption[]{Properties of the HAT-P-3 system}
\centering
\begin{threeparttable}
\begin{tabular}{c c c c c}
\hline
\hline

Parameter & Symbol & Value  &  Unit   &   Origin  \\

\hline 

\multicolumn{5}{c}{\textit{Stellar parameters}} \\ 

\hline

Spectral type & & K1 & & \citet{Grieves2018} \\
 
Stellar temperature & $T_{\rm eff}$ & $5190\pm80$ & K &  \citet{Mancini2018} \\

Stellar radius & $R_\star$ & $0.850\pm0.021$ & $R_\odot$ &  \citet{Mancini2018} \\

Stellar spin inclination & $i_\star$ & 15.8$\stackrel{+6.2}{_{-7.3}}$ & deg & This work \\

Stellar equatorial period & $P_{\rm eq}$ & $19.9\pm1.5$ & d & \citet{Mancini2018} \\

Stellar projected velocity & $v_\mathrm{eq} \sin i_\star$ &    $0.46\stackrel{+0.22}{_{-0.25}}$   & km s$^{-1}$ & This work \\
 
Age & $\tau$ & $2.9^{+1.7}_{-3.7}$ & Gyr & \citet{Mancini2018} \\

Limb-darkening coefficients & $u_1$ & 0.633 & & Exofast \\
 & $u_2$ & 0.141 & &  \\

\hline 

\multicolumn{5}{c}{\textit{Planet b parameters}} \\ 

\hline

Orbital period & $P$& $2.89973797\pm0.00000038$ & d & \citet{Baluev2019} \\

Transit epoch & $T_0$ & $2457237.38678\pm0.00010$ & BJD$_{\rm TDB}$ & \citet{Baluev2019} \\
              &       & $2454218.75960\pm0.00016$ & BJD$_{\rm TDB}$ & This work (STELLA) \\

Eccentricity & $e$ & 0.0 (fixed) & & \citet{Mancini2018} \\
Argument of periastron & $\omega$ & 90 (fixed) & deg & \citet{Mancini2018} \\

Stellar reflex velocity & $K$ & $90.63\pm0.58$ & m s$^{-1}$ & \citet{Mancini2018} \\

Scaled separation & $a/R_\star$ & $9.8105\pm0.2667$ & & \citet{Mancini2018} \\

Orbital inclination & $i$ & $86.31\pm0.19$ & deg & \citet{Mancini2018} \\

Impact parameter & $b$ & $0.615\pm0.012$ & & \citet{Baluev2019} \\

Transit duration & $T_{14}$ & $2.0808\pm0.0082$ & h & \citet{Baluev2019} \\

Planet-to-star radius ratio & $R_{\rm p}/R_\star$ & $0.11091\pm0.00048$ & & \citet{Baluev2019} \\
	 
Projected spin--orbit angle & $\lambda$ & $-25.3\stackrel{+29.4}{_{-22.8}}$    & deg & This work \\

3D spin--orbit angle & $\psi_N$ & 72.0$\stackrel{+8.4}{_{-7.9}}$ & deg & This work \\
               & $\psi_S$ & 79.3$\stackrel{+8.4}{_{-7.9}}$& deg &  \\
               & $\psi$ &   75.7$\stackrel{+8.5}{_{-7.9}}$ & deg &  \\

\hline
\end{tabular}

\begin{tablenotes}[para,flushleft]
Note: The stellar equatorial period is the weighted average of the two consistent values reported by \citet{Mancini2018}. The value of $a/R_\star$ is derived from \citet{Mancini2018}'s $a$ and $R_\star$ values.  
\end{tablenotes}

\end{threeparttable}
\end{table*}

\newpage

%%%%%%%%%%%%%%%%%%%%%%%%%%%%%%%%%%%%%%%%%%%%%%%%%%%%%%%%%%%%%%%%%%%%%%%%%%%%%%%%
% HAT-P-11 b
%%%%%%%%%%%%%%%%%%%%%%%%%%%%%%%%%%%%%%%%%%%%%%%%%%%%%%%%%%%%%%%%%%%%%%%%%%%%%%%%

\begin{table*}  
\caption[]{Properties of the HAT-P-11 system}
\centering
\begin{threeparttable}
\begin{tabular}{c c c c c}
\hline
\hline

Parameter & Symbol & Value  &  Unit   &   Origin  \\

\hline

\multicolumn{5}{c}{\textit{Stellar parameters}} \\ 

\hline

Spectral type & & K4 & & \citet{Bakos2010} \\

Stellar temperature & $T_{\rm eff}$ & $4653^{+33}_{-35}$ & K &  \citet{Morton2016} \\
 
Stellar radius & $R_\star$ & $	0.74\pm0.01$ & $R_\odot$ &  \citet{Morton2016} \\

Stellar spin inclination & $i_{\star \rm S}$ & $160^{+9}_{-19}$ $^\dagger$ & deg & \citet{SanchisOjeda2011} \\

Stellar equatorial period & $P_{\rm eq}$ & $30.5^{+4.1}_{-3.2}$ & d & \citet{SanchisOjeda2011} \\

Stellar projected velocity & $v_\mathrm{eq} \sin i_\star$ & $0.670^{+0.091}_{-0.099}$ & km s$^{-1}$ & This work \\

Age & $\tau$ & $2.69^{+2.88}_{-1.24}$ & Gyr & \citet{Morton2016} \\

 Limb-darkening coefficients & $u_1$ & 0.739 & & Exofast \\
 & $u_2$ & 0.054 & & \\
 
\hline

\multicolumn{5}{c}{\textit{Planet b parameters}} \\ 

\hline
 
Orbital period & $P$ & $4.887802443^{+0.000000034}_{-0.000000030}$ & d & \citet{Huber2017} \\

Transit epoch & $T_0$ & $2454957.8132067^{+0.0000053}_{-0.0000052}$ $^\mathsection$ & BJD$_{\rm TDB}$ & \citet{Huber2017} \\

Eccentricity & $e$ & $0.264353\pm0.000602$ & & \citet{Allart2018} \\

Argument of periastron & $\omega$ & $342.185794\pm0.179084$ & deg & \citet{Allart2018} \\

Stellar reflex velocity & $K$ & $12.01\pm1.38$ & m s$^{-1}$ & \citet{Allart2018} \\

Scaled separation & $a/R_\star$ & $16.50\pm0.18$ & & \citet{Allart2018} \\

Orbital inclination & $i$ & $89.05^{+0.15}_{-0.09}$ & deg & \citet{Huber2017} \\

Impact parameter & $b$ & $0.209^{+0.019}_{-0.032}$ & & \citet{Huber2017} \\

Transit duration & $T_{14}$ & $2.3565^{+0.0015}_{-0.0016}$ & h & \citet{Huber2017} \\

Planet-to-star radius ratio & $R_{\rm p}/R_\star$ & $0.05850^{+0.00009}_{-0.00013}$ & & \citet{Huber2017} \\

Projected spin--orbit angle & $\lambda$ & $133.9^{+7.1}_{-8.3}$ & deg & This work \\

3D spin--orbit angle & $\psi\equiv\psi_{\rm S}$ & $104.9^{+8.6}_{-9.1}$ $^\dagger$ & deg & This work \\

\hline

\multicolumn{5}{c}{\textit{Planet c parameters}} \\ 

\hline

Orbital period & $P$ & $3407^{+360}_{-190}$ & d & \citet{Yee2018} \\

Inferior conjunction epoch & $T_0$ & $2456746^{+24}_{-32}$ & BJD$_{\rm TDB}$ & \citet{Yee2018} \\

Eccentricity & $e$ & $0.601^{+0.032}_{-0.031}$ & & \citet{Yee2018} \\

Argument of periastron & $\omega$ & $143.7^{+4.8}_{-4.9}$ & deg & \citet{Yee2018} \\

Stellar reflex velocity & $K$ & $30.9\pm1.3$ & m s$^{-1}$ & \citet{Yee2018} \\
	    
\hline
\end{tabular}
\begin{tablenotes}[para,flushleft]
Notes : $^\mathsection$ This is the correct value, even if it is different from the value reported in the original paper (see Corrigendum).
$^\dagger$ Our analysis favors the configuration where the stellar south pole is visible, and we use the corresponding stellar inclination derived by \citet{SanchisOjeda2011} from spot-crossing anomalies.
\end{tablenotes}
\end{threeparttable}
\end{table*}

\newpage

%%%%%%%%%%%%%%%%%%%%%%%%%%%%%%%%%%%%%%%%%%%%%%%%%%%%%%%%%%%%%%%%%%%%%%%%%%%%%%%%
% HAT-P-33 b
%%%%%%%%%%%%%%%%%%%%%%%%%%%%%%%%%%%%%%%%%%%%%%%%%%%%%%%%%%%%%%%%%%%%%%%%%%%%%%%%

\begin{table*}  
\caption[]{Properties of the HAT-P-33 system}
\label{tab:HATP33}
\centering
\begin{threeparttable}
\begin{tabular}{c c c c c}
\hline
\hline

Parameter & Symbol & Value  &  Unit   &   Origin  \\

\hline

\multicolumn{5}{c}{\textit{Stellar parameters}} \\ 

\hline

Spectral type & & F4 & & \citet{Luo2018} \\

Stellar temperature & $T_{\rm eff}$ & $6460^{+300}_{-290}$ & K &  \citet{Wang_Y2017} \\
 
Stellar radius & $R_\star$ & $1.91^{+0.26}_{-0.20}$ & $R_\odot$ &  \citet{Wang_Y2017} \\

Stellar spin inclination & $i_\star$ & --- & deg & --- \\

Stellar equatorial period & $P_{\rm eq}$ & --- & d & --- \\

Stellar projected velocity & $v_\mathrm{eq} \sin i_\star$ & $15.57\pm0.31$ & km s$^{-1}$ & This work \\

Age & $\tau$ & $2.3\pm0.3$ & Gyr & \citet{hartman2011} \\

 Limb-darkening coefficients & $u_1$ & 0.355 & & Exofast \\
 & $u_2$ & 0.313 & & \\

\hline

\multicolumn{5}{c}{\textit{Planet b parameters}} \\ 

\hline

Orbital period & $P$ & $3.47447773\pm0.00000066$ & d & This work (STELLA) \\

Transit epoch & $T_0$ & $2456684.86508\pm0.00027$ & BJD$_{\rm TDB}$ & This work (STELLA) \\

Eccentricity & $e$ & $0.180^{+0.110}_{-0.096}$ & & \citet{Wang_Y2017} \\

Argument of periastron & $\omega$ & $88^{+33}_{-34}$ & deg & \citet{Wang_Y2017} \\

Stellar reflex velocity & $K$ & $74.4\pm8.5$ & m s$^{-1}$ & This work \\

Scaled separation & $a/R_\star$ & $5.69^{+0.58}_{-0.59}$ & & \citet{Wang_Y2017} \\

Orbital inclination & $i$ & $88.2^{+1.2}_{-1.3}$ & deg & \citet{Wang_Y2017} \\

Impact parameter & $b$ & $0.151^{+0.100}_{-0.098}$ & & \citet{Wang_Y2017} \\

Transit duration & $T_{14}$ & $4.33800^{+0.02328}_{-0.02136}$ & h & \citet{Wang_Y2017} \\

Planet-to-star radius ratio & $R_{\rm p}/R_\star$ & $0.10097^{+0.00056}_{-0.00052}$ & & \citet{Wang_Y2017} \\

Projected spin--orbit angle & $\lambda$ & $-5.9\pm4.1$ & deg & This work \\

\hline
\end{tabular}
\end{threeparttable}
\end{table*}

\newpage

%%%%%%%%%%%%%%%%%%%%%%%%%%%%%%%%%%%%%%%%%%%%%%%%%%%%%%%%%%%%%%%%%%%%%%%%%%%%%%%%
% HAT-P-49 b
%%%%%%%%%%%%%%%%%%%%%%%%%%%%%%%%%%%%%%%%%%%%%%%%%%%%%%%%%%%%%%%%%%%%%%%%%%%%%%%%

\begin{table*}  
\caption[]{Properties of the HAT-P-49 system}
\label{tab:HATP49}
\centering
\begin{threeparttable}
\begin{tabular}{c c c c c}
\hline
\hline

Parameter & Symbol & Value  &  Unit   &   Origin  \\

\hline

\multicolumn{5}{c}{\textit{Stellar parameters}} \\ 

\hline

Spectral type & & F3 & & Derived \\

Stellar temperature & $T_{\rm eff}$ & $6820\pm52$ & K &  \citet{Bieryla2014} \\
 
Stellar radius & $R_\star$ & $1.833^{+0.138}_{-0.076}$ & $R_\odot$ &  \citet{Bieryla2014} \\

Stellar spin inclination & $i_\star$ & --- & deg & --- \\

Stellar equatorial period & $P_{\rm eq}$ & --- & d & --- \\

Stellar projected velocity & $v_\mathrm{eq} \sin i_\star$ & $10.68^{+0.46}_{-0.47}$ & km s$^{-1}$ & This work \\

Age & $\tau$ & $1.5\pm0.2$ & Gyr & \citet{Bieryla2014} \\

Limb-darkening coefficients & $u_1$ & 0.312 & & Exofast \\
 & $u_2$ & 0.336 & & \\
 
\hline

\multicolumn{5}{c}{\textit{Planet b parameters}} \\ 

\hline

Orbital period & $P$ & $2.6915539\pm0.0000012$ & d & This work (STELLA) \\

Transit epoch & $T_0$ & $2456975.61736\pm0.00050$ & BJD$_{\rm TDB}$ & This work (STELLA) \\

Eccentricity & $e$ & $0.0$ (fixed) & & \citet{Bieryla2014} \\

Argument of periastron & $\omega$ & 90 (fixed) & deg & \citet{Bieryla2014} \\

Stellar reflex velocity & $K$ & $177.6\pm16.0$ & & This work \\

Scaled separation & $a/R_\star$ & $5.13^{+0.19}_{-0.30}$ & & \citet{Bieryla2014} \\

Orbital inclination & $i$ & $86.2\pm1.7$ & deg & \citet{Bieryla2014} \\

Impact parameter & $b$ & $0.340^{+0.119}_{-0.141}$ & & \citet{Bieryla2014} \\

Transit duration & $T_{14}$ & $4.1088\pm0.0456$ & h & \citet{Bieryla2014} \\

Planet-to-star radius ratio & $R_{\rm p}/R_\star$ & $0.0792\pm0.0019$ & & \citet{Bieryla2014} \\

Projected spin--orbit angle & $\lambda$ & $-97.7\pm1.8$ & deg & This work \\

\hline
\end{tabular}
\end{threeparttable}
\end{table*}

\newpage

%%%%%%%%%%%%%%%%%%%%%%%%%%%%%%%%%%%%%%%%%%%%%%%%%%%%%%%%%%%%%%%%%%%%%%%%%%%%%%%%
% HD 89345 b
%%%%%%%%%%%%%%%%%%%%%%%%%%%%%%%%%%%%%%%%%%%%%%%%%%%%%%%%%%%%%%%%%%%%%%%%%%%%%%%%

\begin{table*}  
\caption[]{Properties of the HD\,89345 system}
\label{tab:HD89345}
\centering
\begin{threeparttable}
\begin{tabular}{c c c c c}
\hline
\hline

Parameter & Symbol & Value  &  Unit   &   Origin  \\

\hline

\multicolumn{5}{c}{\textit{Stellar parameters}} \\ 

\hline

Spectral type & & G5 & & \citet{Cannon1993} \\

Stellar temperature & $T_{\rm eff}$ & $5499\pm73$ & K &  \citet{VanEylen2018} \\
 
Stellar radius & $R_\star$ & $1.657^{+0.020}_{-0.004}$ & $R_\odot$ &  \citet{VanEylen2018} \\

Stellar spin inclination & $i_\star$ & $>44$ $^{\dagger}$ & deg & \citet{VanEylen2018} \\

Stellar equatorial period & $P_{\rm eq}$ & --- & d & --- \\

Stellar projected velocity & $v_\mathrm{eq} \sin i_\star$ & $0.58\pm0.28$ & km s$^{-1}$ & This work \\

Age & $\tau$ & $9.4^{+0.4}_{-1.3}$ & Gyr & \citet{VanEylen2018} \\

 Limb-darkening coefficients & $u_1$ & 0.535 & & Exofast \\
 & $u_2$ & 0.215 & & \\
 
\hline

\multicolumn{5}{c}{\textit{Planet b parameters}} \\ 

\hline

Orbital period & $P$ & 11.8144024$\pm$0.0000066 & d & This work (TESS + K2) \\

Transit epoch & $T_0$ & 2458740.81147$\pm$0.00044 & BJD$_{\rm TDB}$ & This work (TESS + K2) \\

Eccentricity & $e$ & $0.208\pm0.039$ & & This work \\

Argument of periastron & $\omega$ & $21.7 \pm 19.1$ & & This work \\

Stellar reflex velocity & $K$ & $9.1\pm0.5$ & & This work \\

Scaled separation & $a/R_\star$ & $13.625\pm0.027$ & & \citet{VanEylen2018} \\

Orbital inclination & $i$ & $87.68\pm0.10$ & deg & This work (TESS + K2) \\

Impact parameter & $b$ & $0.564\pm0.017$ &  & This work (TESS + K2) \\

Transit duration & $T_{14}$ & $5.65\pm0.02$ & h & This work (TESS + K2) \\

Planet-to-star radius ratio & $R_{\rm p}/R_\star$ & $0.03696\pm0.00041$ & & This work (TESS + K2) \\

Projected spin--orbit angle & $\lambda$ & $74.2^{+33.6}_{-32.5}$ & deg & This work \\

3D spin--orbit angle & $\psi_N$ & 78.7$\stackrel{+22.5}{_{-23.0}}$ & deg & This work \\
               & $\psi_S$ & 81.6$\stackrel{+22.2}{_{-23.2}}$& deg &  \\
               & $\psi$ &   80.1$\stackrel{+22.3}{_{-23.1}}$ & deg &  \\

\hline
\end{tabular}
\begin{tablenotes}[para,flushleft]
Notes: $^{\dagger}$ Instead of using this lower limit we reproduced the PDF on $i_{\star}$ from \citet{VanEylen2018} to calculate the PDF on $\psi$ from our results.
\end{tablenotes}
\end{threeparttable}
\end{table*}

\newpage

%%%%%%%%%%%%%%%%%%%%%%%%%%%%%%%%%%%%%%%%%%%%%%%%%%%%%%%%%%%%%%%%%%%%%%%%%%%%%%%%
% HD 106315 c
%%%%%%%%%%%%%%%%%%%%%%%%%%%%%%%%%%%%%%%%%%%%%%%%%%%%%%%%%%%%%%%%%%%%%%%%%%%%%%%%

\begin{table*}  
\caption[]{Properties of the HD 106315 system}
\label{tab:HD106315}
\centering
\begin{threeparttable}
\begin{tabular}{c c c c c}
\hline
\hline

Parameter & Symbol & Value  &  Unit   &   Origin  \\

\hline

\multicolumn{5}{c}{\textit{Stellar parameters}} \\ 

\hline

Spectral type & & F5 & & \citet{Barros2017} \\

Stellar temperature & $T_{\rm eff}$ & $6364\pm87$ & K &  \citet{Kosiarek2021} \\
 
Stellar radius & $R_\star$ & $1.269\pm0.024$ & $R_\odot$ &  \citet{Kosiarek2021} \\

Stellar spin inclination & $i_\star$ & --- & deg & --- \\

Stellar equatorial period & $P_{\rm eq}$ & --- & d & --- \\

Stellar projected velocity & $v_\mathrm{eq} \sin i_\star$ & $9.66^{+0.64}_{-0.65}$ & km s$^{-1}$ & This work \\

Age & $\tau$ & $4.48\pm0.96$ & Gyr & \citet{Barros2017} \\

Limb-darkening coefficients & $u_1$ & 0.340 & & Exofast \\
 & $u_2$ & 0.316 & & \\
 
\hline

\multicolumn{5}{c}{\textit{Planet b parameters}} \\ 

\hline

Orbital period & $P$ & $9.55287\pm0.00021$ & d & \citet{Kosiarek2021} \\

Transit epoch & $T_0$ & $2457586.5476\pm0.0025$ & BJD$_{\rm TDB}$ & \citet{Kosiarek2021} \\

Eccentricity & $e$ & $0.0$ (fixed) & & \citet{Kosiarek2021} \\

Argument of periastron & $\omega$ & 90 (fixed) & deg & \citet{Kosiarek2021} \\

Stellar reflex velocity & $K$ & $2.88^{+0.85}_{-0.84}$ & m s$^{-1}$ & \citet{Kosiarek2021} \\

\hline

\multicolumn{5}{c}{\textit{Planet c parameters}} \\ 

\hline

Orbital period & $P$ & $21.05652\pm0.00012$ & d & \citet{Kosiarek2021} \\

Transit epoch & $T_0$ & $2457569.01767\pm0.00097$ & BJD$_{\rm TDB}$ & \citet{Kosiarek2021} \\

Eccentricity & $e$ & $0.0$ (fixed) & & \citet{Kosiarek2021} \\

Argument of periastron & $\omega$ & $90$ (fixed) & deg & \citet{Kosiarek2021} \\

Stellar reflex velocity & $K$ & $2.53\pm0.79$ & m s$^{-1}$ & \citet{Kosiarek2021} \\

Scaled separation & $a/R_\star$ & $29.5^{+5.7}_{-4.2}$$^{\dagger}$ & & \citet{Kosiarek2021} \\
%    from Spitzer results only; if calculated from RV fit value: 
% aRs = 0.1565*149597870.7/(1.269*695700.0) = 26.52 
%sig_aRs = 26.52*sqrt((0.002/0.1565)**2.+(0.024/1.269)**2.) = 0.60
%Scaled separation & $a/R_\star$ & $25.10\pm0.79$ & & \citet{Guilluy21} \\

Orbital inclination & $i$ & $88.89^{+0.69}_{-0.51}$$^{\dagger}$ & deg & \citet{Kosiarek2021} \\

Impact parameter & $b$ & $0.688^{+0.044}_{-0.094}$  & & \citet{Rodriguez2017} \\

Transit duration & $T_{14}$ & $4.728^{+0.084}_{-0.065}$ & h & \citet{Rodriguez2017} \\

Planet-to-star radius ratio & $R_{\rm p}/R_\star$ & $0.03481\pm0.00099$ & & \citet{Guilluy2021} \\

Projected spin--orbit angle & $\lambda$ & $-2.68^{+2.7}_{-2.6}$ & deg & This work \\

\hline
\end{tabular}
\begin{tablenotes}[para,flushleft]
Notes: $^{\dagger}$ We use the scaled separation and orbital inclination derived by \citet{Kosiarek2021} from Spitzer transits, as they provide a better match to the transit duration.
\end{tablenotes}
\end{threeparttable}
\end{table*}

%%%%%%%%%%%%%%%%%%%%%%%%%%%%%%%%%%%%%%%%%%%%%%%%%%%%%%%%%%%%%%%%%%%%%%%%%%%%%%%%
% K2-105 b
%%%%%%%%%%%%%%%%%%%%%%%%%%%%%%%%%%%%%%%%%%%%%%%%%%%%%%%%%%%%%%%%%%%%%%%%%%%%%%%%

\begin{table*}  
\caption[]{Properties of the K2-105 system}
\label{tab:K2-105}
\centering
\begin{threeparttable}
\begin{tabular}{c c c c c}
\hline
\hline

Parameter & Symbol & Value  &  Unit   &   Origin  \\

\hline

\multicolumn{5}{c}{\textit{Stellar parameters}} \\ 

\hline

Spectral type & & K1 & & \citet{Luo2018} \\

Stellar temperature & $T_{\rm eff}$ & $5636^{+49}_{-52}$ & & \citet{CastroGonzalez2021} \\

Stellar radius & $R_\star$ & $0.97\pm0.01$ & $R_\odot$  & \citet{CastroGonzalez2021} \\

Stellar spin inclination & $i_\star$ & --- & deg & --- \\

Stellar equatorial period & $P_{\rm eq}$ & --- & d & --- \\

Stellar projected velocity & $v_\mathrm{eq} \sin i_\star$ & $2.13_{-0.92}^{+0.96}$ & km s$^{-1}$ & This work \\

Age & $\tau$ & $>0.6$ & Gyr & \citet{Narita2017} \\

 Limb-darkening coefficients & $u_1$ & 0.338 & & Exofast \\
 & $u_2$ & 0.186 & & \\
 
\hline

\multicolumn{5}{c}{\textit{Planet b parameters}} \\ 

\hline
 
Orbital period & $P$ & $8.267043\pm0.000015$ & d & This work (STELLA) \\
              &     & $8.2669897\pm0.0000057$ & d & This work (TESS + K2) \\

Transit epoch & $T_0$ & $2457379.46827\pm0.00114$ & BJD$_{\rm TDB}$ & This work (STELLA) \\
            &       & $2458363.23873^{+0.00069}_{-0.00063}$ & BJD$_{\rm TDB}$ & This work (TESS + K2) \\

Eccentricity & $e$ & 0 & & (fixed) \\

Argument of periastron & $\omega$ & 90 & deg & (fixed) \\

Stellar reflex velocity & $K$ & $9.4\pm5.8$ & m s$^{-1}$ & \citet{Narita2017} \\

Scaled separation & $a/R_\star$ & $17.39\pm0.19$ &  & This work (TESS + K2) \\

Orbital inclination & $i$ & $88.62\pm0.10$ & deg & This work (TESS + K2) \\
                    
Impact parameter & $b$ & $0.42\pm0.03$ &  & This work (TESS + K2) \\

Transit duration & $T_{14}$ & $3.43\pm0.03$ & h & This work (TESS + K2) \\

Planet-to-star radius ratio & $R_{\rm p}/R_\star$ & $0.03332\pm0.00067$ & & This work (TESS + K2) \\  
 
Projected spin--orbit angle & $\lambda$ & $-81^{+50}_{-47}$ $^{\dagger}$ & deg & This work \\

\hline
\end{tabular}
\begin{tablenotes}[para,flushleft]
Notes: $^{\dagger}$ We consider this measurement as marginal, as $\lambda$ is not constrained at the 3$\sigma$ level (see text)
\end{tablenotes}
\end{threeparttable}
\end{table*}

\newpage

%%%%%%%%%%%%%%%%%%%%%%%%%%%%%%%%%%%%%%%%%%%%%%%%%%%%%%%%%%%%%%%%%%%%%%%%%%%%%%%%
% Kepler-25 c
%%%%%%%%%%%%%%%%%%%%%%%%%%%%%%%%%%%%%%%%%%%%%%%%%%%%%%%%%%%%%%%%%%%%%%%%%%%%%%%%

\begin{table*}  
\caption[]{Properties of the Kepler-25 system}
\centering
\begin{threeparttable}
\begin{tabular}{c c c c c}
\hline
\hline

Parameter & Symbol & Value  &  Unit   &   Origin  \\

\hline

\multicolumn{5}{c}{\textit{Stellar parameters}} \\ 

\hline

Spectral type & & F8 & & Derived \\
 
Stellar temperature & $T_{\rm eff}$ & $6354\pm27$ & K &  \citet{Benomar2014} \\

Stellar radius & $R_\star$ & $1.316^{+0.016}_{-0.015}$ & $R_\odot$ & \citet{Mills2019} \\

Stellar Mass & $M_\star$ & $1.165^{+0.029}_{-0.027}$ & $M_\odot$ & \citet{Mills2019} \\

Stellar spin inclination & $i_\star$ & $66.7^{+12.1}_{-7.4}$ $^{\dagger}$ & deg & \citet{Benomar2014} \\

Stellar equatorial period & $P_{\rm eq}$ & --- & d & --- \\

Stellar projected velocity & $v_\mathrm{eq} \sin i_\star$ & $8.89^{+0.59}_{-0.63}$ & & This work \\

Limb-darkening coefficients & $u_1$ & 0.381 & & Exofast \\
 & $u_2$ & 0.302 & & \\
 
Age & $\tau$ & 2.75$\pm$0.3 & Gyr & \citet{Benomar2014} \\

\hline

\multicolumn{5}{c}{\textit{Planet b parameters}} \\ 

\hline

Orbital period & $P$ & $6.2385347882 \pm0.0000001619 $ & d & \citet{Battley2021} \\

Transit epoch & $T_0$ & $2454954.7979391\pm0.0002168$  & BJD$_{\rm TDB}$ & \citet{Battley2021} \\
 &  & $2458648.00807^{+0.00057}_{-0.00051}$ & BJD$_{\rm TDB}$ & This work \\  

Eccentricity & $e$ & $0.0029^{+0.0023}_{-0.0017}$ & & \citet{Mills2019} \\

Argument of periastron & $\omega$ & --- & deg & --- \\

Planetary mass & $M_{\rm p}$ & $0.0275^{+0.0079}_{-0.0073}$ & $M_{\rm Jup}$ & \citet{Mills2019} \\

Orbital inclination & $i$ & $87.173^{+0.083}_{-0.084}$ & deg & \citet{Mills2019} \\

\hline

\multicolumn{5}{c}{\textit{Planet c parameters}} \\ 

\hline

Orbital period & $P$ & $12.720370495\pm0.000001703$ & d & \citet{Battley2021} \\

Transit epoch & $T_0$ & $2454960.6467450\pm0.0001144$ & BJD$_{\rm TDB}$ & \citet{Battley2021} \\
             &  & $2458649.55482^{+0.00028}_{-0.00026}$ & BJD$_{\rm TDB}$ & This work \\  

Eccentricity & $e$ & $0.0061^{+0.0049}_{-0.0041}$ & & \citet{Mills2019} \\

Argument of periastron & $\omega$ & --- & deg & --- \\

Planetary mass & $M_{\rm p}$ & $0.0479^{+0.0051}_{-0.0041}$ & $M_{\rm Jup}$ & \citet{Mills2019} \\

Scaled separation & $a/R_\star$ & $18.336\pm0.27$ & & \citet{Mills2019} \\

Orbital inclination & $i$ & $87.236^{+0.039}_{-0.042}$ & deg & \citet{Mills2019} \\

Impact parameter & $b$ & $0.8826\pm0.0018$ & & \citet{Benomar2014} \\

Transit duration & $T_{14}$ & $2.862\pm0.006$ & h & \citet{Benomar2014} \\

Planet-to-star radius ratio & $R_{\rm p}/R_\star$ & $0.03637\pm0.00012$ & & \citet{Mills2019} \\

Projected spin--orbit angle & $\lambda$ & $-0.9\stackrel{+7.7}{_{-6.4}}$    & deg & This work \\

3D spin--orbit angle & $\psi_N$ & $21.4\stackrel{+8.9}{_{-9.2}}$  & deg & This work \\
               & $\psi_S$ & $26.8\stackrel{+9.5}{_{-9.4}}$   & deg &  \\
               & $\psi$ &   $24.1\stackrel{+9.2}{_{-9.3}}$  & deg &  \\

\hline

\multicolumn{5}{c}{\textit{Planet d parameters}} \\ 

\hline

Orbital period & $P$ & $122.4^{+0.80}_{-0.71}$ & d & \citet{Mills2019} \\

Transit epoch & $T_0$ & $2455715^{+6.8}_{-7.2}$ & BJD$_{\rm TDB}$ & \citet{Mills2019} \\

Eccentricity & $e$ & $0.13^{+0.13}_{-0.09}$ & & \citet{Mills2019} \\

Argument of periastron & $\omega$ & --- & deg & --- \\

Minimum planetary mass & $M_{\rm p} \sin i$ & $0.226\pm0.031$ & $M_{\rm Jup}$ & \citet{Mills2019} \\

\hline
\end{tabular}

\begin{tablenotes}[para,flushleft]
Notes: For consistency with our framework we brought the orbital inclinations published by \citet{Mills2019} within 90-180$^{\circ}$ back to within 0-90$^{\circ}$. Arguments of periastron were not derived by \citet{Mills2019}, but are not required here as circular orbital models were used for the RM analysis. The stellar reflex motion induced by each planet was calculated using the stellar and planetary masses and orbital inclination. The scaled separation was derived by reconstructing PDFs for P$_{\rm c}$, $R_\star$, and $M_\star$ from \citet{Mills2019} and using Kepler's third law. $^{\dagger}$ We use the value derived from asteroseismology alone (Fig.~9 in \citealt{Benomar2014}).
\end{tablenotes}

\end{threeparttable}
\end{table*}

\newpage

%%%%%%%%%%%%%%%%%%%%%%%%%%%%%%%%%%%%%%%%%%%%%%%%%%%%%%%%%%%%%%%%%%%%%%%%%%%%%%%%
% Kepler-63 b
%%%%%%%%%%%%%%%%%%%%%%%%%%%%%%%%%%%%%%%%%%%%%%%%%%%%%%%%%%%%%%%%%%%%%%%%%%%%%%%%

\begin{table*}  
\caption[]{Properties of the Kepler-63 system}
\centering
\begin{threeparttable}
\begin{tabular}{c c c c c}
\hline
\hline

Parameter & Symbol & Value  &  Unit   &   Origin  \\

\hline

\multicolumn{5}{c}{\textit{Stellar parameters}} \\ 

\hline

Spectral type & & G5 & & \citet{Frasca2016} \\

Stellar temperature & $T_{\rm eff}$ & $5576\pm50$ & K &  \citet{SanchisOjeda2013_K63b} \\
 
Stellar radius & $R_\star$ & $0.901^{+0.027}_{-0.022}$ & $R_\odot$ &  \citet{SanchisOjeda2013_K63b} \\

Stellar spin inclination & $i_\star$ & $138\pm7$ $^{\dagger}$ & deg & \citet{SanchisOjeda2013_K63b} \\

Stellar equatorial period & $P_{\rm eq}$ & $5.401\pm0.014$ & d & \citet{SanchisOjeda2013_K63b} \\

Stellar projected velocity & $v_\mathrm{eq} \sin i_\star$ & $7.47^{+2.6}_{-2.7}$ & km s$^{-1}$ & This work \\

Age & $\tau$ & $0.210\pm0.045$ & Gyr & \citet{SanchisOjeda2013_K63b} \\

 Limb-darkening coefficients & $u_1$ & 0.527 & & Exofast \\
 & $u_2$ & 0.217 & & \\
 
\hline

\multicolumn{5}{c}{\textit{Planet b parameters}} \\ 

\hline
 
Orbital period & $P$ & $9.4341503479\pm0.0000003339$ & d & \citet{Gajdos2019} \\

Transit epoch & $T_0$ & $2455010.84340000\pm0.00002768$ & BJD$_{\rm TDB}$ & \citet{Gajdos2019} \\

Eccentricity & $e$ & 0 (fixed)$^{\dagger\dagger}$ & & --- \\

Argument of periastron & $\omega$ & 90 (fixed)$^{\dagger\dagger}$ & deg & --- \\

Stellar reflex velocity & $K$ & $40\pm20^{\dagger\dagger}$ & m s$^{-1}$ & \citet{SanchisOjeda2013_K63b} \\

Scaled separation & $a/R_\star$ & $19.12\pm0.08$ & & \citet{SanchisOjeda2013_K63b} \\

Orbital inclination & $i$ & $87.806^{+0.018}_{-0.019}$ & deg & \citet{SanchisOjeda2013_K63b} \\

Impact parameter & $b$ & $0.732\pm0.003$ & & \citet{SanchisOjeda2013_K63b} \\

Transit duration & $T_{14}$ & $2.903\pm0.003$ & h & \citet{SanchisOjeda2013_K63b} \\

Planet-to-star radius ratio & $R_{\rm p}/R_\star$ & $0.0622\pm0.0010$ & & \citet{SanchisOjeda2013_K63b} \\

Projected spin--orbit angle & $\lambda$ & $-135^{+21.2}_{-26.8}$  & deg & This work \\

3D spin--orbit angle & $\psi\equiv\psi_{\rm S}$ & $114.6^{+16.6}_{-12.5}$ $^{\dagger}$ & deg & This work \\

\hline
\end{tabular}
\begin{tablenotes}[para,flushleft]
Notes: $^{\dagger}$ The analysis from \citet{SanchisOjeda2013_K63b} favors the configuration where the southern stellar pole is visible.

$^{\dagger\dagger}$ The RV data from \citet{SanchisOjeda2013_K63b} are too strongly affected by stellar activity to accurately constrain the orbital properties. The reflex velocity semi-amplitude we adopt comes from the RM effect and seems compatible with the apparent value from RVs. The orbit is fixed to circular, as \citet{SanchisOjeda2013_K63b} set a 3$\sigma$ upper limit of 0.45 on $e$.  

\end{tablenotes}
\end{threeparttable}
\end{table*}

\newpage

%%%%%%%%%%%%%%%%%%%%%%%%%%%%%%%%%%%%%%%%%%%%%%%%%%%%%%%%%%%%%%%%%%%%%%%%%%%%%%%%
% Kepler-68 b
%%%%%%%%%%%%%%%%%%%%%%%%%%%%%%%%%%%%%%%%%%%%%%%%%%%%%%%%%%%%%%%%%%%%%%%%%%%%%%%%

\begin{table*}  
\caption[]{Properties of the Kepler-68 system}
\centering
\begin{threeparttable}
\begin{tabular}{c c c c c}
\hline
\hline

Parameter & Symbol & Value  &  Unit   &   Origin  \\

\hline

\multicolumn{5}{c}{\textit{Stellar parameters}} \\ 

\hline

Spectral type & & G1 & & \citet{Grieves2018} \\

Stellar temperature & $T_{\rm eff}$ & $5793\pm74$ & K &  \citet{Gilliland2013} \\

Stellar radius & $R_\star$ & $1.243\pm0.019$ & $R_\odot$ &  \citet{Gilliland2013} \\

Stellar spin inclination & $i_\star$ & --- & deg & --- \\

Stellar equatorial period & $P_{\rm eq}$ & --- & d & --- \\

Stellar projected velocity & $v_\mathrm{eq} \sin i_\star$ & $0.5\pm0.5$ & km s$^{-1}$ & \citet{Gilliland2013} \\
 
Age & $\tau$ & 6.3$\pm$1.7 & Gyr & \citet{Gilliland2013} \\

Limb-darkening coefficients & $u_1$ & 0.449 & & Exofast \\
 & $u_2$ & 0.267 & &  \\

\hline

\multicolumn{5}{c}{\textit{Planet b parameters}} \\ 

\hline

Orbital period & $P$ & $5.3987525913\pm0.0000005231$ & d & \citet{Gajdos2019} \\

Transit epoch & $T_0$ & $2455006.85878000\pm0.00007639$ & BJD$_{\rm TDB}$ & \citet{Gajdos2019} \\
 
Eccentricity & $e$ & 0.0 (fixed) & & \citet{Mills2019} \\

Argument of periastron & $\omega$ & 90 (fixed) & deg & \citet{Mills2019} \\

Stellar reflex velocity & $K$ & $2.70^{+0.48}_{-0.46}$ & m s$^{-1}$ & \citet{Mills2019} \\

Scaled separation & $a/R_\star$ & $10.68\pm0.14$ & & \citet{Gilliland2013} \\

Orbital inclination & $i$ & $87.60\pm0.90$ & deg & \citet{Gilliland2013} \\

Impact parameter & $b$ & $0.45\pm0.17$ & & \citet{Gilliland2013} \\

Transit duration & $T_{14}$ & $3.459\pm0.009$ & h & \citet{Gilliland2013} \\

Planet-to-star radius ratio & $R_{\rm p}/R_\star$ & $0.01700\pm0.00046$ & & \citet{Gilliland2013} \\

Projected spin--orbit angle & $\lambda$ & Nondetection &  & This work \\

\hline

\multicolumn{5}{c}{\textit{Planet c parameters}} \\ 

\hline

Orbital period & $P$ & $9.60502738150\pm0.0000132365$ & d & \citet{Gajdos2019} \\

Transit epoch & $T_0$ & $2454969.38207000\pm0.00110495$ & BJD$_{\rm TDB}$ & \citet{Gajdos2019} \\

Eccentricity & $e$ & 0.0 (fixed) & & \citet{Mills2019} \\

Argument of periastron & $\omega$ & 90 (fixed)  & deg & \citet{Mills2019} \\

Stellar reflex velocity & $K$ & $0.59^{+0.50}_{-0.52}$ & m s$^{-1}$ & \citet{Mills2019} \\

\hline

\multicolumn{5}{c}{\textit{Planet d parameters}} \\ 

\hline

Orbital period & $P$ & $634.6^{+4.1}_{-3.7}$ & d & \citet{Mills2019} \\

Transit epoch & $T_0$ & $2455878\pm11$ & BJD$_{\rm TDB}$ & \citet{Mills2019} \\

Eccentricity & $e$ & $0.112^{+0.035}_{-0.034}$ & & \citet{Mills2019} \\

Argument of periastron & $\omega$ & $-1.13^{+0.36}_{-0.45}$ & deg & \citet{Mills2019} \\

Stellar reflex velocity & $K$ & $17.75^{+0.50}_{-0.49}$ & m s$^{-1}$ & \citet{Mills2019} \\

\hline
\end{tabular}
\end{threeparttable}
\end{table*}

\newpage

%%%%%%%%%%%%%%%%%%%%%%%%%%%%%%%%%%%%%%%%%%%%%%%%%%%%%%%%%%%%%%%%%%%%%%%%%%%%%%%%
% WASP-47 d
%%%%%%%%%%%%%%%%%%%%%%%%%%%%%%%%%%%%%%%%%%%%%%%%%%%%%%%%%%%%%%%%%%%%%%%%%%%%%%%%

\begin{table*}  
\caption[]{Properties of the WASP-47 system}
\centering
\begin{threeparttable}
\begin{tabular}{c c c c c}
\hline
\hline

Parameter & Symbol & Value  &  Unit   &   Origin  \\

\hline

\multicolumn{5}{c}{\textit{Stellar parameters}} \\ 

\hline

Spectral type & & G9 & & \citet{Hellier2012} \\

Stellar temperature & $T_{\rm eff}$ & $5552\pm75$ & K &  \citet{Vanderburg2017} \\

Stellar mass & $M_\star$ & $1.040\pm0.031$ & $M_\odot$ &  \citet{Vanderburg2017} \\
 
Stellar radius & $R_\star$ & $1.137\pm0.013$ & $R_\odot$ &  \citet{Vanderburg2017} \\

Stellar spin inclination & $i_\star$ & $69.8^{+11.0}_{-9.2}$ & deg & This work \\

Stellar equatorial period & $P_{\rm eq}$ & $39.4^{+2.2}_{-4.5}$ & d & \citet{Bryant2022} \\

Stellar projected velocity & $v_\mathrm{eq} \sin i_\star$ & $1.80^{+0.24}_{-0.16}$ & km s$^{-1}$ & \citet{SanchisOjeda2015} \\

Age & $\tau$ & $6.5^{+2.6}_{-1.2}$ & Gyr & \citet{Almenara2016} \\

 Limb-darkening coefficients & $u_1$ & 0.540 & & Exofast \\
 & $u_2$ & 0.209 & & \\
 
\hline

\multicolumn{5}{c}{\textit{Planet b parameters}} \\ 

\hline

Orbital period & $P$ & $4.1591492\pm0.0000006$ & d & \citet{Bryant2022} \\

Transit epoch & $T_0$ & $2457007.932103\pm0.000019$ & BJD$_{\rm TDB}$ & \citet{Bryant2022} \\

Eccentricity & $e$ & 0 (fixed) & & \citet{Bryant2022} \\
Argument of periastron & $\omega$ & 90 (fixed) & deg & \citet{Bryant2022} \\

Stellar reflex velocity & $K$ & $140.84\pm0.40$ & m s$^{-1}$ & \citet{Bryant2022} \\

Orbital inclination & $i$ & $88.98^{+0.20}_{-0.17}$ & deg & \citet{Bryant2022} \\

Projected spin--orbit angle & $\lambda$ & 0$\pm$24 & deg & \citet{SanchisOjeda2015} \\

3D spin--orbit angle & $\psi_N$ & 28.8$\stackrel{+11.1}{_{-13.3}}$ & deg & This work \\
               & $\psi_S$ & 29.5$\stackrel{+11.0}{_{-13.3}}$& deg &  \\
               & $\psi$ &   29.2$\stackrel{+11.1}{_{-13.3}}$ & deg &  \\

\hline

\multicolumn{5}{c}{\textit{Planet c parameters}} \\ 

\hline

Orbital period & $P$ & $588.8\pm2.0$ & d & \citet{Bryant2022} \\

Inferior conjunction epoch & $T_0$ & $2457763.1\pm4.3$ & BJD$_{\rm TDB}$ & \citet{Bryant2022} \\

Eccentricity & $e$ & $0.295\pm0.016$ & & \citet{Bryant2022} \\

Argument of periastron & $\omega$ & $112.0\pm4.3$ & deg & \citet{Bryant2022} \\

Stellar reflex velocity & $K$ & $31.04\pm0.40$ & m s$^{-1}$ & \citet{Bryant2022} \\
 
\hline

\multicolumn{5}{c}{\textit{Planet d parameters}} \\ 

\hline

Orbital period & $P$ & $9.03052118\pm0.00000753$ & d & This work \\

Transit epoch & $T_0$ & $2459426.5437\pm0.0028$ $^{\dagger}$ & BJD$_{\rm TDB}$ & This work \\

Eccentricity & $e$ & $0.01^{+0.011}_{-0.007}$ & & \citet{Bryant2022} \\
Argument of periastron & $\omega$ & $16.5^{+84.2}_{-98.6}$ & deg & \citet{Bryant2022} \\

Stellar reflex velocity & $K$ & $4.26\pm0.37$ & m s$^{-1}$ & \citet{Bryant2022} \\

Scaled separation & $a/R_\star$ & $16.34^{+0.08}_{-0.11}$ & & \citet{Bryant2022} \\

Orbital inclination & $i$ & $89.55^{+0.30}_{-0.27}$ & deg & \citet{Bryant2022} \\

Impact parameter & $b$ & $0.128^{+0.076}_{-0.085}$ & & \citet{Bryant2022} \\ 

Transit duration & $T_{14}$ & $4.288\pm0.039$ & h & \citet{Vanderburg2017} \\

Planet-to-star radius ratio & $R_{\rm p}/R_\star$ & $0.02876\pm0.00017$ &   & \citet{Bryant2022} \\

Projected spin--orbit angle & Nondetection & deg & This work \\

\hline

\multicolumn{5}{c}{\textit{Planet e parameters}} \\ 

\hline

Orbital period & $P$ & $0.7895933\pm0.0000044$ & d & \citet{Bryant2022} \\

Transit epoch & $T_0$ & $2457011.34862\pm0.00030$ & BJD$_{\rm TDB}$ & \citet{Bryant2022} \\

Eccentricity & $e$ & 0 (fixed)  & & \citet{Bryant2022} \\
Argument of periastron & $\omega$ & 90 (fixed) & deg & \citet{Bryant2022} \\

Stellar reflex velocity & $K$ & $4.55\pm0.37$ & m s$^{-1}$ & \citet{Bryant2022} \\

Transit duration & $T_{14}$ & $1.899\pm0.013$ & h & \citet{Vanderburg2017} \\

\hline
\end{tabular}
\begin{tablenotes}[para,flushleft]
$^\dagger$ We caution that WASP-47 d displays strong TTV with an amplitude of $\sim$6\,min. We report here the mid-transit time calculated with a TTV model at the specific epoch of our RM observations (see Sect.~\ref{sec:WASP47d}).
\end{tablenotes}
\end{threeparttable}
\end{table*}

%%%%%%%%%%%%%%%%%%%%%%%%%%%%%%%%%%%%%%%%%%%%%%%%%%%%%%%%%%%%%%%%%%%%%%%%%%%%%%%%
% WASP-107 b
%%%%%%%%%%%%%%%%%%%%%%%%%%%%%%%%%%%%%%%%%%%%%%%%%%%%%%%%%%%%%%%%%%%%%%%%%%%%%%%%

\begin{table*}  
\caption[]{Properties of the WASP-107 system}
\centering
\begin{threeparttable}
\begin{tabular}{c c c c c}
\hline
\hline

Parameter & Symbol & Value  &  Unit   &   Origin  \\

\hline

\multicolumn{5}{c}{\textit{Stellar parameters}} \\ 

\hline

Spectral type & & K6 & & \citet{Anderson2017} \\

Stellar temperature & $T_{\rm eff}$ & $4425\pm70$ & K &  \citet{Piaulet2021} \\
 
Stellar radius & $R_\star$ & $0.67\pm0.02$ & $R_\odot$ &  \citet{Piaulet2021} \\

Stellar spin inclination & $i_\star$ & $15.1^{+2.1}_{-2.6}$ & deg & This work \\

Stellar equatorial period & $P_{\rm eq}$ & $17.1\pm1.0$ & d & \citet{Anderson2017} \\

Stellar projected velocity & $v_\mathrm{eq} \sin i_\star$ & $0.507^{+0.072}_{-0.086}$ & km s$^{-1}$ & This work \\

Age & $\tau$ & $3.4\pm0.7$ & Gyr & \citet{Piaulet2021} \\

 Limb-darkening coefficients & $u_1$ & 0.771 & & Exofast \\
 & $u_2$ & 0.023 & & \\

\hline

\multicolumn{5}{c}{\textit{Planet b parameters}} \\ 

\hline

Orbital period & $P$ & $5.7214742\pm0.0000043$ & d & \citet{Dai2017} \\

Transit epoch & $T_0$ & $2457584.329897\pm0.000032$ & BJD$_{\rm TDB}$ & \citet{Dai2017} \\

Eccentricity & $e$ & $0.0$ (fixed) & & \citet{Allart2019} \\ 

Argument of periastron & $\omega$ & 90 (fixed) & deg & \citet{Allart2019} \\ 

Stellar reflex velocity & $K$ & $14.1\pm0.8$ & m s$^{-1}$ & \citet{Piaulet2021} \\

Scaled separation & $a/R_\star$ & $18.02\pm0.27$ & & \citet{Allart2019} \\

Orbital inclination & $i$ & $89.560\pm0.078$ & & \citet{Mocnik2017} \\

Impact parameter & $b$ & $0.139\pm0.024$ & & \citet{Mocnik2017} \\ 

Transit duration & $T_{14}$ & $2.7528\pm0.0072$ & h & \citet{Anderson2017} \\

Planet-to-star radius ratio & $R_{\rm p}/R_\star$ & $0.142988\pm0.00012 $ & & \citet{Spake2018} \\

Projected spin--orbit angle & $\lambda$ & $-158.0^{+15.2}_{-18.5}$ & deg & This work \\

3D spin--orbit angle & $\psi_N$ & 103.0$\stackrel{+1.7}{_{-1.8}}$ & deg & This work \\
               & $\psi_S$ & 103.9$\stackrel{+1.7}{_{-1.8}}$& deg &  \\
               & $\psi$ &   103.5$\stackrel{+1.7}{_{-1.8}}$ & deg &  \\

\hline

\multicolumn{5}{c}{\textit{Planet c parameters}} \\ 

\hline

Orbital period & $P$ & $1088^{+15}_{-16}$ & d & \citet{Piaulet2021} \\

Transit epoch & $T_0$ & $2458520^{+60}_{-70}$ & BJD$_{\rm TDB}$ & \citet{Piaulet2021} \\

Eccentricity & $e$ & $0.28\pm0.07$ & & \citet{Piaulet2021} \\

Argument of periastron & $\omega$ & $-120^{+30}_{-20}$ & deg & \citet{Piaulet2021} \\

Stellar reflex velocity & $K$ & $9.6^{+1.1}_{-1.0}$ & m s$^{-1}$ & \citet{Piaulet2021} \\
	    
\hline
\end{tabular}
\begin{tablenotes}[para,flushleft]
\end{tablenotes}
\end{threeparttable}
\end{table*}

\newpage

%%%%%%%%%%%%%%%%%%%%%%%%%%%%%%%%%%%%%%%%%%%%%%%%%%%%%%%%%%%%%%%%%%%%%%%%%%%%%%%%
% WASP-156 b
%%%%%%%%%%%%%%%%%%%%%%%%%%%%%%%%%%%%%%%%%%%%%%%%%%%%%%%%%%%%%%%%%%%%%%%%%%%%%%%%

\begin{table*}  
\caption[]{Properties of the WASP-156 system}
\centering
\begin{threeparttable}
\begin{tabular}{c c c c c}
\hline
\hline

Parameter & Symbol & Value  &  Unit   &   Origin  \\

\hline

\multicolumn{5}{c}{\textit{Stellar parameters}} \\ 

\hline

Spectral type & & K3 & & \citet{Demangeon2018} \\

Stellar temperature & $T_{\rm eff}$ & $4910\pm61$ & K &  \citet{Demangeon2018} \\
 
Stellar radius & $R_\star$ & $0.76\pm0.03$ & $R_\odot$ &  \citet{Demangeon2018} \\

Stellar mass & $M_\star$ & $0.842\pm0.052$ & $M_\odot$ &  \citet{Demangeon2018} \\

Stellar spin inclination & $i_\star$ & --- & deg & --- \\

Stellar equatorial period & $P_{\rm eq}$ & --- & d & --- \\

Stellar projected velocity & $v_\mathrm{eq} \sin i_\star$ & $3.17^{+0.73}_{-0.84}$ & km s$^{-1}$ & This work \\

Age & $\tau$ & $6.4\pm4.0$ & Gyr & \citet{Demangeon2018} \\

 Limb-darkening coefficients & $u_1$ & 0.693 & & Exofast \\
 & $u_2$ & 0.092 & & \\
 
\hline

\multicolumn{5}{c}{\textit{Planet b parameters}} \\ 

\hline

Orbital period & $P$ & $3.8361672\pm0.0000019$ & d & This work (STELLA) \\

Transit epoch & $T_0$ & $2458644.30467\pm0.00045$ & BJD$_{\rm TDB}$ & This work (STELLA) \\

Eccentricity & $e$ & $	<0.007$ & & \citet{Demangeon2018} \\

Argument of periastron & $\omega$ & --- & deg & \citet{Demangeon2018} \\

Stellar reflex velocity & $K$ & $19\pm1$ & m s$^{-1}$ & \citet{Demangeon2018} \\

Scaled separation & $a/R_\star$ & $12.748^{+0.025}_{-0.027}$ & & \citet{Saha2021} \\

Orbital inclination & $i$ & $88.902^{+0.033}_{-0.028}$ & & \citet{Saha2021} \\

Impact parameter & $b$ & $0.2442^{+0.0061}_{-0.0073}$ & & \citet{Saha2021} \\

Transit duration & $T_{14}$ & $2.3926^{+0.0049}_{-0.0055}$ & & \citet{Saha2021} \\

Planet-to-star radius ratio & $R_{\rm p}/R_\star$ & $0.0685^{+0.0012}_{-0.0008}$ & & \citet{Demangeon2018} \\
 & & $0.067654^{+0.000082}_{-0.000060}$ & & \citet{Saha2021} \\

Projected spin--orbit angle & $\lambda$ & $105.7^{+14.0}_{-14.4}$ & deg & This work \\

\hline
\end{tabular}
\end{threeparttable}
\end{table*}

\newpage

%%%%%%%%%%%%%%%%%%%%%%%%%%%%%%%%%%%%%%%%%%%%%%%%%%%%%%%%%%%%%%%%%%%%%%%%%%%%%%%%
% WASP-166 b
%%%%%%%%%%%%%%%%%%%%%%%%%%%%%%%%%%%%%%%%%%%%%%%%%%%%%%%%%%%%%%%%%%%%%%%%%%%%%%%%

\begin{table*}  
\caption[]{Properties of the WASP-166 system}
\centering
\begin{threeparttable}
\begin{tabular}{c c c c c}
\hline
\hline

Parameter & Symbol & Value  &  Unit   &   Origin  \\

\hline

\multicolumn{5}{c}{\textit{Stellar parameters}} \\ 

\hline

Spectral type & & F9 & & \citet{Hellier2019} \\

Stellar temperature & $T_{\rm eff}$ & $	6050\pm50$ & K &  \citet{Hellier2019} \\
 
Stellar radius & $R_\star$ & $1.22\pm0.06$ & $R_\odot$ &  \citet{Hellier2019} \\

Stellar spin inclination & $i_\star$ & $87.9^{+22.9}_{-19.3}$ & deg & This work \\

Stellar equatorial period & $P_{\rm eq}$ & $12.3\pm1.9$ & d & \citet{Hellier2019} \\

Stellar projected velocity & $v_\mathrm{eq} \sin i_\star$ & $5.4\pm0.14$ & km s$^{-1}$ & This work \\

Age & $\tau$ & $2.1\pm0.9$ & Gyr & \citet{Hellier2019} \\

Limb-darkening coefficients & $u_1$ & 0.416 & & Exofast \\
 & $u_2$ & 0.287 & & \\
 
\hline

\multicolumn{5}{c}{\textit{Planet b parameters}} \\ 

\hline

Orbital period & $P$ & $5.4435402^{+0.0000027}_{-0.0000030}$ & d & \citet{Bryant2020} \\

Transit epoch & $T_0$ & $2458540.739389^{+0.000721}_{-0.000670}$ & BJD$_{\rm TDB}$ & \citet{Bryant2020} \\

Eccentricity & $e$ & $0.0$ (fixed) & & \citet{Hellier2019} \\

Argument of periastron & $\omega$ & --- & deg & \citet{Hellier2019} \\

Stellar reflex velocity & $K$ & $10.4\pm0.4$ & m s$^{-1}$ & \citet{Hellier2019} \\

Scaled separation & $a/R_\star$ & $11.14^{+0.42}_{-0.50}$ & & \citet{Bryant2020} \\

Orbital inclination & $i$ & $87.95^{+0.62}_{-0.59}$ & deg & \citet{Bryant2020} \\

Impact parameter & $b$ & $0.398^{+0.093}_{-0.111}$ & & \citet{Bryant2020} \\

Transit duration & $T_{14}$ & $3.600\pm0.024$ & h & \citet{Hellier2019} \\

Planet-to-star radius ratio & $R_{\rm p}/R_\star$ & $0.0515\pm0.0011$ & & \citet{Bryant2020} \\

Projected spin--orbit angle & $\lambda$ & $-0.7\pm1.6$ & deg & This work \\

3D spin--orbit angle & $\psi$ $^{\dagger}$ & $<$22.0 (1$\sigma$)  & deg & This work  \\
						&   &	    $<$56.2 (3$\sigma$) &  &  \\

\hline
\end{tabular}
\begin{tablenotes}[para,flushleft]
$^{\dagger}$ WASP-166 is seen edge-on with $\lambda\sim$0\,$^{\circ}$, so that $\psi_N \equiv \psi_S$
\end{tablenotes}
\end{threeparttable}
\end{table*}

\section{Residual CCF maps}
\label{apn:CCFres}

\begin{figure*}
\begin{minipage}[tbh!]{\textwidth}
\includegraphics[trim=0cm 0cm 0cm 0cm,clip=true,width=\columnwidth]{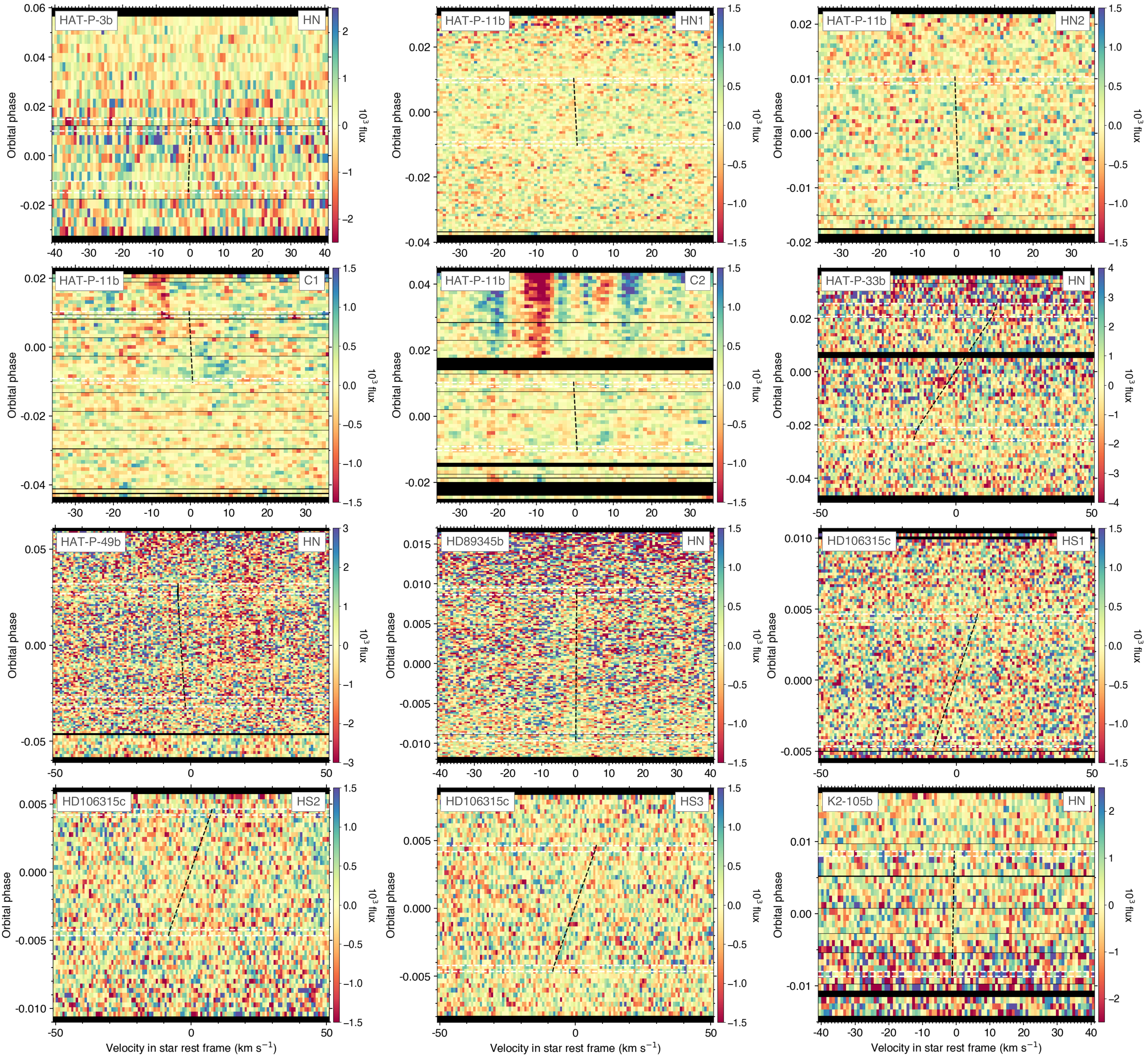}  
\centering
\end{minipage}
\caption[]{Residual maps between the master-out and individual CCF$_{\rm DI}$ (outside of transit) and between the CCF$_{\rm Intr}$ and their best-fit RMR model (during transit). Transit contacts are shown as white dashed lines. Values are colored as a function of the residual flux and plotted as a function of RV in the star rest frame (in abscissa) and orbital phase (in ordinate). The black dashed line shows the stellar surface RV model from the RMR best fit.}
\label{fig:RMR_resA}
\end{figure*}

\begin{figure*}
\ContinuedFloat
\begin{minipage}[tbh!]{\textwidth}
\includegraphics[trim=0cm 0cm 0cm 0cm,clip=true,width=\columnwidth]{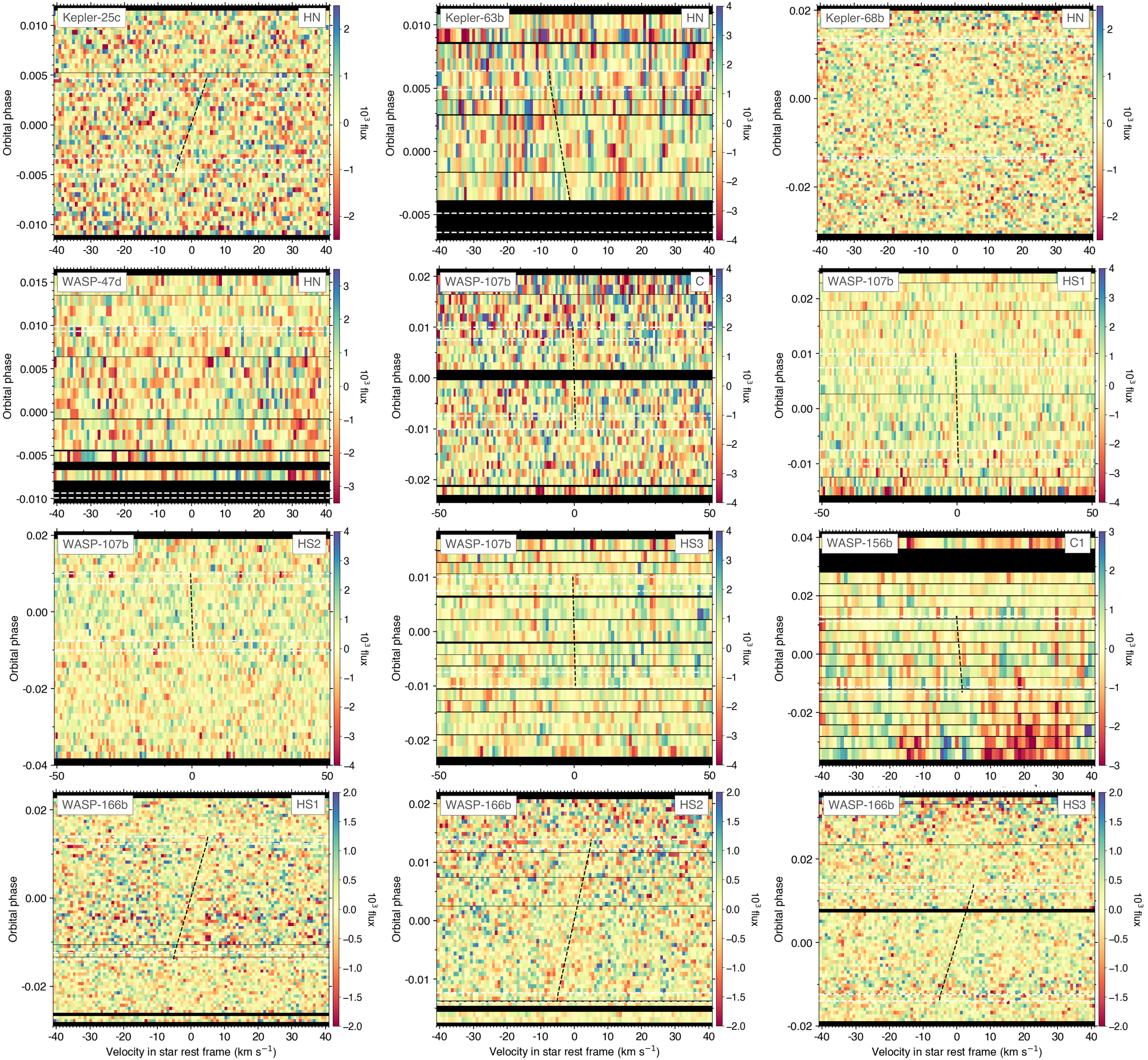}   %plus leger
\centering
\end{minipage}
\caption[]{Continued}
\label{fig:RMR_resB}
\end{figure*}

\end{appendix}

\end{document}